\documentclass[12pt,a4paper, fleqn]{article}
\usepackage[textheight=23.5cm, textwidth=16cm]{geometry}
\usepackage[american]{babel}
\usepackage[utf8]{inputenc}
\usepackage{amssymb}
\usepackage{amsfonts}
\usepackage{amsmath}
\usepackage{bm}
\usepackage{graphicx}
\usepackage{cite}
\usepackage{rotating}
\usepackage{wrapfig}
\usepackage{hyperref}

\usepackage{dcolumn} 
\newcolumntype{d}{D{.}{.}{2}}
\newcolumntype{e}{D{.}{.}{3}}
\newcolumntype{f}{D{.}{.}{4}}

\def\braket#1{\mathinner{\langle{#1}\rangle}}

\begin{document}
\begin{center}

{\LARGE\bf
  Spin in Density-Functional Theory
}


{\large 
Christoph R. Jacob\footnote{E-Mail: christoph.jacob@kit.edu} and
Markus Reiher\footnote{E-Mail: markus.reiher@phys.chem.ethz.ch}
}\\[1ex]

$^1$Karlsruhe Institute of Technology (KIT), \\Center for Functional Nanostructures 
and Institute of Physical Chemistry,\\
Wolfgang-Gaede-Stra\ss{}e 1a, 76131 Karlsruhe, Germany\\[2ex]

$^2$ETH Zurich, Laboratorium f\"ur Physikalische Chemie, \\
Wolfgang-Pauli-Strasse 10, 8093 Zurich, Switzerland
\end{center}


\begin{abstract}

The accurate description of open-shell molecules, in particular of transition metal complexes 
and clusters, is still an important challenge for quantum chemistry. While density-functional
theory (DFT) is widely applied in this area, the sometimes severe limitations of its currently
available approximate realizations often preclude its application as a predictive theory.
Here, we review the foundations of DFT applied to open-shell systems, both within the 
nonrelativistic and the relativistic framework. In particular,  we provide an in-depth 
discussion  of the exact theory, with
a focus on the role of the spin density and possibilities for targeting specific spin states.
It turns out that different options exist for setting up Kohn--Sham DFT schemes for open-shell 
systems, which imply different definitions of the exchange--correlation energy functional and 
lead to different exact conditions on this functional. Finally, we suggest possible 
directions for future developments.
 
\end{abstract}

\vfil

\begin{tabbing}
Date:   \quad \= October 25, 2012 \\
Status:       \>  published in \textit{Int. J. Quantum Chem.} \textbf{112}, 3661--3684 (2012).\\
DOI: \> \url{http://dx.doi.org/10.1002/qua.24309}
\end{tabbing}

\newpage

\vspace*{6ex}
\begin{center}
\textbf{\Large Table of Contents Graphics and Text}

\vspace{2ex}
\includegraphics{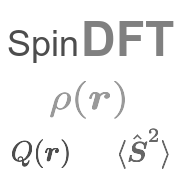}
\end{center}

In this Tutorial Review, we outline the foundations of density-functional theory (DFT) applied
to open-shell systems, both in the non-relativistic case and within the relativistic theory. The role 
of the spin density as well as possibilities for targeting specific spin states are discussed, and
we suggest some possible future directions for Spin-DFT.

\hspace{1cm}

relativistic quantum chemistry, magnetic interactions

\newpage
\section{Introduction}

Open-shell molecules such as, for example, radicals or transition metal complexes and clusters, feature 
a measurable magnetic moment that originates from their electronic structure. In fact, the electronic spin 
gives rise to a magnetic moment that makes such molecular systems functional for various purposes. For 
instance, organic radicals can be employed as spin probes in biomolecules \cite{jeschke_distance_2007}
and are of interest as building blocks for molecular spintronics devices \cite{herrmann_organic_2010,
herrmann_designing_2011}, single-molecule magnets have the potential to act as molecular 
qubits for quantum information processing \cite{timco_engineering_2009}, and
open-shell transition metal compounds serve as catalytic centers in (bio-)inorganic 
chemistry \cite{mcevoy_water-splitting_2006,vignais_occurrence_2007,stiebritz_hydrogenases_2012,
hu_decoding_2010}, where a change in the spin state can be an essential step in the catalytic 
cycle \cite{schrder_two-state_2000}.

Consequently, a first-principles theory that is useful for descriptive and analytic purposes and that  
has the potential to be a predictive tool in theoretical studies on such chemical systems \textit{must} 
consider the spin properties of the electronic structure. While for closed-shell systems, quantum
chemical methods --- both wavefunction theory for accurate calculations on small molecules \cite{helgaker-book}
and density-functional theory (DFT) for studies on complex chemical systems \cite{koch-holthausen} --- offer
such predictive tools, the situation is less satisfactory for open-shell systems, in particular for transition 
metal complexes and clusters \cite{reiher_theoretical_2009}.

With wavefunction based methods, a multi-reference treatment is in general mandatory for open-shell systems.
In particular, the complete active space self-consistent field (CASSCF) method, usually in combination with second-order
perturbation theory (CASPT2),  has been employed to study transition metal complexes (for examples, see
Refs.~\cite{roos_not_2008,rado_binding_2008,rado_electronic_2010,sala_water-oxidation_2010,
planas_electronic_2011,vigara_experimental_2012}). 
However, the factorial scaling with the size of the active space puts rather
severe limits on the size of the active space, which prevents most applications to polynuclear transition
metal complexes and clusters. Novel approaches, such as the density matrix renormalization group (DMRG)
algorithm \cite{chan_introduction_2008,marti_density_2010} and its generalizations \cite{marti_new_2011}
might make it possible to overcome this limitation, although the molecular sizes that
can be studied are clearly much smaller compared to those accessible to DFT methods.

Therefore, DFT is usually the method of choice in theoretical studies of transition-metal
catalysis as well as molecular and spectroscopic properties of open-shell molecular systems \cite{frenking_nature_2000,
ziegler_theoretical_2005,neese_prediction_2009,podewitz_spin_2010}. Despite much success, it has also
become clear that for open-shell systems, DFT with the currently available approximate functionals shows
a number of shortcomings. In addition to inaccuracies in predicting energies, geometries, and molecular 
properties (for a case study, see, e.g., Ref.~\cite{podewitz_density_2011} and for overviews, see, e.g., 
Refs.~\cite{koch-holthausen,ghosh_just_2006,cramer_density_2009}), a severe limitation are unsystematic errors 
in the prediction  of the relative energies of different spin states \cite{reiher_reparameterization_2001,reiher_theoretical_2002,
ghosh_high-level_2003,harvey_dft_2004,herrmann_spin_2006,swart_accurate_2008,ye_accurate_2010,
swart_spin_2012}.
Moreover, the spin density --- which serves as an additional fundamental quantity in the spin-DFT formalism 
commonly employed for open-shell systems --- is qualitatively incorrect in some cases \cite{conradie_dft_2007,
rado_electronic_2010,boguslawski_can_2011,boguslawski_accurate_2012}. 
To make things even worse, the treatment of low-spin states 
usually requires the use of a broken-symmetry description \cite{noodleman_valence_1981,jonkers_broken_1982,
noodleman_models_1985,noodleman_ligand_1986}, which provides an unphysical spin  density by construction 
(see, e.g., Refs.~\cite{reiher_definition_2007,neese_prediction_2009} for a discussion). This precludes the simple 
prediction of spectroscopic properties depending on the spin density (for schemes to address this difficulty, see, 
e.g., Refs.~\cite{noodleman_orbital_1995,van_wllen_broken_2009,pantazis_new_2009,schinzel_density_2010}).

Consequently, the development of better approximate DFT methods for open-shell systems is currently still 
one of the most important and challenging topics in theoretical chemistry \cite{reiher_theoretical_2009,
cramer_density_2009,cohen_challenges_2012}. To make progress is this area, it is important to understand
the exact theory underlying DFT for open-shell systems. While for the closed-shell case, exhaustive presentations
of this theory exists in several textbooks \cite{yang-parr,gross_density_1990,fiolhais_primer_2003,
engel_density_2011}, this is not the case for open-shell systems, which are often only mentioned in 
passing in these accounts. Here, we attempt to close this gap by reviewing the foundations of DFT for 
open-shell systems. In our presentation, we will pay particular attention to the role of the spin-density 
in DFT and to possibilities for targeting different spin states within the exact theory. 
Even though we will not discuss the currently available approximations in detail, we believe that for the future 
development of better approximations, it is crucial to know which exact theory is to be approximated. This is
also a prerequisite for deriving exact conditions on the approximate functionals, for setting up model systems
that can be treated exactly, and for obtaining benchmark results from accurate wave-function theory calculations.

This work is organized as follows. First, we introduce spin in the context of nonrelativistic quantum chemistry
in Section~\ref{sec:spin-nonrel}. This is followed by a discussion of Hohenberg--Kohn (HK) DFT, highlighting the role
of the spin density and of spin states for open-shell systems in Section~\ref{sec:hk-dft}. Next, the treatment of spin
in the Kohn--Sham (KS) framework of DFT is reviewed in Section~\ref{sec:ks-dft}. It turns out that different options
exist for deriving KS-DFT for open-shell systems, which are discussed and compared in detail.
For completeness, in Sec.~\ref{sec:relativistic-dft} we discuss DFT within the relativistic framework where
spin is no longer a good quantum number. Finally, some possible future directions for DFT applied to open-shell 
systems are outlined in Section~\ref{sec:future}.

\section{Spin in Nonrelativistic Quantum Chemistry}
\label{sec:spin-nonrel}

\subsection{Spin Structure of the One-Electron Wavefunction}

The nonrelativistic quantum-mechanical equation of  motion for a single electron in an external 
electrostatic potential $v_\text{ext}(\boldsymbol{r})$ is provided by the time-dependent Schr\"odinger 
equation (SE), which in Gaussian units reads,
\begin{equation}
  \label{eq:sgl-td}
  \hat{h} \, \psi(\boldsymbol{r},t) = \bigl[ \hat{T} + q_e \, v_\text{ext}(\boldsymbol{r} )\bigr] \psi(\boldsymbol{r},t)
  = {\rm i}\hbar \frac{\partial}{\partial t} \psi(\boldsymbol{r},t),
\end{equation}
with the kinetic energy operator $\hat{T} = \dfrac{\hat{\boldsymbol{p}}^2}{2 m_e} = -\dfrac{\hbar^2}{2 m_e}\Delta$, 
where $\hat{\boldsymbol{p}} = -{\rm i}\hbar \boldsymbol{\nabla}$ is the momentum operator and $m_e$ and $q_e$ are
the mass and the charge of the electron, respectively. Stationary states can then be obtained from the time-independent 
Schr\"odinger equation,
\begin{equation}
\label{eq:sgl-stat}
  \hat{h} \, \psi(\boldsymbol{r}) = \bigl[ \hat{T} + q_e \, v_\text{ext}(\boldsymbol{r} )\bigr] \psi(\boldsymbol{r})
  = E \, \psi(\boldsymbol{r}).
\end{equation}

In a nonrelativistic framework, spin is introduced in an \textit{ad hoc} fashion by employing a two-component
representation for the wavefunction \cite{pauli_zur_1927,cohen-tannoudji-1}, i.e.,
\begin{equation}
  \label{eq:wavefunction-twocomponent}
  \psi(\boldsymbol{r}) = \begin{pmatrix} \psi_\alpha(\boldsymbol{r}) \\ 
                                         \psi_\beta(\boldsymbol{r}) \end{pmatrix}.
\end{equation}
For a rigorous introduction of spin in quantum chemistry, it is necessary to start from relativistic quantum mechanics, 
where spin is naturally included in the Dirac equation. This will be discussed later on in Section~\ref{sec:relativistic-dft}.

In the nonrelativistic two-component picture, spin-independent operators --- such as the one-electron Hamiltonian $\hat{h}$ in
Eqs.~\eqref{eq:sgl-td} and~\eqref{eq:sgl-stat} --- act on both of these components, i.e., they are proportional to the
$2 \times 2$ unit matrix $1_2$. The two-component structure of the wavefunction is only probed by 
operators expressed in terms of the Pauli matrices,
\begin{equation}
  \label{eq:pauli-matrices}
  \sigma_x =
  \begin{pmatrix}
    0 & 1 \\
    1 & 0
  \end{pmatrix},
  \quad
  \sigma_y =
  \begin{pmatrix}
    0 & -{\rm i} \\
    {\rm i} & 0
  \end{pmatrix},
  \quad
  \sigma_z =
  \begin{pmatrix}
    1 & 0 \\
    0 & -1
  \end{pmatrix}.
\end{equation}
In particular, the operator corresponding to the electron spin is 
\begin{equation}
  \hat{\boldsymbol{s}} = \frac{\hbar}{2} \boldsymbol{\sigma} = \frac{\hbar}{2} \bigl( \sigma_x,\sigma_y, \sigma_z\bigr)^T
\end{equation}
The three components of this spin operator fulfill the same commutation relations as those of the angular momentum 
operator $\hat{\boldsymbol{l}}$, i.e.,
\begin{equation}
  [\hat{s}_x, \hat{s}_y] = {\rm i}\hbar \, \hat{s}_z, \quad
  [\hat{s}_y, \hat{s}_z] = {\rm i}\hbar \, \hat{s}_x, \quad
  [\hat{s}_x, \hat{s}_z] = {\rm i}\hbar \, \hat{s}_y,
\end{equation}
which is the basis for considering spin as an intrinsic angular momentum vector $\hat{\boldsymbol{s}}$.
For the squared magnitude of the electron spin, one obtains the diagonal operator
\begin{equation}
  \hat{\boldsymbol{s}}^2 = \hat{s}_x^2 + \hat{s}_y^2 + \hat{s}_z^2 = \frac{3}{4} \hbar^2 \, 1_2
\end{equation}
and as for the angular momentum, this operator commutes with each component of the spin 
(i.e., $[\hat{\boldsymbol{s}}^2, \hat{s}_\alpha] = 0$ for $\alpha = x,y,z$).

In nonrelativistic quantum mechanics, one postulates that the spin operator $\hat{\boldsymbol{s}}$
is related to an intrinsic magnetic moment of the electron \cite{pauli_zur_1927,mcweeny_spins_2004,
reiher_relativistic_2009}. This spin magnetic moment is described by the operator
\begin{equation}
  \hat{\boldsymbol{\mu}}_s = -\frac{2 \mu_B}{\hbar} \hat{\boldsymbol{s}} = -\mu_B \, \boldsymbol{\sigma},
\end{equation}
where $\mu_B = \dfrac{|q_e| \hbar}{2 m_e c}$ is the Bohr magneton, $c$ is the speed of light in vacuum, and 
the factor two is the electron  $g$-factor when neglecting quantum electrodynamical effects.

Instead of explicitly writing two-component wavefunctions and $2 \times 2$ matrix operators, in quantum chemistry
it is common to employ a different notation, which will turn out to be particularly convenient for handling 
many-electron systems. Namely, as a shorthand notation, one introduces the orthonormal \textit{spin functions} $\alpha(s)$ 
and $\beta(s)$, which depend on a \textit{spin variable} $s$ \cite{pauli_zur_1927,mcweeny_spins_2004,szabo-ostlund}.
This spin variable can only assume the values $+\tfrac{1}{2}$ and $-\tfrac{1}{2}$, and the spin functions
are defined such that
\begin{align}
  \alpha(+\tfrac{1}{2}) = 1 \quad &\text{and} \quad \alpha(-\tfrac{1}{2}) = 0, \\
  \beta(+\tfrac{1}{2})  = 0 \quad &\text{and} \quad \beta(-\tfrac{1}{2})  = 1. 
\end{align}
Then, the wavefunction of Eq.~\eqref{eq:wavefunction-twocomponent} can be expressed as
\begin{equation}
  \label{eq:wavefunction-spinorbit}
  \psi(\boldsymbol{r}, s) = \psi_\alpha(\boldsymbol{r}) \alpha(s) + \psi_\beta(\boldsymbol{r}) \beta(s), 
\end{equation}
with the first component given by $\psi(\boldsymbol{r}, +\tfrac{1}{2}) = \psi_\alpha(\boldsymbol{r})$ 
and the second component given by $\psi(\boldsymbol{r}, -\tfrac{1}{2}) = \psi_\beta(\boldsymbol{r})$.
It is important to realize that the spin functions $\alpha$ and $\beta$ are merely a way of expressing
two-component wavefunctions, in which the spin variable $s$ has the role of labeling the different components.
Spin-independent operators are then given by a one-component operator acting only on the
parts of the wavefunction that depend on the spatial coordinate $\boldsymbol{r}$, while the spin 
operators $\hat{s}_x$, $\hat{s}_y$, $\hat{s}_z$, and $\hat{\boldsymbol{s}}^2$ act only on the parts 
depending on the spin variable $s$.

The nonrelativistic one-electron Hamiltonian $\hat{h}$ is spin-independent and hence commutes with 
all spin operators, in particular $[\hat{h}, \hat{s}_z]=[\hat{h}, \hat{\boldsymbol{s}}^2] = 0$. Therefore, its eigenfunctions 
can be chosen as eigenfunctions of $\hat{\boldsymbol{s}}^2$ and of $\hat{s}_z$.  The eigenfunctions of $\hat{s}_z$ 
are given by $\alpha(s)$ and $\beta(s)$ and the wavefunctions of an electron which are also 
eigenfunctions of $\hat{s}_z$ are thus of the form
\begin{equation}
  \psi(\boldsymbol{r}, s) = \psi(\boldsymbol{r}) \alpha(s) 
  \quad \text{or} \quad
  \psi(\boldsymbol{r}, s) = \psi(\boldsymbol{r}) \beta(s). 
\end{equation}
For any spatial part of the wavefunction, these two different total wavefunctions are possible, i.e., each 
eigenvalue of the nonrelativistic one-electron Hamiltonian is two-fold degenerate. The first is identified with an $\alpha$- or
``spin-up'' electron ($s_z = +\hbar/2$), whereas the second one corresponds to a $\beta$- or ``spin-down'' 
electron ($s_z = -\hbar/2$). Of course, also linear combinations of these two degenerate eigenfunctions 
are solutions of the Schr\"odinger equation. Nevertheless, for a single electron any eigenfunction can be 
expressed as a product of a spatial part and a spin part.

\subsection{Spin Structure of the Many-Electron Wavefunction}

Within the two-component picture introduced in the previous section, the one-electron Hilbert 
space $^{1}\mathcal{H}$ is spanned by all admissible one-electron wavefunctions (which have two 
components related to the spin of the electron). For an $N$-electron system, the wavefunction 
is an element of the corresponding $N$-electron Hilbert space $^{N}\mathcal{H}$, which is the tensor 
product space of the Hilbert spaces of each electron, i.e.,
\begin{equation}
  ^{N}\mathcal{H} = {^{1}\mathcal{H}(1)} \otimes {^{1}\mathcal{H}(2)} \otimes \dotsm \otimes {^{1}\mathcal{H}(N)},
\end{equation}
where the number given in parentheses designates the corresponding electron. As the one-particle wavefunctions
each have two components, the $N$-electron wavefunctions have $2^N$-components, and operators have the 
dimension $2^N \times 2^N$. For a more detailed discussion of the tensor structure of the many-electron wavefunction,
see, e.g., Refs.~\cite{chan_introduction_2008,marti_density_2010,marti_new_2011} and chapter~8.4 in 
Ref.~\cite{reiher_relativistic_2009}.

As an alternative to explicitly handling many-component wavefunctions, it is again convenient 
to introduce spin coordinates $s_i$ that can be used to distinguish the different 
components \cite{dirac_quantum_1929,mcweeny_methods_1969,mcweeny_spins_2004}. 
Then the $N$-electron wavefunction depends on $N$ spin coordinates in addition to the $N$ spatial coordinates,
\begin{equation}
  \Psi = \Psi(\boldsymbol{r}_1, s_1, \boldsymbol{r}_2, s_2, \dotsc, \boldsymbol{r}_N, s_N)
          = \Psi(\boldsymbol{x}_1, \boldsymbol{x}_2, \dotsc, \boldsymbol{x}_N),
\end{equation}
where $\boldsymbol{x}_i = (\boldsymbol{r}_i, s_i)$ denotes the combination of spatial and spin coordinates.
Each spin coordinate can assume the values $-1/2$ and $+1/2$. The possible combination of 
values for these spin variables each corresponds to one component of the many-electron wavefunction. In total,
 $2^N$ different combinations of values are possible, that are thus used to label the $2^N$ components of the 
 many-electron wavefunction.

The nonrelativistic Hamiltonian describing $N$ electrons in an external electrostatic potential 
$v_\text{ext}(\boldsymbol{r})$ is (in Gaussian units) given by,
\begin{equation}
  \label{eq:many-electron-hamiltonian}
  \hat{H} =  \sum_{i=1}^N \left[ -\frac{\hbar^2}{2 m_e} \Delta_i + q_e \, v_\text{ext}(\boldsymbol{r}_i)  \right] + 
  \sum_{i=1}^N\sum_{j=i+1}^N \frac{q_e^2}{r_{ij}},
\end{equation}
where the Laplace operator $\Delta_i$ acts on the coordinate of the $i$th electron and $r_{ij} = |\boldsymbol{r}_i - \boldsymbol{r}_j|$ 
is the distance between electrons $i$ and $j$.  This Hamiltonian does not contain 
any spin-dependent terms and only acts on the spatial coordinates. 

The operator of the total spin $\hat{\boldsymbol{S}}$ of a many-electron system is obtained \cite{szabo-ostlund,mcweeny_spins_2004} 
by summing the spins of the individual electrons $\hat{\boldsymbol{S}} = \sum_{1=1}^N \hat{\boldsymbol{s}}(s_i)$.
In particular, one has for the $z$-component of the total spin operator,
\begin{equation}
  \hat{S}_z = \sum_{i=1}^N \hat{s}_z(s_i),
\end{equation}
and for the square of the total spin
\begin{align}
  \label{eq:operator-S2}
  \hat{\boldsymbol{S}}^2 = \sum_{i=1}^N \sum_{j=1}^N \hat{\boldsymbol{s}}(s_i) \cdot \hat{\boldsymbol{s}}(s_j)
                         &= \frac{3}{4} N \hbar^2 + 2 \sum_{i=1}^N\sum_{j=i+1}^N \hat{\boldsymbol{s}}(s_i) \cdot \hat{\boldsymbol{s}}(s_j),  
\end{align}
where the first term emerges because any many-electron wavefunction is an eigenfunction of $\hat{\boldsymbol{s}}^2(s_i)$
with eigenvalue $(3/4) \hbar^2$ (i.e., electrons are spin-$1/2$ particles).
Note that $\hat{\boldsymbol{S}}^2$ couples different electrons, i.e., it is a two-electron operator \cite{lwdin_quantum_1955}.

Both $\hat{\boldsymbol{S}}^2$ and $\hat{S}_z$ commute with the Hamiltonian of Eq.~\eqref{eq:many-electron-hamiltonian}
and with each other, i.e.,
\begin{equation}
  [\hat{H}, \hat{\boldsymbol{S}}^2] = [\hat{H}, \hat{S}_z] = [\hat{\boldsymbol{S}}^2, \hat{S}_z] = 0.
\end{equation}
Therefore, the eigenfunctions $\Psi$ of the Hamiltonian can always be chosen as eigenfunctions of $\hat{S}_z$ 
and $\hat{\boldsymbol{S}}^2$ with,
\begin{align}
  \hat{\boldsymbol{S}}^2 \Psi &= S(S+1) \hbar^2 \, \Psi \\
  \hat{S}_z \Psi &= M_S \hbar \, \Psi \quad \text{with} \quad M_S = -S, \dotsc, +S.
\end{align}
In general, eigenfunctions of the Hamiltonian belonging to different eigenvalues of $\hat{\boldsymbol{S}}^2$ 
have different energies, while for each energy eigenvalue there are always $2S+1$ degenerate eigenfunctions 
differing in $M_S$. Since all three components of $\hat{\boldsymbol{S}}$ commute with
the Hamiltonian, any choice of the quantization axis is possible and will lead to identical results.

Finally, the spin structure of the many-electron wavefunction \cite{dirac_quantum_1929} is also determined 
by the Pauli principle \cite{heisenberg-pauli-1,heisenberg-pauli-2}. It requires that the wavefunction is 
antisymmetric (i.e., it has to change sign upon exchange of two electrons). This can be expressed with 
the permutation operator,
\begin{equation}
  \hat{\mathcal{P}}_{ij} \Psi(\dotsc,\boldsymbol{r}_i, s_i, \dotsc, \boldsymbol{r}_j, s_j, \dotsc) 
  = \Psi(\dotsc, \boldsymbol{r}_j, s_j, \dotsc, \boldsymbol{r}_i, s_i, \dotsc),
\end{equation}
as $\hat{\mathcal{P}}_{ij} \Psi = - \Psi$. Here, $\hat{\mathcal{P}}_{ij}$ exchanges both the spatial and the
spin coordinates of electrons $i$ and $j$. To express this requirement in a different form, one can introduce
the antisymmetrizer $\hat{\mathcal{A}}$ defined as 
\begin{equation}
  \hat{\mathcal{A}} = \frac{1}{\sqrt{N!}} \sum_{p=1}^{N!} (-1)^p \ \hat{\mathcal{P}}_p,
\end{equation}
where the permutation operators $\hat{\mathcal{P}}_p$ are ordered such that even numbers $p$ are
assigned to those that are generated by an even number of pair permutations, and odd numbers
denote those generated by an odd number of pair permutations. Then, the requirement that the 
wavefunction is antisymmetric with respect to any pair permutation is equivalent to requiring
$\hat{\mathcal{A}}\Psi = \sqrt{N!} \, \Psi$, i.e., $\Psi$ has to be an eigenfunction of $\hat{\mathcal{A}}$. 

Since this Hamiltonian does not contain terms that couple the spatial and the spin coordinates, one could 
naively expect that --- as in the one-electron case --- the many-electron wavefunction can always be written 
as a product of a part depending on the spatial coordinates and of a part depending on the spin coordinates. 
However, for systems with more than two electrons this is in general not the case.
The antisymmetrizer $\hat{\mathcal{A}}$ then contains a sum of permutation operators each acting on both spatial 
and spin coordinates. Hence, it couples spin and spatial coordinates so that its eigenfunctions cannot be expressed 
as a product of a spatial and a spin part. Thus, the structure of the many-electron wavefunction with respect to 
the exchange of spatial coordinates is dependent on the spin structure \cite{matsen_spin-free_1964,
pauncz_spin_1979,pauncz_symmetric_1995}. This is the most important consequence of the presence of spin in a 
nonrelativistic theory.

\subsection{Spin Structure of the Electron Density and Spin Density}
\label{sec:spinstruct-density}

The (total) electron density $\rho(\boldsymbol{r})$ describes the probability density for finding any 
electron of a many-electron system at position $\boldsymbol{r}$. It can be calculated from the 
wavefunction as
\begin{equation}
  \rho(\boldsymbol{r}) 
       =  N \int |\Psi(\boldsymbol{r}, s_1, \boldsymbol{x}_2, \dotsc, \boldsymbol{x}_N) |^2 
                    \ {\rm d}s_1 {\rm d} \boldsymbol{x}_2 \cdots {\rm d}\boldsymbol{x}_N,
\end{equation}
i.e., by integrating the squared absolute value of the wavefunction over all but one spatial
coordinate. By writing out the integration over the corresponding spin variable explicitly,
\begin{align}
    \rho(\boldsymbol{r})
    =& \, N \int | \Psi(\boldsymbol{r}, +\tfrac{1}{2}, \boldsymbol{r}_2, s_2, \dotsc, \boldsymbol{r}_N, s_N)|^2 
             \ {\rm d}^3r_2 {\rm d}s_2 \dotsm {\rm d}^3r_N {\rm d}s_N \nonumber \\
    &+ \, N \int | \Psi(\boldsymbol{r}, -\tfrac{1}{2}, \boldsymbol{r}_2, s_2, \dotsc, \boldsymbol{r}_N, s_N)|^2 
             \ {\rm d}^3r_2 {\rm d}s_2 \dotsm {\rm d}^3r_N {\rm d}s_N \nonumber \\[1ex]
    =& \, \rho_\alpha(\boldsymbol{r}) + \rho_\beta(\boldsymbol{r}),         
\end{align}
one notices that it is a sum of components $\rho_\alpha(\boldsymbol{r})$ and $\rho_\beta(\boldsymbol{r})$ 
that can be interpreted as the probability densities of finding an $\alpha$- or a $\beta$-spin 
electron \cite{mcweeny_methods_1969,mcweeny_spins_2004}. Their integrals,
\begin{equation}
  N_\alpha = \int \rho_\alpha(\boldsymbol{r}) \, {\rm d}^3r
  \quad \text{and} \qquad
  N_\beta = \int \rho_\beta(\boldsymbol{r}) \, {\rm d}^3r,
\end{equation}
give the number of $\alpha$- and $\beta$-electrons. 

It is then natural to define the spin density, which gives the excess of $\alpha$-electrons at a given point, as
\begin{align}
   Q(\boldsymbol{r}) &= N \int \Psi^*(\boldsymbol{r}, s_1, \boldsymbol{x}_2, \dotsc, \boldsymbol{x}_N)
                    \, \sigma_z(s_1) \, \Psi(\boldsymbol{r}, s_1, \boldsymbol{x}_2, \dotsc, \boldsymbol{x}_N)
                    \ {\rm d}s_1 {\rm d} \boldsymbol{x}_2 \cdots {\rm d}\boldsymbol{x}_N 
   \nonumber \\
   &=  \rho_\alpha(\boldsymbol{r}) - \rho_\beta(\boldsymbol{r}).
\end{align}
Here the Pauli matrix $\sigma_z(s_1)$ operates on the first spin coordinate only. Just as the electron density,
which is probed by X-ray diffraction experiments, the spin density is an observable. The spin density
$Q(\boldsymbol{R}_I)$ at the position of a nucleus $I$ is accessible from electron paramagnetic resonance
(EPR) experiments, where it determines the nuclear hyperfine coupling constants \cite{mcweeny_spins_2004,
jeschke-book,kaupp-book}. Similarly, the spin density at a nucleus can give rise to shifts in paramagnetic 
nuclear magnetic resonance (pNMR) \cite{kaupp-book,rastrelli_predicting_2009,autschbach_calculation_2011,
aquino_scalar_2012}. Full spatially resolved spin densities can be determined in neutron scattering 
experiments \cite{gillet_determination_2007,zheludev_spin_1994,baron_spin-density_1996,
pontillon_magnetization_1999,claiser_combined_2005,zaharko_spin-density_2010}.

The expectation value of a multiplicative one-electron operator such as $\hat{V} = \sum_{i=1}^N v(\boldsymbol{r}_i)$ 
can be calculated directly from the electron density,
\begin{equation}
  \braket{\hat{V}} = \braket{\Psi|\hat{V}|\Psi} = \int \rho(\boldsymbol{r}) v(\boldsymbol{r}) \, {\rm d}^3r,
\end{equation}
i.e., the full wavefunction is not needed. Similarly, expectation values of spin-dependent operators 
expressed only in terms of $\sigma_z$ can be obtained from the spin density. In particular, the expectation 
value of $\hat{S}_z$ is given by
\begin{equation}
  \braket{\hat{S}_z} = \frac{\hbar}{2} \int Q(\boldsymbol{r}) \, {\rm d}^3r,
\end{equation}
and for eigenfunctions of $\hat{S}_z$, one has 
\begin{equation}
 M_S = \frac{1}{2} \int Q(\boldsymbol{r}) \, {\rm d}^3r = \frac{1}{2} (N_\alpha - N_\beta).
\end{equation}
The contribution of the electron spin magnetic moments to the interaction of a molecule with 
an inhomogeneous external magnetic field $\boldsymbol{B}_\text{ext}(\boldsymbol{r})$, the so-called 
spin Zeeman interaction, is determined by the operator \cite{mcweeny_spins_2004}
\begin{equation}
  \label{eq:zeeman-op-nonrel}
  \hat{H}_Z = - \sum_{i=1}^N \boldsymbol{B}_\text{ext}(\boldsymbol{r}_i) \cdot \hat{\boldsymbol{\mu}}_{s}(s_i) 
                    =  \mu_B \sum_{i=1}^N \boldsymbol{B}_\text{ext}(\boldsymbol{r}_i) \cdot \boldsymbol{\sigma}(s_i).
\end{equation}
Hence, for an inhomogeneous  magnetic field in $z$-direction, i.e., 
$\boldsymbol{B}_\text{ext}(\boldsymbol{r}) = \big(0,0,B_z(\boldsymbol{r})\big)^T$, 
the expectation value of the spin Zeeman interaction can be evaluated directly from the spin density as
\begin{equation}
  \label{eq:zeeman-nonrel}
  \braket{\hat{H}_Z} = \mu_B \int Q(\boldsymbol{r} )B_z(\boldsymbol{r}) \, {\rm d}^3r.
\end{equation}

If one considers an eigenfunction of $\hat{S}^2$ with eigenvalue $S(S+1) \hbar^2$, this eigenvalue is $(2S+1)$-fold
degenerate and one can construct a set of $2S+1$ eigenstates of $\hat{S}_z$ with eigenvalues $M_S \hbar$, where
$M_S= -S, \dotsc, +S$. 
The total electron densities $\rho^{M_S}$ and the spin densities $Q^{M_S}(\boldsymbol{r})$ of these $\hat{S}_z$ eigenstates are 
related to each other\cite{mcweeny_density_1961,mcweeny_methods_1969,davidson_reduced_1976}: All $2S+1$ states
share the same total electron density, 
\begin{equation}
  \label{eq:dens-ms-scaling}
   \rho^{M_S}(\boldsymbol{r}) =  \rho^{M_S= S}(\boldsymbol{r})
\end{equation}
and the spin densities are given by
\begin{equation}
  \label{eq:spindens-ms-scaling}
   Q^{M_S}(\boldsymbol{r}) = \left( \frac{M_S}{S} \right) Q^{M_S = S}(\boldsymbol{r})
\end{equation}
where $\rho^{M_S=S}(\boldsymbol{r})$ and $Q^{M_S = S}(\boldsymbol{r})$ are the total electron density and the 
spin density of the state with highest $M_S$, respectively. Hence,  the spin densities have the same functional form and
are connected by a simple scaling. It immediately follows that $Q^{M_S}(\boldsymbol{r}) = - Q^{-M_S}(\boldsymbol{r})$ 
and that the spin density vanishes for states with $M_S = 0$.

\section{Spin in Hohenberg--Kohn DFT}
\label{sec:hk-dft}

Traditionally, quantum chemistry sets out to calculate approximations to the many-electron 
wavefunction $\Psi$ of a molecule in its ground state by minimizing the energy expectation value 
with respect to $\Psi$, under the constraint that $\Psi$ represents a normalized and antisymmetric 
$N$-electron wavefunction, i.e.,
\begin{equation}
  \label{eq:wf-var}
  E_0 = \min_{\Psi^N} \langle \Psi^N | \hat{H} | \Psi^N \rangle
  \qquad \text{with} \qquad \langle\Psi^N|\Psi^N\rangle = 1
  \quad \text{and} \quad \hat{\mathcal{A}} \Psi^N = \sqrt{N!} \, \Psi^N,
\end{equation}
where the nonrelativistic Hamiltonian $\hat{H}$ within the Born--Oppenheimer approximation was given 
in Eq.~\eqref{eq:many-electron-hamiltonian}. In molecular systems, the external potential $v_\text{ext}(\boldsymbol{r})$ 
is given by the Coulomb potential of the nuclei, i.e., $v_\text{ext}(\boldsymbol{r}) = v_\text{nuc}(\boldsymbol{r}) 
= -q_e \sum_I Z_I / |\boldsymbol{r}-\boldsymbol{R}_I|$, where the sum runs over all nuclei with charges $Z_I$ 
at positions $\boldsymbol{R}_I$. Thus, the nonrelativistic molecular Hamiltonian assumes the form,
\begin{equation}
  \hat{H} =  \sum_{i=1}^N -\frac{\hbar^2}{2 m_e} \Delta_i + 
                  \sum_{i=1}^N\sum_{j=i+1}^N \frac{q_e^2}{r_{ij}} + \sum_{i=1}^Nq_e \, v_\text{nuc}(\boldsymbol{r}_i) 
              = \hat{T} + \hat{V}_{ee} + \hat{V}_\text{nuc}
\end{equation}

According to this structure of the Hamiltonian, the energy expectation value in the above minimization is usually split up as
\begin{equation}
  \langle \Psi | \hat{H} | \Psi \rangle = \langle \Psi | \hat{T} | \Psi \rangle + \langle \Psi | \hat{V}_{ee} | \Psi \rangle + \langle \Psi | \hat{V}_\text{nuc} | \Psi \rangle.
\end{equation}
However, the wavefunction itself is not directly needed for calculating these expectation values. The evaluation of the first 
term, corresponding to the kinetic energy, only requires the one-electron reduced density matrix (1-RDM), whereas the second
term describing the electron--electron interaction can be calculated from the diagonal two-electron reduced density (2-RDM)
matrix. Finally, the electron--nuclear attraction energy can be evaluated directly from the multiplicative operator of the 
electron--nuclei Coulomb interaction and from the electron density only as,
\begin{equation}
  \langle \Psi | \hat{V}_\text{nuc} | \Psi \rangle=  q_e \int \rho(\boldsymbol{r}) v_\text{nuc}(\boldsymbol{r}) \, {\rm d}^3r.
\end{equation}
This can be exploited  by performing the minimization with respect to the 2-RDM directly \cite{mazziotti_two-electron_2011}, but 
difficulties arise in enforcing that the 2-RDM corresponds to an actual antisymmetric $N$-electron wavefunction. 

Density-functional theory (DFT) \cite{yang-parr,gross_density_1990,fiolhais_primer_2003,engel_density_2011} provides 
the theoretical framework for calculating the energy expectation value directly from the electron density $\rho(\boldsymbol{r})$ 
only. The foundations of this exact theory will be outlined in the following, focussing on its application to open-shell systems.

\subsection{Hohenberg--Kohn Theorems}

\begin{figure}
  \caption{Connection between the electron density $\rho(\boldsymbol{r})$ and the 
           total energy as stated by the first Hohenberg--Kohn theorem.}
  \label{fig:hk-theorem}

  \vspace{5ex}
  \begin{center}
  \includegraphics{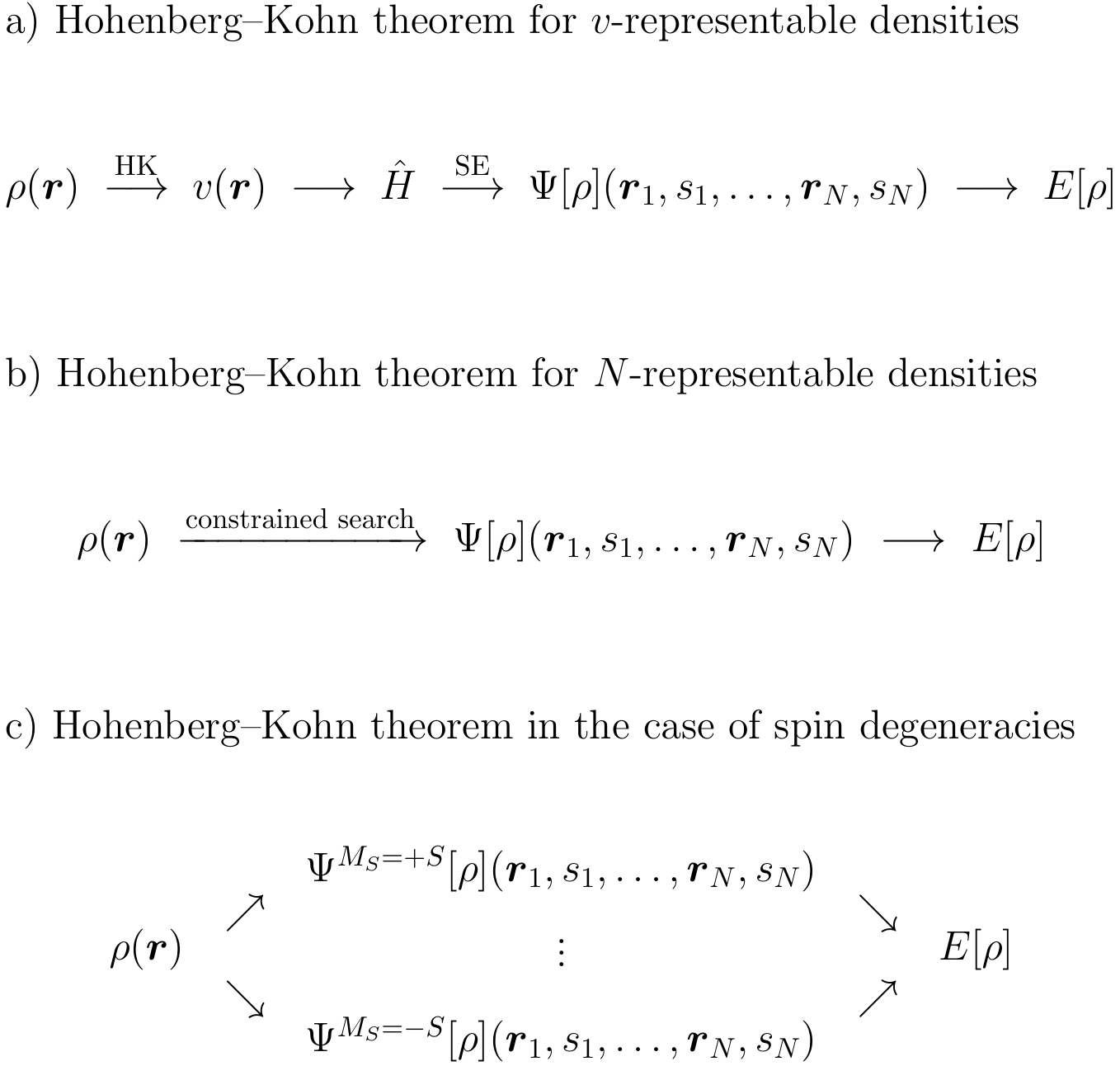}
   \end{center}
   
\end{figure}

The first Hohenberg--Kohn theorem \cite{hohenberg-kohn-1964} states that 
for each electron density $\rho(\boldsymbol{r})$ that can be obtained from a 
ground-state wavefunction (such densities are called $v$-representable densities), 
the external potential $v_\text{ext}(\boldsymbol{r})$ that yields this electron density as the ground state 
when employed in the Hamiltonian of Eq.~\eqref{eq:many-electron-hamiltonian}
is unique up to a constant. Therefore, this potential is a functional of the 
electron density and since it completely determines the Hamiltonian,
also the ground-state wavefunction --- which can in turn be determined by solving the corresponding 
Schr\"odinger equation --- is a functional of the electron density. Furthermore, all observables
of the system, in particular the total energy in a given nuclear potential 
$v_\text{nuc}(\boldsymbol{r})$, can be obtained from this wavefunction. 
This connection between the electron density and the total energy is illustrated
in Fig.~\ref{fig:hk-theorem}a. Therefore, there exists an energy functional $E[\rho]$ 
that relates the electron density to the total energy. 

According to the second HK theorem \cite{hohenberg-kohn-1964}, the ground state energy $E_0$
of a system of electrons in a given nuclear potential $v_\text{nuc}(\boldsymbol{r})$ can be 
determined by minimizing the total energy functional
\begin{equation}
  E_0 = \min_\rho E[\rho],
\end{equation}
with 
\begin{equation}
  \label{eq:hk-energy}
  E[\rho] = q_e \!\! \int \rho(\boldsymbol{r}) \, v_\text{nuc}(\boldsymbol{r}) \, {\rm d}^3r
              + F_\text{HK}[\rho],
\end{equation}
under the constraint that $\rho(\boldsymbol{r})$ integrates to $N$ electrons. The electron density for which 
this minimum is achieved is the ground-state electron density $\rho_0(\boldsymbol{r})$. In this equation, the total 
energy functional has been split into a system-specific part (the first term), depending on the nuclear potential, 
and a system-independent part (the second term), which is called the \textit{universal Hohenberg--Kohn 
functional} $F_\text{HK}[\rho]$.

Following the Levy constrained-search formulation of DFT \cite{levy-1979,levy-1982},
this universal HK functional is given by
\begin{equation}
  \label{eq:hk-levy}
  F_\text{HK}[\rho] = \min_{\Psi \rightarrow \rho} 
                     \left< \Psi \left| \hat{T} + \hat{V}_\text{ee} \right| \Psi \right>,
\end{equation}
where $\hat{T}$ and $\hat{V}_\text{ee}$ are the operators of the kinetic energy and of the
electron--electron repulsion energy, respectively. The minimization runs over all
wavefunctions $\Psi$ that yield the target electron density $\rho$. From these wavefunctions, 
the one with the lowest expectation value of $\hat{T}+\hat{V}_{ee}$ is chosen.
The minimization of $E[\rho]$ will lead to the exact ground-state electron density $\rho_0$,
and the exact ground-state wavefunction $\Psi_0$ is the one for which the minimum in
Eq.~\eqref{eq:hk-levy} is obtained. 
This (nonrelativistic) ground-state wavefunction $\Psi_0$ has to be an eigenfunction of $\hat{\boldsymbol{S}}^2$. 
Therefore, it is sufficient to restrict the constrained search to 
wavefunctions that are eigenfunctions of $\hat{\boldsymbol{S}}^2$. 

Using the Levy constrained search for defining the HK functional also extends  
the domain in which the above functionals are defined from $v$-representable densities 
(i.e., densities that are obtained from a ground state wavefunction) to $N$-representable densities (i.e., 
densities that are obtained from \textit{any} wavefunction, not necessarily a ground state). The resulting
generalization of the first HK theorem is illustrated in Fig.~\ref{fig:hk-theorem}b. More details on the 
foundations of HK-DFT and on the more general definition of the universal
HK functional introduced by Lieb \cite{lieb_density_1983} can be found in dedicated reviews 
on these topics \cite{gross_density_1990,van_leeuwen_density_2003,eschrig_fundamentals_2003}.

Initially, Hohenberg and Kohn explicitly excluded degeneracies in their derivation of the HK theorems\cite{hohenberg-kohn-1964}. 
However, a state with $S>0$ will be $(2S+1)$-fold degenerate, with the different degenerate wavefunctions $\Psi^{M_S}$ corresponding 
to $M_S = -S, \dotsc, S$. The necessary generalization of the HK theorems is possible in a straightforward way\cite{kohn_density_1985}. 
The degenerate wavefunctions $\Psi^{M_S}$ and all their linear combinations share the same electron density. For a given
$\rho$, the minimum in Eq.~\eqref{eq:hk-levy} is achieved for all degenerate wavefunctions spanned by the $\hat{S}_z$-eigenfunctions 
$\Psi^{M_S}$. This is illustrated in Figure~\ref{fig:hk-theorem}c. Therefore, the minimization of the total energy functional $E[\rho]$ will 
still lead to a unique ground-state density $\rho_0$. If it is necessary to obtain a unique minimizing wavefunction, the 
constrained search can be restricted to wavefunctions corresponding to a specific value of $M_S$. In this case, the 
wavefunction is again uniquely determined by the electron density.

\subsection{Spin Density in Hohenberg--Kohn DFT}
\label{sec:hk-spindens}

According to the HK theorem only the total electron density is required for obtaining the 
exact ground-state energy and electron density \cite{hohenberg-kohn-1964,perdew_fundamental_2009}. 
Therefore, irrespective of the spin state,
the spin density $Q(\boldsymbol{r})$ or the individual $\alpha$-electron and $\beta$-electron 
densities $\rho_{\alpha}(\boldsymbol{r})$ and $\rho_{\beta}(\boldsymbol{r})$ are not required 
during the minimization of the total energy functional, and the ground-state spin density 
$Q_0(\boldsymbol{r})$ is not directly available.
However, if a suitable value of $M_S$ is chosen, the wavefunction $\Psi^{M_S}$ is uniquely
determined by the total density $\rho(\boldsymbol{r})$, i.e., $\Psi^{M_S} = \Psi^{M_S}[\rho]$. 
From this wavefunction, the ground-state $\alpha$-electron and $\beta$-electron densities,
\begin{align}
  \rho_\alpha^{M_S}[\rho](\boldsymbol{r}) 
                 &= N \int \left| \Psi^{M_S}[\rho](\boldsymbol{r}, +\tfrac{1}{2}, \boldsymbol{r}_2, s_2, \dotsc) \right|^2 
                    \ {\rm d}^3r_2 {\rm d}s_2 \cdots {\rm d}^3r_N {\rm d}s_N \\
  \rho_\beta^{M_S}[\rho](\boldsymbol{r})  
                &= N \int \left| \Psi^{M_S}[\rho](\boldsymbol{r}, -\tfrac{1}{2}, \boldsymbol{r}_2, s_2, \dotsc) \right|^2
                    \ {\rm d}^3r_2 {\rm d}s_2 \cdots {\rm d}^3r_N {\rm d}s_N,
\end{align}
as well as the corresponding spin density
\begin{equation}
  Q^{M_S}[\rho](\boldsymbol{r}) =  \rho_\alpha^{M_S}[\rho](\boldsymbol{r}) - \rho_\beta^{M_S}[\rho](\boldsymbol{r})
\end{equation}
can be calculated. In these expressions, the superscript $M_S$ indicates that a specific value of $M_S$ has 
to be selected if one is interested in individual $\alpha$- and $\beta$-densities or in the spin density. 
The chosen value of $M_S$ fixes the number of $\alpha$- and $\beta$-electrons by $M_S = \frac{1}{2} (N^\alpha - N^\beta)$.

In order to generalize HK-DFT to use the individual $\alpha$- and $\beta$-densities instead of the total
density only, the minimization of the total energy can be rewritten as \cite{perdew_self-interaction_1981,
yang-parr},
\begin{align}
  E_0  = \min_{\rho} E[\rho] 
      &= \min_{\rho} \left\{
                  q_e \!\! \int \rho(\boldsymbol{r}) v_\text{nuc}(\boldsymbol{r}) \ {\rm d}^3r 
               +  \min_{\Psi \rightarrow \rho} 
                     \left< \Psi \left| \hat{T} + \hat{V}_\text{ee} \right| \Psi \right> 
                                \right\} \nonumber \\
 & = \min_{\rho} \left\{ q_e \!\! \int \rho(\boldsymbol{r}) v_\text{nuc}(\boldsymbol{r}) \ {\rm d}^3r 
        +  \min_{\rho^\alpha, \rho^\beta \rightarrow \rho} \left[ \min_{\Psi \rightarrow \rho^\alpha,\rho^\beta} 
                     \left< \Psi \left| \hat{T} + \hat{V}_\text{ee} \right| \Psi \right> 
                                \right] \right\},
\end{align}
and by removing the outer minimization with respect to $\rho$, one obtains
\begin{equation}
\label{eq:min-toten-spinres}
  E_0 
      = \min_{\rho^\alpha, \rho^\beta} \left\{ 
        q_e \!\! \int \bigl[\rho^\alpha(\boldsymbol{r}) + \rho^\beta(\boldsymbol{r})\bigr] 
              v_\text{nuc}(\boldsymbol{r}) \ {\rm d}^3r 
        + \min_{\Psi \rightarrow \rho^\alpha,\rho^\beta} 
                     \left< \Psi \left| \hat{T} + \hat{V}_\text{ee} \right| \Psi \right> \right\}.
\end{equation}
This defines a universal HK functional in terms of the $\alpha$-electron
 and $\beta$-electron densities,
\begin{equation}
  \label{eq:hk-alpha-beta}
  F_\text{HK}[\rho^\alpha, \rho^\beta] 
     = \min_{\Psi \rightarrow \rho^\alpha, \rho^\beta} 
       \left< \Psi \left| \hat{T} + \hat{V}_\text{ee} \right| \Psi \right>.
\end{equation}
In contrast to the HK functional of Eq.~\eqref{eq:hk-levy}, the constrained search in this spin-resolved HK functional 
now runs over all wavefunctions corresponding to a given pair of $\rho^\alpha$ and $\rho^\beta$.
By minimizing the generalized total energy functional [cf. Eq.~\eqref{eq:min-toten-spinres}],
\begin{equation}
  \label{eq:toten-spinresolved}
  E[\rho^\alpha, \rho^\beta] = 
      q_e \!\! \int \bigl[\rho^\alpha(\boldsymbol{r}) + \rho^\beta(\boldsymbol{r})\bigr]
           v_\text{nuc}(\boldsymbol{r}) \ {\rm d}^3r 
      + F_\text{HK}[\rho^\alpha, \rho^\beta],
\end{equation}
under the constraint that $\rho^\alpha$ and $\rho^\beta$ integrate to $N^\alpha$ and $N^\beta$ electrons, 
respectively, it is then possible to obtain the ground-state $\alpha$-electron and $\beta$-electron 
densities $\rho^\alpha_0$ and $\rho^\beta_0$. Again, in this minimization a specific value 
of $M_S = \frac{1}{2}(N^\alpha - N^\beta)$ has to be selected.

Instead of using $\rho^\alpha$ and $\rho^\beta$, it is also possible to employ the total electron 
density $\rho = \rho^\alpha + \rho^\beta$ and the spin-density $Q = \rho^\alpha - \rho^\beta$
as variables of the spin-resolved HK functional. This gives
\begin{equation}
  \label{eq:fhk-spin-resolved}
  F_\text{HK}[\rho,Q] 
            = \min_{\Psi \rightarrow \rho, Q} 
               \left< \Psi \left| \hat{T} + \hat{V}_\text{ee} \right| \Psi \right>.
\end{equation}
While usually --- in particular in practical applications of spin-DFT --- it is more common to employ
the $\alpha$- and $\beta$-densities as basic variables, in the following such a formulation in 
terms of $\rho$ and $Q$ will often turn out to be useful, because it allows for an easier comparison 
to the spin-independent functionals defined only in terms of the density $\rho$.
We will switch between these two representations whenever suitable.

With the spin-resolved HK functional $F_\text{HK}[\rho,Q]$, it is now also possible to give a simpler
prescription for obtaining the spin-density corresponding to a given total density: $Q^{M_S}[\rho]$ 
is the spin density for which $F_\text{HK}[\rho, Q]$ is minimized, under the constraint that $Q^{M_S}$ integrates 
to twice the chosen value of $M_S$, i.e.,
\begin{equation}
  \label{eq:F-spinindep-from-spindep}
  Q^{M_S}[\rho] = \arg \min_{Q^{M_S}} F_\text{HK}[\rho,Q^{M_S}] 
  \qquad \text{with} \quad \frac{1}{2} \int Q^{M_S}(\boldsymbol{r}) \, {\rm d}^3r = M_S.
\end{equation}
Therefore, the spin-independent HK functional $F[\rho]$ can be obtained from the spin-dependent HK 
functional $F_\text{HK}[\rho,Q]$ as
\begin{equation}
  \label{eq:F-spinindep-from-spindep}
  F_\text{HK}[\rho] = F_\text{HK}[\rho,Q[\rho]] = \min_Q F_\text{HK}[\rho,Q],
\end{equation}
where $Q[\rho]$ is \textit{any} of the spin densities that minimize $F_\text{HK}[\rho,Q]$ for the given
total density $\rho$. Of course, as long as no specific $M_S$ is chosen, $Q[\rho]$ is not unique,
but any admissible choice must lead to the same energy.

Finally, we have to consider whether there exist extensions of the HK theorems that justify
the use of $\rho$ and $Q$ (or of $\rho^\alpha$ and $\rho^\beta$) as basic variables.
For a generalization of this kind, it is not sufficient to consider wavefunctions generated by an external 
potential $v_\text{ext}(\boldsymbol{r})$, but also wavefunctions obtained in the presence of an 
additional external magnetic field $B_z(\boldsymbol{r})$ have to be taken into account. 
Such an extension of the HK theorems was first given by von Barth and Hedin \cite{von_barth_local_1972}, 
and was only recently put on more firm ground by extensions of Lieb's formulation of 
DFT \cite{ayers_legendre-transform_2006, holas_comment_2006}. Similar to the external potential, which 
is only known up to a constant, it is also possible to add a constant shift to the external magnetic field 
$B_z(\boldsymbol{r})$ without changing the wavefunction or the \mbox{(spin-)}density \cite{eschrig_density_2001,
capelle_nonuniqueness_2001}. This leads to a number of peculiarities related to the differentiability of
the spin-dependent energy functional \cite{gl_derivative_2010,gl_energy_2010}. However, most of these 
issues do not appear if the treatment is restricted to eigenfunctions of $S_z$ (the case of interest 
here) \cite{gidopoulos_potential_2007} or can be addressed by constraining $M_S$ to a fixed value 
(as we are always requiring here) \cite{gl_differentiability_2007,gl_nonuniqueness_2009}.

\subsection{Fractional Spin Conditions on the HK Functional}
\label{sec:fractional-spin}

For states that are not a singlet (i.e., for $S>0$), there are different degenerate wavefunctions $\Psi^{M_S}[\rho]$. 
These wavefunctions all share the same electron density, but correspond to different spin densities $Q^{M_S}[\rho]$. 
These different spin-densities are related by [cf. Eq.~\eqref{eq:spindens-ms-scaling}]
\begin{equation}
  Q^{M_S}[\rho](\boldsymbol{r}) = \left(\frac{M_S}{S}\right) Q^{M_S=S}[\rho](\boldsymbol{r}),
\end{equation}
and it follows that $Q^{M_S}(\boldsymbol{r}) = - Q^{-M_S}(\boldsymbol{r})$ and that one
obtains $Q^{M_S=0}(\boldsymbol{r}) = 0$ for even values of $S$. For all these degenerate 
spin-densities, the exact total energy functional $E[\rho,Q]$ must yield the same value.

This statement can be generalized to linear combinations of these spin-densities. Within an ensemble formulation 
of spin-DFT, it can be shown \cite{YZA-2000} that for any properly normalized linear combination of the
spin-densities $Q^{M_S}[\rho]$, the same energy should be obtained. Therefore, one finds that
\begin{equation}
  E\big[\rho, \alpha \, Q^{M_S=S}[\rho]\big] = \text{const.} \quad \text{for} \ -1 \le \alpha \le +1,
\end{equation}
which implies for the spin-dependent HK functional
\begin{equation}
  \label{eq:fractional-spin}
  F_\text{HK}\big[\rho, \alpha \, Q^{M_S=S}[\rho]\big] = \text{const.} \quad \text{for} \ -1 \le \alpha \le +1.
\end{equation}
Such ensemble spin densities correspond to a situation with a non-integer number
of $\alpha$- and $\beta$-electrons (``\textit{fractional spins}'').
As has been pointed out by Yang and co-workers, the above \textit{constancy conditions for fractional 
spins}, which is a property of the exact energy functional, is violated by all contemporary approximations.
Therefore, it was suggested that many problems appearing in practical DFT calculations with such 
approximations might be connected to this violation \cite{cohen_insights_2008,cohen_fractional_2008,
cohen_challenges_2012}.

The use of fractional spins makes it possible to further simplify the relation between the 
spin-dependent and spin-independent HK functionals given in Eq.~\eqref{eq:F-spinindep-from-spindep},
because one realizes that
\begin{equation}
  \label{eq:fhk-spin-independent}
  F_\text{HK}[\rho] = F_\text{HK}[\rho, Q=0].
\end{equation}
This equation holds both for systems with an even number of electrons (where one can
always choose $M_S=0$, corresponding to $Q(\boldsymbol{r})=0$) and for systems with an odd number
of electrons (where $Q(\boldsymbol{r})=0$ is only possible if one allows for fractional spins).

\subsection{Spin States in Hohenberg--Kohn DFT}
\label{sec:spin-states-hk}

In their initial formulation \cite{hohenberg-kohn-1964}, the HK theorems were applicable only to the ground 
state. In particular, minimizing the total energy functional $E[\rho]$ only yields the ground-state energy and
electron density (and if the spin-dependent energy functional is employed, also the spin density). 
However, as was first shown by Gunnarson and Lundquist, HK-DFT can be generalized to the lowest-energy 
states of a given symmetry \cite{gunnarsson_exchange_1976}. Of particular importance is the 
calculation of the lowest state of a particular spin symmetry, i.e., of the lowest state 
with a particular eigenvalue of $\hat{\boldsymbol{S}}^2$. Such a generalization is most easily
presented within the constrained-search formulation of DFT. 

In order to obtain the energy and electron density of the lowest state corresponding to a given 
value of $S$, one has to define the spin-state specific energy functional,
\begin{equation}
  \label{eq:hk-energy-state-specific}
  E^S[\rho] = q_e \!\! \int \rho(\boldsymbol{r}) \, v_\text{nuc}(\boldsymbol{r}) \, {\rm d}^3r
              + F_\text{HK}^S[\rho],
\end{equation}
with the spin-state specific HK functional,
\begin{equation}
  \label{eq:fhk-spinstates}
  F_\text{HK}^S[\rho] = \min_{\Psi^S \rightarrow \rho}
                       \left< \Psi^S \left| \hat{T} + \hat{V}_\text{ee} \right| \Psi^S \right>
  \quad \text{with} \quad \hat{\boldsymbol{S}}^2 \Psi^S = S(S+1) \hbar^2 \Psi^S,                
\end{equation}
where the constrained search now only includes wavefunctions $\Psi^S$ which are 
eigenfunctions of $\hat{\boldsymbol{S}}^2$ with the proper eigenvalue. Therefore, one obtains a different 
HK functional and thus a different total energy functional for each value of $S$.

Within spin-DFT, the simplest way of obtaining a functional that at least partly allows one to select certain spin states
is by choosing an appropriate value for $M_S$. This way, only states with $S \ge M_S$ are accessible because for
spin states corresponding to a smaller values of $S$, the chosen value of $M_S$ is not admissible.
By minimizing the energy functional $E_v[\rho, Q]$ under the constraint that $Q(\boldsymbol{r})$ 
integrates to $2M_S$, the lowest-energy state with $S \ge M_S$ is obtained, i.e., 
\begin{equation}
  E^{S \ge M_S}[\rho] = E\big[\rho, Q^{M_S}[\rho]\big].
\end{equation}
Equivalently, one can, of course, also
minimize $E_v[\rho^\alpha, \rho^\beta]$ under appropriate constraints for $N^\alpha$ and $N^\beta$.
However, fixing $M_S$ to target a specific spin-state is not completely general
since the minimization is only restricted to states with $S \ge M_S$, not to states with a specific $S$. 
While it is, for instance, possible to calculate the lowest triplet ($S=1$) state if the ground-state 
is a singlet ($S=0$), it is not possible to target the lowest singlet state if the ground-state is a 
triplet. Therefore, to be able to calculate the lowest state of a given $S$, it would in general be necessary
to employ the true spin-state specific energy functional of Eq.~\eqref{eq:hk-energy-state-specific}.

\section{Spin in Kohn--Sham DFT}
\label{sec:ks-dft}

While the Hohenberg--Kohn formulation of DFT is exact, it is very difficult to set up computationally
feasible, but nevertheless accurate approximate realizations of it. This is mainly rooted in the difficulty
of approximating the kinetic-energy contribution to the HK functional as a functional of the electron 
density only \cite{carter-kin-review,xia_can_2012}. 

A possible way out of this dilemma, that forms the basis of almost every present-day application of (approximate) 
DFT calculations, was suggested by Kohn and Sham \cite{kohn-sham-1965}. Instead of considering the
kinetic energy of the true system of interacting electrons, they proposed to calculate the kinetic
energy of a reference system of noninteracting electrons with the same electron density instead. This then already
 accounts for the largest part of the kinetic energy, and only a small remainder has to be approximated. The
KS approach still allows for the formulation of an exact theory, which will be outlined in this section.

\begin{figure}
  \caption{Relationship between wave function, total electron density, and spin density of the 
  system of fully interacting electrons and of the spin-restricted and spin-unrestricted Kohn--Sham 
  reference systems of noninteracting electrons.}
  \label{fig:ks-dft}

  \vspace{5ex}
  \begin{center}
  \includegraphics{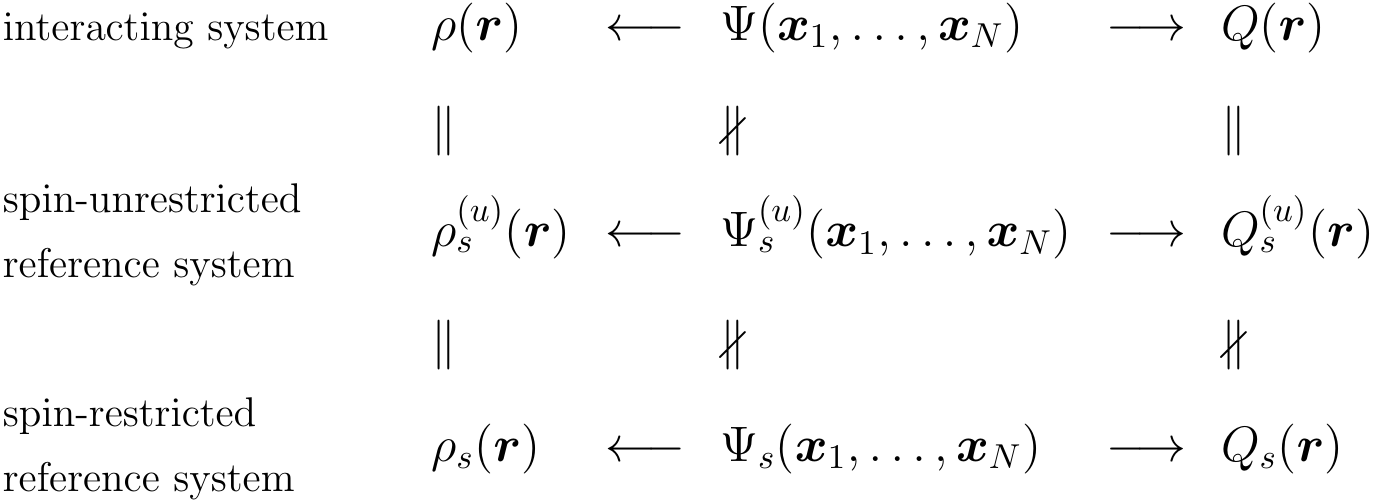}
  \end{center}

\end{figure}

In KS-DFT one considers two different quantum-mechanical systems at the same time:
The true molecular system of interacting electrons and a reference system of noninteracting electrons.
The link between these two systems is established by requiring that their electron densities $\rho(\boldsymbol{r})$
and $\rho_s(\boldsymbol{r})$
are equal (see Fig.~\ref{fig:ks-dft}). Their wavefunctions, however, will in general be different. For open-shell 
molecules, different options exist for introducing such a reference system: The first option is to require only 
that the electron densities of the interacting and the noninteracting systems agree. This leads to a 
\textit{spin-restricted KS-DFT formulation}. The second option is to require
that in addition to the total electron densities, also the spin densities of the two systems agree. This
results in a \textit{spin-unrestricted formulation of KS-DFT}. Of course, these two options are equivalent
for closed-shell systems (i.e., for singlet states with $S=0$).

Note that any version of KS-DFT relies on the assumption that such a noninteracting reference system with
the same electron density (and possibly also the same spin density) as the interacting system exists. In practice, 
it  is always assumed that this so-called $v_s$-representability condition is fulfilled, even though this is not guaranteed 
and several counter-examples are known \cite{levy_electron_1982,lieb_density_1983,englisch_exact_1984,
schipper_one_1998,morrison_electron_2002,katriel_study_2004}. 
For a detailed discussion of these subtle issues, see, e.g., Refs.~\cite{van_leeuwen_density_2003}.

\subsection{Spin-Restricted Kohn--Sham DFT}

\subsubsection{Noninteracting Reference System}

First, we consider a system of $N$ noninteracting electrons in an external potential $v_s(\boldsymbol{r})$.
Such a system is described by the Hamiltonian
\begin{equation}
  \label{eq:hamiltonian-nonint}
  \hat{H}_s = \hat{T} + \hat{V}_s = \sum_{i=1}^N{-\frac{\hbar^2}{2 m_e}} \Delta_i + \sum_{i=1}^N q_e \, v_s(\boldsymbol{r}_i),
\end{equation}
where the subscript $s$ (for ``single-particle'') is introduced to indicate that the quantities
refer to a system of noninteracting electrons. Because this Hamiltonian does not contain
terms that couple different electrons, an exact wavefunction is given by an antisymmetrized 
product of one-electron functions (i.e., by a single Slater determinant), for which the short-hand notation
\begin{equation}
  \Phi_s(\boldsymbol{x}_1,\dotsc,\boldsymbol{x}_N) 
  = \Bigl| \varphi_1\alpha, \varphi_1\beta, \varphi_2\alpha, \varphi_2\beta, \dotsc \Bigr| 
\end{equation}
can be used. The spatial parts of the one-electron functions (orbitals) can be obtained as the 
solutions of the one-electron equation
\begin{equation}
  \left[ - \frac{\hbar^2}{2 m_e} \Delta + q_e \, v_s(\boldsymbol{r}) \right] \varphi_i(\boldsymbol{r}) = \epsilon_i \varphi_i(\boldsymbol{r}),
\end{equation}
which trivially emerge from the energy eigenvalue equation $\hat{H}_s\Phi_s = E_s \Phi_s$ with
$E_s = \sum_i\epsilon_i$.
Since this one-electron Hamiltonian does not depend on the spin of the electron, each energy 
eigenvalue is two-fold degenerate (i.e., each spatial orbital can be combined with an $\alpha$- or 
with a $\beta$-spin function). In particular, the spatial orbitals are identical for $\alpha$- and for
$\beta$-electrons, and the resulting Slater determinant is therefore \textit{spin-restricted}.

In such a spin-restricted Slater determinant, each spatial orbital $\varphi_i$ can either be doubly occupied
(i.e., it appears both in combination with an $\alpha$- and with a $\beta$-spin function) or it can
be singly occupied with either an $\alpha$- or a $\beta$-electron. For the ground state, such singly
occupied orbitals can only occur for the highest occupied molecular orbital (HOMO), and more than 
one singly occupied orbital can only be present if the HOMO is degenerate (for a detailed discussion,
see, e.g., Ref.~\cite{schipper_one_1998} and chapters~3.3 and~3.4 in Ref.~\cite{engel_density_2011}). 

For a single, spin-restricted Slater determinant, the electron density is given by
\begin{equation}
  \rho(\boldsymbol{r}) = \sum_i^{\text{occ.}} f_i \, |\varphi_i(\boldsymbol{r})|^2,
\end{equation}
where the occupation numbers $f_i$ are either $1$ or $2$. The spin density is determined
only by the singly occupied orbitals, and can be calculated as 
\begin{equation}
  \label{eq:spindens-restr}
  Q_s(\boldsymbol{r}) = \sum_i^\text{singly occ.} s_i |\varphi_i(\boldsymbol{r})|^2,
\end{equation}
where the sum only runs over the singly occupied orbitals and where $s_i = +1$ for $\alpha$-spin 
and $s_i = -1$ for $\beta$-spin orbitals, as all doubly-occupied orbital contributions drop out because
of the identical spatial distribution of these $\alpha,\beta$-pairings.
Finally, the kinetic energy of a spin-restricted Slater determinant can be calculated as
  \begin{equation}
    T_s
    = \sum_i^{\text{occ.}} f_i \, \big\langle \varphi_i \big| \hat{T} \big| \varphi_i \big\rangle
    = -\frac{\hbar^2}{2 m_e} \sum_i^{\text{occ.}} f_i \int \varphi_i(\boldsymbol{r}) \Delta \varphi_i(\boldsymbol{r}) \, {\rm d}^3r.
  \end{equation}
Thus, it turns out that spin-restricted Slater determinants in which the occupation numbers $f_i$ are
identical share the same electron density and have the same kinetic energy. Such determinants
--- which can only differ in the spin of the singly occupied orbitals --- are degenerate with respect to 
the noninteracting Hamiltonian of Eq.~\eqref{eq:hamiltonian-nonint}.

The noninteracting Hamiltonian $\hat{H}_s$ commutes with both $\hat{S}_z$ and with 
$\hat{\boldsymbol{S}}^2$. Therefore, it is always possible to combine
degenerate eigenfunctions such that they are also eigenfunctions of $\hat{S}_z$ and $\hat{\boldsymbol{S}}^2$.
A restricted Slater determinant is always an eigenfunction of $\hat{S}_z$ with eigenvalue 
$M_S \hbar = (N_\alpha-N_\beta) \hbar/2$ \cite{mcweeny_methods_1969,szabo-ostlund}. 
If all singly occupied orbitals are either $\alpha$- or
$\beta$-spin orbitals, then it is also an eigenfunction of $\hat{\boldsymbol{S}}^2$ with $S=|M_S|$. In all
other cases, an eigenfunction of $\hat{\boldsymbol{S}}^2$ can be constructed as a linear combination of 
(degenerate) determinants in which the same orbitals are singly occupied and which correspond
to the same value of $M_S$. Such linear combinations are known as
\textit{configuration state functions} (CSF) \cite{helgaker-book}.
It is important to understand that for any eigenfunction of the noninteracting Hamiltonian that is
an eigenfunction of $\hat{\boldsymbol{S}}^2$ with eigenvalue $S_\text{max}(S_\text{max}+1) \hbar^2$, 
degenerate eigenfunctions with $S=0,1,\dotsc,S_\text{max}$ (for an even number of electrons) or with
$S=\tfrac{1}{2}, \tfrac{3}{2},\dotsc,S_\text{max}$ (for an odd number of electrons) can also 
be constructed. Thus, for each energy eigenvalue, there is one CSF corresponding to $S=0$ 
(i.e., a singlet state) or $S=\tfrac{1}{2}$ (i.e., a doublet state) for an even or odd number of
electrons, respectively.


The Hohenberg--Kohn theorems still hold for a system of noninteracting electrons. Thus, the 
ground-state density of $\hat{H}_s$ can be determined by minimizing the noninteracting energy 
functional,
\begin{align}
  \label{eq:toten-nonint}
  E_s[\rho] 
            &= T_s[\rho] + q_e \!\!\int \rho(\boldsymbol{r}) v_s(\boldsymbol{r}) \, {\rm d}^3r,
\end{align}
where the noninteracting kinetic-energy functional $T_s[\rho]$ can be defined in the
Levy con\-strained-search formalism as
\begin{equation}
  \label{eq:def-ts}
  T_s[\rho] =  \min_{\Psi_s \rightarrow \rho} \langle \Psi_s | \hat{T} | \Psi_s \rangle.
\end{equation}
In this definition, the constrained search includes all wavefunctions $\Psi_s$ that correspond to a
system of noninteracting electrons with density $\rho$. As discussed above, this could be further 
restricted to singlet or doublet wavefunctions. In this definition, $T_s[\rho]$ is independent of the spin 
density.

The ground-state electron density $\rho_0$ is obtained from minimizing $E_s[\rho]$ under the constraint 
that the number of electrons is preserved, and the corresponding ground-state wavefunction $\Psi_{s,0}$ is 
the one for which the minimum in Eq.~\eqref{eq:def-ts} is achieved. Again, this wavefunction could 
always be chosen as a singlet or a doublet for an even or odd number of electrons, respectively.

\subsubsection{Interacting Energy Functional and Exchange--Correlation Energy}

The (spin-resolved) HK functional of the true system of interacting electrons [Eq.~\eqref{eq:fhk-spin-resolved}] 
can now be decomposed as
\begin{equation}
  F_\text{HK}[\rho,Q] = T_s[\rho] + J[\rho] + E_{xc}[\rho,Q],
\end{equation}
where $T_s[\rho]$ is the noninteracting kinetic energy introduced in the previous section, $J[\rho]$
is the classical Coulomb interaction of the electron density with itself,
\begin{equation}
  J[\rho] = \frac{q_e^2}{2} \int \frac{\rho(\boldsymbol{r})\rho(\boldsymbol{r}')}{|\boldsymbol{r} - \boldsymbol{r}'|} \,
            {\rm d}^3r {\rm d}^3r',
\end{equation}
and the exchange--correlation energy $E_\text{xc}[\rho,Q]$ is defined to account for the remaining
energy contributions
\begin{equation}
  \label{eq:exc-restr-spinresolved}
  E_\text{xc}[\rho,Q] = F_\text{HK}[\rho,Q] - T_s[\rho] - J[\rho].
\end{equation}
This exchange--correlation functional could also be expressed in terms of the $\alpha$- and 
$\beta$-electron  densities.
The noninteracting kinetic energy $T_s[\rho]$ is different from the true kinetic energy of
the fully interacting system $T[\rho]$. Therefore, the exchange--correlation energy also contains the 
difference $T_c[\rho] = T[\rho] - T_s[\rho]$ between the kinetic energy of the true interacting  
system and the  kinetic energy of the noninteracting reference system.

With these definitions, the total energy functional of the interacting system can be expressed as
\begin{align}
  \label{eq:toten-restr}
  E[\rho,Q] = T_s[\rho] + J[\rho] + E_{xc}[\rho,Q] + q_e \!\! \int \rho(\boldsymbol{r}) v_\text{nuc}(\boldsymbol{r}) {\rm d}^3r.
\end{align}
The spin-independent analogues of the exchange--correlation and the total energy functional, 
$E_\text{xc}[\rho]$ and $E[\rho]$, are recovered from these definitions when setting $Q(\boldsymbol{r}) = 0$ 
[cf. Eq.~\eqref{eq:fhk-spin-independent}].

The ground-state density of the true interacting system can be determined by minimizing the total energy 
functional $E[\rho]$ with respect to $\rho$, under the constraint that it integrates to the correct number of electrons.
With the exact functionals, this will lead to the exact ground-state electron density $\rho_0(\boldsymbol{r})$. 
The corresponding exact wavefunction $\Psi_0$ is the one for which the minimum in the
constraint search in Eq.~\eqref{eq:fhk-spin-resolved} is obtained. This ground-state wavefunction has to be an eigenfunction 
of $\hat{\boldsymbol{S}}^2$ and $\hat{S_z}$, and the corresponding value of $S$ determines the spin multiplicity 
of the ground-state, whereas the different $M_S$-states are degenerate. In contrast to the noninteracting case,
the ground-state wavefunction is not necessarily a singlet or a doublet wavefunction, but could have
a higher spin multiplicity.

\subsubsection{Kohn--Sham Potential}

Minimization of the noninteracting energy functional [Eq.~\eqref{eq:toten-nonint}] with respect 
to the total density $\rho$, under the constraint that $\rho$ integrates to $N$ electrons, yields the 
following Euler--Lagrange equation \cite{yang-parr},
\begin{equation}
  \label{eq:min-nonint}  
  0 = \frac{\delta E_s[\rho]}{\delta\rho(\boldsymbol{r})} - \mu
    = \frac{\delta T_s[\rho]}{\delta\rho(\boldsymbol{r})}  + q_e v_s(\boldsymbol{r}) - \mu.
\end{equation}

On the other hand, the total energy functional of the interacting system is given by 
Eq.~\eqref{eq:toten-restr}, which upon minimization leads to the condition
\begin{equation}
     \label{eq:min-int}  
  0 =  \frac{\delta E[\rho]}{\delta\rho(\boldsymbol{r})} - \mu
      = \frac{\delta T_s[\rho]}{\delta\rho(\boldsymbol{r})} 
       + q_e \big( v_\text{nuc}(\boldsymbol{r}) + v_\text{Coul}[\rho](\boldsymbol{r})
       + v_\text{xc}[\rho](\boldsymbol{r}) \big) - \mu,
\end{equation}
where $v_\text{Coul}[\rho](\boldsymbol{r}) = q_e \int \rho(\boldsymbol{r}')/|\boldsymbol{r}-\boldsymbol{r}'| \, {\rm d}^3r'$ 
is the classical Coulomb potential of the electrons and $v_\text{xc}[\rho](\boldsymbol{r}) = (1/q_e) \,
\delta E_\text{xc}[\rho]/\delta\rho(\boldsymbol{r})$ is the exchange--correlation potential.

Since the definition of the noninteracting kinetic energy $T_s[\rho]$ is the same in both
minimizations, and because we require that the total densities obtained for the interacting 
and the noninteracting system agree, we obtain for the KS potential
\begin{equation}
  \label{eq:vs-restricted}
  v_s[\rho](\boldsymbol{r}) = v_\text{ext}(\boldsymbol{r}) + v_\text{Coul}[\rho](\boldsymbol{r})
                        + v_\text{xc}[\rho](\boldsymbol{r}).
\end{equation}
The ground-state electron density of the fully interacting system can thus be
determined by solving the Schr\"odinger equation of a noninteracting system 
[i.e. with the Hamiltonian of Eq.~\eqref{eq:hamiltonian-nonint}) including the external 
potential $v_s(\boldsymbol{r})$ given by Eq.~\eqref{eq:vs-restricted}], and the
orbitals of this noninteracting system can be obtained from the KS equations
\begin{equation}
  \label{eq:ks-restricted}
      \left[ - \frac{\hbar^2}{2 m_e} \Delta + q_e \big( v_\text{nuc}(\boldsymbol{r}) + v_\text{Coul}[\rho](\boldsymbol{r}) + v_\text{xc}[\rho](\boldsymbol{r}) \big) \right] \varphi_i(\boldsymbol{r}) 
    = \epsilon_i \varphi_i(\boldsymbol{r}).
\end{equation} 
Thus, the ground-state electron density of the true interacting system is obtained from the 
wavefunction $\Psi_{s,0}$ of the noninteracting reference system.
However, this ground-state wavefunction $\Psi_{s,0}$ of the noninteracting 
reference system does \textbf{not} agree with the ground-state wavefunction $\Psi_0$ 
of the interacting system. Moreover, $\Psi_{s,0}$ can always be chosen as a singlet or
doublet state (i.e., $S=0$ or $S=\tfrac{1}{2}$), whereas $\Psi_0$ can correspond to any
value of $S$. Therefore, the spin multiplicities of the ground-states of the noninteracting and
of the interacting system can be different.

\subsubsection{Spin Density}
\label{sec:spindens-restr}

For $S>0$, the ground state of the true interacting system is degenerate and there is a set 
of eigenfunctions of $\hat{S}_z$ with different eigenvalues $M_s$. These
wavefunctions $\Psi_0^{M_S}$ have different spin-densities $Q_0^{M_S}(\boldsymbol{r})$
that are related by Eq.~\eqref{eq:spindens-ms-scaling}.
By construction, the ground-state wavefunction $\Psi_{s,0}$ of the noninteracting reference 
system and the ground-state wavefunction $\Psi_0$ of the fully interacting system only share the
same electron density. However, the corresponding spin densities $Q(\boldsymbol{r})$ and
$Q_s(\boldsymbol{r})$ are in general not equal and the true ground-state spin density 
cannot be calculated from $\Psi_{s,0}$.

From Eq.~\eqref{eq:spindens-restr} it is obvious that $Q(\boldsymbol{r})$ and $Q_s(\boldsymbol{r})$
have to be different: For the spin-restricted noninteracting reference system, the spin density
is determined only by the singly occupied orbitals and will thus have the same sign at every 
point in space (i.e., $Q_s(\boldsymbol{r}) > 0$ for $M_S > 0$). However, it is known both from
accurate calculations and from experiment, that for the interacting system the spin density 
has different signs in different regions in space \cite{pople_spin-unrestricted_1995,
chipman_theoretical_1983,chipman_spin_1992}.

To obtain the spin density in a restricted KS-DFT formulation, one has to minimize the
spin-resolved HK functional $F_\text{HK}[\rho_0, Q]$ defined in Eq.~\eqref{eq:fhk-spin-resolved}
with respect to $Q(\boldsymbol{r})$, under the constraint that the spin density integrates 
to $2 M_s$. 
Since the exchange--correlation energy is the only part of this functional that depends on 
the spin density, the minimization with respect to $Q(\boldsymbol{r})$ leads to the condition
\begin{equation}
  \label{eq:spindens-min-restricted}
   \frac{\delta F_\text{HK}[\rho_0,Q]}{\delta Q(\boldsymbol{r})} - \lambda
    = \frac{\delta E_\text{xc}[\rho_0,Q]}{\delta Q(\boldsymbol{r})} = 0,
\end{equation}
where the Lagrange multiplier $\lambda$ is zero because of Eq.~\eqref{eq:spindens-ms-scaling}.
This suggests a two-step procedure for determining the spin density in restricted KS-DFT. First,
the total ground-state density $\rho_0(\boldsymbol{r})$ is determined by solving the KS equations. 
Subsequently, the corresponding ground-state spin density $Q_0(\boldsymbol{r})$ can be calculated 
from the above minimization condition for a chosen value of $M_S$.

\subsection{Spin-Unrestricted Kohn--Sham DFT}

\subsubsection{Noninteracting Reference System}

The choice of a reference system of noninteracting electrons with the same
total electron density as the interacting system is not the only option.
Alternatively, it is also possible to envisage a reference system of noninteracting electrons 
that has the same $\alpha$-electron and $\beta$-electron densities as the interacting 
system \cite{von_barth_local_1972,gunnarsson_exchange_1976,pople_spin-unrestricted_1995}. 
In this case, a reference system with the Hamiltonian 
\begin{align}
  \label{eq:hamiltonian-nonint-mag}
  \hat{H}_s^{(u)} &= \hat{T}_s + \hat{V}_s^{\text{tot}} + \hat{V}_s^\text{spin} 
          = \sum_{i=1}^N{\frac{\hbar^2}{2 m_e}\Delta_i} 
             + q_e \sum_{i=1}^N \Bigl[ v_s^\text{tot}(\boldsymbol{r}_i) + v_s^\text{spin}(\boldsymbol{r}_i)\sigma_z(s_i) \Bigr]
\nonumber \\
  &= \hat{T}_s + \hat{V}_s^\alpha + \hat{V}_s^\beta 
          = \sum_{i=1}^N{\frac{\hbar^2}{2 m_e}\Delta_i} 
             + q_e \sum_{i=1}^N \Bigl[ v_s^\alpha(\boldsymbol{r}_i)\alpha(s_i) - v_s^\beta(\boldsymbol{r}_i)\beta(s_i) \Bigr]
\end{align}
is used. To distinguish them from those introduced earlier for a spin-restricted reference system,
the superscript ``$(u)$'' will be used for quantities referring to this spin-unrestricted reference system.
Different potentials $v_s^\alpha(\boldsymbol{r}) = v_s^\text{tot}(\boldsymbol{r}) + v_s^\text{spin}(\boldsymbol{r})$
and $v_s^\beta(\boldsymbol{r}) = v_s^\text{tot}(\boldsymbol{r}) - v_s^\text{spin}(\boldsymbol{r})$
for the $\alpha$- and $\beta$-electrons, respectively, are now needed in order to allow 
the reference system of noninteracting electrons to have the same spin density as the interacting one.
This corresponds to introducing an inhomogeneous external magnetic field in $z$-direction 
$B_z(\boldsymbol{r}) = - \dfrac{q_e}{\mu_B} v_s^\text{spin}(\boldsymbol{r})$ that only interacts 
with the electronic spins (i.e., the interaction due to orbital angular momentum is ignored)
[cf. Eq.~\eqref{eq:zeeman-op-nonrel}].

An exact solution to the corresponding Schr\"odinger equation has the form of a single Slater 
determinant, but in contrast to the spin-restricted case the spatial orbitals now differ for $\alpha$- 
and $\beta$-electrons, i.e.,
\begin{equation}
  \Phi_s^{(u)}(\boldsymbol{x}_1,\dotsc,\boldsymbol{x}_N) 
  = \Bigl| \varphi^\alpha_1\alpha, \varphi^\beta_1\beta, \varphi^\alpha_2\alpha, \varphi^\beta_2\beta, \dots \Bigr|
\end{equation}
The spatial parts of the orbitals can be obtained from two separate sets of one-electron equations
\begin{align}
  & \left[ - \frac{\hbar^2}{2 m_e} \Delta + q_e \, v^\alpha_s(\boldsymbol{r}) \right] \varphi^\alpha_i(\boldsymbol{r}) 
  = \epsilon^\alpha_i \varphi^\alpha_i(\boldsymbol{r}) \qquad \text{and} \nonumber \\
  & \left[ -\frac{\hbar^2}{2 m_e} \Delta + q_e \, v^\beta_s(\boldsymbol{r}) \right] \varphi^\beta_i(\boldsymbol{r}) 
  = \epsilon^\beta_i \varphi^\beta_i(\boldsymbol{r}).
\end{align}
Both the resulting $\alpha$- and $\beta$-orbitals form an orthonormal set,
$\langle\varphi_i^\alpha|\varphi_j^\alpha\rangle = \delta_{ij}$ and $\langle\varphi_i^\beta|\varphi_j^\beta\rangle = \delta_{ij}$, 
but  $\alpha$- and $\beta$-orbitals are in general not orthogonal to each other, i.e., 
$\langle \varphi_i^\alpha|\varphi_j^\beta\rangle \ne 0$.

The noninteracting Hamiltonian $H_s^{(u)}$ still commutes with $\hat{S}_z$, and
any spin-unrestricted Slater determinant is an eigenfunction of $\hat{S}_z$, with the eigenvalue
$M_S$ being determined by the number of $\alpha$- and $\beta$-electrons. However, in contrast 
to the spin-restricted case, the different $\hat{S}_z$ eigenstates are not degenerate anymore. For 
constructing the ground-state wavefunction, the $N$ orbitals with the lowest orbital energies 
have to be occupied, which automatically fixes $M_S$. By occupying other ($\alpha$- or 
$\beta$-electron) orbitals, excited state wavefunctions for the noninteracting reference system 
corresponding to different values of $M_S$ can be obtained.

However, $H_s^{(u)}$ does in general not commute with $\hat{\boldsymbol{S}^2}$, and 
the ground-state wavefunction is thus not an eigenfunction of $\hat{\boldsymbol{S}^2}$ anymore. Instead, 
the expectation value of $\hat{\boldsymbol{S}^2}$ can be calculated as (assuming $N_\alpha \ge N_\beta$) 
\cite{lwdin_quantum_1955,wang_evaluation_1995}
\begin{equation}
  \braket{\hat{\boldsymbol{S}^2}} 
     = M_S(M_S+1)\hbar^2 + \hbar^2 N_\beta  
       - \hbar^2 \sum_{i=1}^{N_\alpha}\sum_{i=1}^{N_\beta} \left| \int \varphi_i^\alpha(\boldsymbol{r}) \varphi_j^\beta(\boldsymbol{r}) \, {\rm d}^3r \right|^2.
\end{equation}
In the spin-restricted case, where the $\alpha$- and $\beta$-orbitals are equal and therefore mutually orthogonal, 
the last term equals the number of doubly occupied orbitals, and one obtains $\braket{\hat{\boldsymbol{S}^2}} = M_S(M_S+1) \hbar^2$.
In the unrestricted case, this cancellation is only partial and a larger expectation value is obtained. This is often referred 
to as \textit{spin contamination}.

For an unrestricted Slater determinant the total electron density is given by,
\begin{equation}
  \rho(\boldsymbol{r}) = \sum_{i=1}^{N_\alpha}  |\varphi_i^\alpha(\boldsymbol{r})|^2 + \sum_{i=1}^{N_\beta}  |\varphi_i^\beta(\boldsymbol{r})|^2,
\end{equation}
and the spin density can be calculated as
\begin{equation}
  Q(\boldsymbol{r}) = \sum_{i=1}^{N_\alpha}  |\varphi_i^\alpha(\boldsymbol{r})|^2 - \sum_{i=1}^{N_\beta}  |\varphi_i^\beta(\boldsymbol{r})|^2.
\end{equation}
In contrast to the spin-restricted case, the spin density can now have different signs at different points in space.
The kinetic energy of the unrestricted Slater determinant $\Phi_s^{(u)}$ is
\begin{align}
  T_s^{(u)} = \langle \Psi_s | \hat{T} | \Psi_s \rangle
    =& - \frac{\hbar^2}{2m_e} \sum_{i=1}^{N_\alpha} \int \varphi_i^\alpha(\boldsymbol{r}) \Delta \varphi_i^\alpha(\boldsymbol{r}) \, {\rm d}^3r
    \nonumber \\
       & \quad - \frac{\hbar^2}{2m_e} \sum_{i=1}^{N_\beta} \int \varphi_i^\beta(\boldsymbol{r}) \Delta \varphi_i^\beta(\boldsymbol{r}) \, {\rm d}^3r,
\end{align}
and a noninteracting kinetic-energy functional can now be defined as
\begin{equation}
    T_s^{(u)}[\rho_\alpha,\rho_\beta] 
    =  \min_{\Psi_s^{(u)} \rightarrow \rho_\alpha,\rho_\beta} \big\langle \Psi_s^{(u)} \big| \hat{T} \big| \Psi_s^{(u)} \big\rangle
\end{equation}
or as
\begin{equation}
    T_s^{(u)}[\rho,Q] 
    =  \min_{\Psi_s^{(u)} \rightarrow \rho,Q} \big\langle \Psi_s^{(u)} \big| \hat{T} \big| \Psi_s^{(u)} \big\rangle.
\end{equation}
In contrast to the spin-restricted case, this functional depends not only on the total electron density 
$\rho(\boldsymbol{r}) = \rho_\alpha(\boldsymbol{r}) + \rho_\beta(\boldsymbol{r})$, but also on the spin density
$Q(\boldsymbol{r}) = \rho_\alpha(\boldsymbol{r}) - \rho_\beta(\boldsymbol{r})$. Usually, $T_s^{(u)}[\rho, Q]$ 
yields different kinetic energies for systems that share the same total electron density, but have different spin densities.

In terms of $\alpha$- and $\beta$-electron densities, the total energy functional of the noninteracting system
is given by 
\begin{align}
  \label{eq:toten-nonint-ab}
  E_s^{(u)}[\rho_\alpha,\rho_\beta] 
  = T_s^{(u)}[\rho_\alpha,\rho_\beta] 
     + q_e \!\! \int \rho_\alpha(\boldsymbol{r}) v_s^\alpha(\boldsymbol{r}) \, {\rm d}^3r 
     + q_e \!\! \int \rho_\beta(\boldsymbol{r}) v_s^\beta(\boldsymbol{r}) \, {\rm d}^3r.
\end{align}
The ground-state $\alpha$- and $\beta$-densities $\rho^\alpha_0(\boldsymbol{r})$ and $\rho^\beta_0(\boldsymbol{r})$ 
of the noninteracting system can then be determined by minimizing this energy functional with respect to $\rho_\alpha$ 
and $\rho_\beta$, under the constraint that these integrate to the correct number of $\alpha$- and $\beta$-electrons.

\subsubsection{Exchange--Correlation Energy Functional}
  
The spin-resolved HK functional (cf. Eq.~\eqref{eq:hk-alpha-beta}) of the true system of interacting 
electrons can now be decomposed as
\begin{align}
  \label{eq:hk-alpha-beta-2}
  F_\text{HK}[\rho_\alpha, \rho_\beta] 
     &= T_s^{(u)}[\rho_\alpha, \rho_\beta] + J[\rho] + E_\text{xc}^{(u)}[\rho_\alpha, \rho_\beta],   
\end{align}
where the spin-resolved exchange--correlation energy is defined as
\begin{equation}
  \label{eq:exc-unrestr}
  E_\text{xc}^{(u)}[\rho_\alpha, \rho_\beta] 
  = F_\text{HK}[\rho_\alpha, \rho_\beta] - T_s^{(u)}[\rho_\alpha, \rho_\beta] - J[\rho].
\end{equation}
Since this functional $E_\text{xc}^{(u)}[\rho_\alpha, \rho_\beta]$ has been defined via the spin-unrestricted reference system, it is
in general different from the functional $E_\text{xc}[\rho_\alpha, \rho_\beta] $ defined in the previous section. This difference arises 
because different definitions of the noninteracting kinetic energy are used in the two cases. 

With this definition of the exchange--correlation energy, the total energy functional of the true
system of interacting electrons is given by
\begin{align}
  \label{eq:toten-unrestr}
  E[\rho_\alpha,\rho_\beta] 
  = T_s^{(u)}[\rho_\alpha,\rho_\beta] + J[\rho] + E_{xc}^{(u)}[\rho_\alpha,\rho_\beta] 
  + q_e \int \rho(\boldsymbol{r}) v_\text{nuc}(\boldsymbol{r}) \, {\rm d}^3r.
\end{align}
This total energy functional is identical to the spin-resolved version of the total energy functional 
derived in the spin-restricted case [cf. Eq.~\eqref{eq:toten-restr}], even though it is decomposed 
in a different fashion.

The ground-state $\alpha$- and $\beta$-electron densities $\rho^\alpha_0(\boldsymbol{r})$ and $\rho^\beta_0(\boldsymbol{r})$ 
can then be determined by minimizing this total energy functional with respect to $\rho_\alpha$ and $\rho_\beta$,
under the constraint that these integrate to $N_\alpha$ and $N_\beta$ electrons, respectively.
Note that by choosing $N_\alpha$ and $N_\beta$, a specific value of $M_S$ is selected. As 
long as this $M_S$ can be realized for the exact ground state, the resulting total ground-state 
density $\rho_0(\boldsymbol{r}) = \rho^\alpha_0(\boldsymbol{r}) + \rho^\beta_0(\boldsymbol{r})$ 
will be independent of the choice of $M_S$. In addition, the minimization will then yield 
the exact ground-state spin density $Q_0^{M_S}(\boldsymbol{r}) = \rho^\alpha_0(\boldsymbol{r}) - \rho^\beta_0(\boldsymbol{r})$.

\subsubsection{Kohn--Sham Potential}

The minimization of the total energy functional $E_s^{(u)}[\rho_\alpha,\rho_\beta]$ of the 
spin-unrestricted noninteracting reference system leads to these Euler--Lagrange equations
\begin{align}
  \frac{\delta E_s^{(u)}[\rho_\alpha,\rho_\beta]}{\delta\rho_\alpha(\boldsymbol{r})} - \mu_\alpha
  &= \frac{\delta T_s^{(u)}[\rho_\alpha,\rho_\beta]}{\delta\rho_\alpha(\boldsymbol{r})}
       + q_e \, v_s^\alpha(\boldsymbol{r}) - \mu_\alpha = 0 \\
  \frac{\delta E_s^{(u)}[\rho_\alpha,\rho_\beta]}{\delta\rho_\beta(\boldsymbol{r})} - \mu_\beta
  &= \frac{\delta T_s^{(u)}[\rho_\alpha,\rho_\beta]}{\delta\rho_\beta(\boldsymbol{r})}
       + q_e \, v_s^\beta(\boldsymbol{r}) - \mu_\beta = 0.
\end{align}
For the interacting system, the minimization of $E[\rho_\alpha,\rho_\beta]$ with respect to 
the $\alpha$- and $\beta$-electron densities yields
\begin{align}
  \label{eq:euler-unrestricted-1}
  \frac{\delta E[\rho_\alpha,\rho_\beta]}{\delta\rho_\alpha(\boldsymbol{r})} - \mu_\alpha
      &= \frac{\delta T_s^{(u)}[\rho_\alpha,\rho_\beta]}{\delta\rho_\alpha(\boldsymbol{r})}
       + q_e \big( v_\text{nuc}(\boldsymbol{r}) + v_\text{Coul}[\rho](\boldsymbol{r})
       + v_{\text{xc}}^\alpha[\rho_\alpha,\rho_\beta](\boldsymbol{r}) \big) - \mu_\alpha = 0,\\
  \label{eq:euler-unrestricted-2}
  \frac{\delta E[\rho_\alpha,\rho_\beta]}{\delta\rho_\beta(\boldsymbol{r})} - \mu_\beta
      &= \frac{\delta T_s^{(u)}[\rho_\alpha,\rho_\beta]}{\delta\rho_\beta(\boldsymbol{r})}
       + q_e \big(v_\text{nuc}(\boldsymbol{r}) + v_\text{Coul}[\rho](\boldsymbol{r})
       + v_{\text{xc}}^\beta[\rho_\alpha,\rho_\beta](\boldsymbol{r}) \big) - \mu_\beta = 0.
\end{align}
with the spin components of the exchange--correlation potential
\begin{equation}
  v_{\text{xc}}^\alpha[\rho_\alpha,\rho_\beta](\boldsymbol{r}) = \frac{1}{q_e}
   \frac{\delta{E}_\text{xc}^{(u)}[\rho_\alpha, \rho_\beta]}{\delta\rho_{\alpha}(\boldsymbol{r})} 
   \qquad \text{and} \qquad
  v_{\text{xc}}^\beta[\rho_\alpha,\rho_\beta](\boldsymbol{r}) = \frac{1}{q_e}
  \frac{\delta{E}_\text{xc}^{(u)}[\rho_\alpha, \rho_\beta]}{\delta\rho_{\beta}(\boldsymbol{r})}.
\end{equation}

If we require that $\rho_\alpha(\boldsymbol{r})$ and $\rho_\beta(\boldsymbol{r})$ --- and thus 
both the total and the spin density of the ground state --- are the same for the noninteracting reference
system and the true interacting system, we find that the spin components of the KS potential are given by
\begin{align}
  v_{s}^\alpha[\rho^\alpha,\rho^\beta](\boldsymbol{r}) &=
     v_\text{ext}(\boldsymbol{r}) + v_\text{Coul}[\rho](\boldsymbol{r})
     + v_{\text{xc}}^\alpha[\rho^\alpha,\rho^\beta](\boldsymbol{r}) \\
  v_{s}^\beta[\rho^\alpha,\rho^\beta](\boldsymbol{r}) &=
     v_\text{ext}(\boldsymbol{r}) + v_\text{Coul}[\rho](\boldsymbol{r})
     + v_{\text{xc}}^\beta[\rho^\alpha,\rho^\beta](\boldsymbol{r}).
\end{align}

Therefore, the exact ground-state $\alpha$- and $\beta$-electron densities of the true interacting
system can be calculating by solving the Schr\"odinger equation of an auxilliary system of noninteracting
electrons with the Hamiltonian of Eq.~\eqref{eq:hamiltonian-nonint-mag}. The ground-state
wavefunction $\Psi_{s,0}^{(u)}$ of this KS reference system is given by an unrestricted
Slater determinant, constructed from the orbitals obtained from the KS equations,
\begin{align}
  \label{eq:ks-unrestricted}
      \left[ - \frac{\hbar^2}{2 m_e} \Delta + q_e \big(v_\text{nuc}(\boldsymbol{r}) + v_\text{Coul}[\rho](\boldsymbol{r}) + v_\text{xc}^{\alpha}[\rho_\alpha,\rho_\beta](\boldsymbol{r}) \big) \right]
      \varphi_i^\alpha(\boldsymbol{r}) &= \epsilon_i^\alpha \varphi_i^\alpha(\boldsymbol{r}) 
    \nonumber \\
      \left[ - \frac{\hbar^2}{2 m_e} \Delta + q_e \big(v_\text{nuc}(\boldsymbol{r}) + v_\text{Coul}[\rho](\boldsymbol{r}) + v_\text{xc}^{\beta}[\rho_\alpha,\rho_\beta](\boldsymbol{r}) \big) \right]
      \varphi_i^\beta(\boldsymbol{r}) &= \epsilon_i^\beta \varphi_i^\beta(\boldsymbol{r}).
\end{align} 
Here, the equations for the $\alpha$- and $\beta$-orbitals are coupled through the 
Coulomb and exchange--correlation potentials.

Equivalently, the KS potential can be expressed as a component that acts on the total electron density,
\begin{equation}
  v_s^\text{tot}[\rho,Q](\boldsymbol{r}) = \frac{1}{2} \Big( v_s^\alpha(\boldsymbol{r}) + v_s^\beta(\boldsymbol{r})\Big)
  = v_\text{ext}(\boldsymbol{r}) + v_\text{Coul}[\rho](\boldsymbol{r}) + v_{\text{xc}}^\text{tot}[\rho,Q](\boldsymbol{r}),
\end{equation}
and one that acts on the spin density,
\begin{equation}
  v_s^\text{spin}[\rho,Q](\boldsymbol{r}) =  \frac{1}{2} \Big( v_s^\alpha(\boldsymbol{r}) - v_s^\beta(\boldsymbol{r})\Big) 
  = v_{\text{xc}}^\text{spin}[\rho,Q](\boldsymbol{r}),
\end{equation}
where the total and spin exchange--correlation potential are given by
\begin{equation}
  v_\text{xc}^\text{tot}[\rho,Q](\boldsymbol{r}) = \frac{1}{q_e} \frac{\delta E_\text{xc}^{(u)}[\rho,Q]}{\delta\rho(\boldsymbol{r})}
   \qquad \text{and} \qquad
  v_\text{xc}^\text{spin}[\rho,Q](\boldsymbol{r}) = \frac{1}{q_e} \frac{\delta E_\text{xc}^{(u)}[\rho,Q]}{\delta Q(\boldsymbol{r})}.
\end{equation}
These expressions will be used in the following section for comparing with the spin-restricted theory.

Even though the electron density and the spin density calculated from this $\Psi_{s,0}^{(u)}$ are equal
to those of the fully interacting system,
it is important to realize that $\Psi_{s,0}^{(u)}$ does \textbf{not} agree with the ground-state wavefunction $\Psi_0$ 
of the true interacting system. In particular, $\Psi_0$ can always be chosen as an eigenfunction
of $\hat{\boldsymbol{S}}^2$, whereas by construction, $\Psi_{s,0}^{(u)}$ is not an eigenfunction of $\hat{\boldsymbol{S}}^2$ 
for $S>0$. Thus, within an exact formulation of unrestricted KS-DFT, the wavefunction of the
KS reference system is always spin contaminated for $S>0$.

\subsection{Comparison of Restricted and Unrestricted Formulation}

The restricted and the unrestricted formulation of KS-DFT are based on different definitions of the noninteracting
reference system for open-shell systems. In the spin-restricted case, the reference system is chosen such that its total electron 
density $\rho_s(\boldsymbol{r})$ agrees with the one of the fully interacting system, while its spin density $Q_s(\boldsymbol{r})$
usually differs from the one of the interacting system. On the other hand, in the spin-unrestricted case the reference
system is defined such that both its total electron density and its spin density agree with those of the fully interacting
system.

These different definitions of the noninteracting reference system have implications for the treatment of spin in KS-DFT.
In the spin-restricted case, the wavefunction $\Psi_s$ of the noninteracting reference system can always be chosen as
an eigenfunction of $\hat{\boldsymbol{S}}^2$. Nevertheless, the corresponding eigenvalue $\langle \hat{\boldsymbol{S}}^2 \rangle 
= S(S+1) \hbar^2$ does not necessarily agree with the one obtained for the true interacting system. However, it is possible to
require this equality with an additional constraint on the noninteracting reference system. In the spin-unrestricted case,
the wavefunction $\Psi_s^{(u)}$ of the reference system is not an eigenfunction of $\hat{\boldsymbol{S}}^2$, i.e., it is spin-contaminated.
This is a direct consequence of the requirement that the correct spin density is obtained. Thus, the expectation value
of $\hat{\boldsymbol{S}}^2$ becomes a complicated functional of the electron density \cite{wang_evaluation_1995,
cohen_evaluation_2007}.  Of course, the exact ground state density will still correspond to an interacting wavefunction
that is an eigenfunction of $\hat{\boldsymbol{S}}^2$.

In both the restricted and in the
unrestricted case, the wavefunction of the noninteracting reference system is an eigenfunction of $\hat{S}_z$.
Only in the spin-unrestricted case it is guaranteed that the corresponding eigenvalue $M_S$ is the same as for the fully
interacting system, but also in the spin-restricted case it can be chosen accordingly.
These differences between the restricted and the unrestricted formulation are summarized 
in Table~\ref{tab:comparison-properties}. One important observation is that it is impossible to set up
a KS-DFT formalism such that for the noninteracting reference system one obtains both the
correct spin density \textit{and} a wavefunction that is an eigenfunction of $\hat{\boldsymbol{S}}^2$ 
(see also the discussion of this issue in Refs.~\cite{pople_spin-unrestricted_1995,perdew_fundamental_2009}).

The different definitions of the noninteracting reference system in the restricted and the unrestricted formulations of 
KS-DFT also imply different definitions of the noninteracting kinetic energy, the exchange--correlation energy, and 
the exchange--correlation potential. These definitions are collected in Table~\ref{tab:comparison-functionals}. First of all,
the use of different reference systems leads to different definitions of the noninteracting kinetic energy. In the spin-restricted
case, $T_s[\rho]$ is defined as the kinetic energy of a system of noninteracting electrons with the total electron density 
$\rho(\boldsymbol{r})$ and is independent of the spin density $Q(\boldsymbol{r})$. In contrast, in the spin-unrestricted 
case $T_s^{(u)}[\rho,Q]$ is defined as the kinetic energy of a system of noninteracting electrons with the total electron 
density $\rho(\boldsymbol{r})$ and the spin density $Q(\boldsymbol{r})$. These differ by the ``unrestricted'' contribution
to the noninteracting kinetic energy,
\begin{equation}
    T_u[\rho,Q] = T_s[\rho] - T_s^{(u)}[\rho,Q].
\end{equation}
Only if the spin density vanishes, the restricted and the unrestricted definitions of the noninteracting kinetic energy are 
identical, i.e., $T_u[\rho, Q=0] = 0$.
  
Because of these different definitions of the noninteracting kinetic energy, a different decomposition of the HK functional
is introduced in the restricted and unrestricted formalisms, respectively, which in turn leads to different definitions of
the exchange--correlation energy. These are related by
\begin{equation}
  E_\text{xc}^{(u)}[\rho,Q] 
                                         = E_\text{xc}[\rho,Q] + T_u[\rho,Q].
\end{equation}

One important difference between the two formalisms is that in spin-restricted KS-DFT, the exchange--correlation
energy $E_\text{xc}[\rho,Q]$ is the only contribution to the HK functional that depends on the spin density, whereas in 
the spin-unrestricted theory both the exchange--correlation energy $E_\text{xc}^{(u)}[\rho,Q]$ and the noninteracting 
kinetic energy $T_s^{(u)}[\rho,Q]$ depend on the spin density. Therefore, the fractional spin condition of 
Eq.~\eqref{eq:fractional-spin}, formulated for the HK functional in Sec.~\ref{sec:fractional-spin}, 
leads to different exact conditions for the exchange--correlation functional. In the spin-restricted case, the
fractional spin condition applies directly to the exchange--correlation energy,
\begin{equation}
  E_\text{xc}\big[\rho, \alpha \, Q^{M_S=S}[\rho]\big] = \text{const.} \quad \text{for} \ -1 \le \alpha \le +1,
\end{equation}
whereas in the spin-unrestricted case it applies to the sum of the exchange--correlation energy and the
noninteracting kinetic energy,
\begin{equation}
  T_s^{(u)}\big[\rho, \alpha \, Q^{M_S=S}[\rho]\big] + E_\text{xc}^{(u)}\big[\rho, \alpha \, Q^{M_S=S}[\rho]\big]= \text{const.} \quad \text{for} \ -1 \le \alpha \le +1.
\end{equation}

Finally, the exchange--correlation potential (and thus also the resulting KS potential) differs in the two
formalisms. In spin-restricted KS-DFT, the exchange--correlation potential $v_\text{xc}[\rho]$ depends only on the
total electron density and acts on electrons of both spin. On the other hand, in spin-unrestricted KS-DFT the 
exchange--correlation potential is different for $\alpha$- and $\beta$-electrons, i.e., it has two distinct components.
The component of the exchange--correlation potential acting on the total electron density is given by,
\begin{align}
  v_\text{xc}^{\text{tot}}[\rho,Q](\boldsymbol{r}) 
  &= \frac{1}{q_e} \frac{\delta E_\text{xc}^{(u)}[\rho,Q]}{\delta\rho(\boldsymbol{r})}
    =  \frac{1}{q_e} \left( \frac{\delta E_\text{xc}[\rho,Q]}{\delta\rho(\boldsymbol{r}) } + \frac{\delta T_u[\rho,Q]}{\delta\rho(\boldsymbol{r}) }  \right)
  \nonumber \\
  &= v_\text{xc}[\rho,Q](\boldsymbol{r}) + v_u[\rho,Q](\boldsymbol{r}).
\end{align}
Here, the first term is the exchange--correlation potential in the spin-restricted formalism, while the
second term $v_u[\rho,Q](\boldsymbol{r}) = (1/q_e) \, \delta T_u[\rho,Q] / \delta\rho(\boldsymbol{r})$ is given by the 
functional derivative of $T_u[\rho,Q]$. It appears because of the different definitions of the exchange--correlation
energy in the restricted and unrestricted theories.
The component of the exchange--correlation potential acting on the spin density is given by 
\begin{align}
  v_\text{xc}^{\text{spin}}[\rho,Q](\boldsymbol{r})
  &= \frac{1}{q_e} \frac{\delta E_\text{xc}^{(u)}[\rho,Q]}{\delta Q(\boldsymbol{r})}
    =  \frac{1}{q_e} \left( \frac{\delta E_\text{xc}[\rho,Q]}{\delta Q(\boldsymbol{r}) } + \frac{\delta T_u[\rho,Q]}{\delta Q(\boldsymbol{r})} \right).
\end{align}
For the ground-state electron and spin densities, the first term vanishes according to Eq.~\eqref{eq:spindens-min-restricted}, 
and the above expression for $v_\text{xc}^{\text{spin}}[\rho,Q](\boldsymbol{r})$ reduces to the Euler--Lagrange equation for 
the spin density [cf. Eqs.~\eqref{eq:euler-unrestricted-1} and~\eqref{eq:euler-unrestricted-2}].

In summary, the KS potential in the spin-unrestricted case differs from the spin-restricted KS potential 
$v_\text{xc}[\rho](\boldsymbol{r})$ by (a) an additional component $v_\text{xc}^{\text{spin}}[\rho,Q](\boldsymbol{r}) \, \sigma_z$
acting on the spin density, and (b) a correction to the spin-independent potential $v_u[\rho,Q](\boldsymbol{r})$. Thus, 
starting from a spin-restricted reference system  with the total electron density $\rho(\boldsymbol{r})$ and with 
a spin density that differs from the spin density $Q_s(\boldsymbol{r})$ of the interacting system, the KS potential 
is modified such that its spin density becomes equal to the one of the fully interacting system $Q(\boldsymbol{r})$. 
To achieve this, the spin potential $v_\text{xc}^{\text{spin}}[\rho,Q](\boldsymbol{r}) \, \sigma_z$ has 
to be introduced. However, with the total potential kept fixed, this would lead to a change of the 
total electron density, and in order to keep $\rho(\boldsymbol{r})$ unchanged, the correction 
$v_u[\rho,Q](\boldsymbol{r})$ is needed.

\subsection{Spin States in KS-DFT}

So far, we have only considered a spin-state independent theory, i.e., with the exact exchange--correlation
functionals, the spin-restricted and spin-unrestricted KS-DFT formalism discussed above will lead to the
ground-state, irrespective of its spin symmetry. As discussed in Section~\ref{sec:spin-states-hk}, targeting
the lowest state of a given spin symmetry (i.e., with a specific eigenvalue $S(S+1)$ of $\hat{\boldsymbol{S}}^2$)
requires a spin-state specific Hohenberg--Kohn functional $F_\text{HK}^S[\rho]$ as defined in 
Eq.~\eqref{eq:fhk-spinstates}.

In the spin-restricted case, this is formally possible by employing the spin-state independent definition of
the noninteracting kinetic energy, which results in a spin-state specific exchange--correlation functional. In
practice, a different strategy is followed: The noninteracting reference system is defined such that it is
described by a single Slater determinant with $M_S=S$. This can be achieved by defining the spin-state
specific noninteracting kinetic energy as
\begin{equation}
  T_s^S[\rho] = \min_{\Phi^{M_S=S}\rightarrow \rho} \big\langle \Phi^{M_S=S} | \hat{T} | \Phi^{M_S=S} \rangle,
\end{equation}
where $\Phi^{M_S=S}$ is a Slater determinant with $M_S=S$. Such a Slater determinant is always an
eigenfunction of $\hat{\boldsymbol{S}}^2$ with eigenvalue $S(S+1)$. Thus, it is ensured that the
wavefunction of the noninteracting reference system has the same spin symmetry as the wavefunction 
of the true interacting system. However, as always in spin-restricted KS-DFT, for $S>0$ the spin 
density $Q_s^{M_S=S}$ of the noninteracting reference system will differ from the one of the fully
interacting system.

With this definition of a spin-state specific noninteracting kinetic energy, the spin-state specific 
exchange--correlation energy is given by
\begin{equation}
  E_\text{xc}^S[\rho] = F_\text{HK}^S[\rho] - T_s^S[\rho] - J[\rho] .
\end{equation}
One protocol for constructing this spin-state dependent exchange--correlation functional then
proceeds by using the spin density $Q_s^{M_S=S}$ of the noninteracting reference system to
distinguish the different spin states. To this end, $E_\text{xc}^S[\rho]$ is expressed as
\begin{equation}
  E_\text{xc}^S[\rho] = E_\text{xc}^{(ss)}\big[\rho, Q_s^{M_S=S}[\rho]\big].
\end{equation} 
Here, $E_\text{xc}^{(ss)}[\rho,Q]$ is \textit{not} equal to the spin-resolved exchange--correlation
functional $E_\text{xc}[\rho,Q]$ defined in Eq.~\eqref{eq:exc-restr-spinresolved}. Its dependence on 
$Q$ does not describe the spin-density dependence of the exchange--correlation energy, but instead 
introduces the spin-state dependence. This is indicated by the superscript ``${(ss)}$''. The multiplet-DFT
scheme of Daul  \cite{daul_density_1994,daul_calculation_1995} and the restricted open-shell KS
(ROKS) scheme \cite{filatov_spin-restricted_1998,filatov_spin-restricted_1999,illas_spin_2006,
frank_molecular_1998,grimm_restricted_2003,nonnenberg_restricted_2003} as well as related 
approaches \cite{della_sala_open-shell_2003,vitale_open-shell_2005} proceed along 
these lines, but usually include additional ideas originating from Hartree--Fock 
theory \cite{bagus_singlettriplet_1975,ziegler_calculation_1977}.

It appears that if applied in a spin-restricted formalism, all available approximate exchange--correlation
functionals have to be understood as approximations to $E_\text{xc}^{(ss)}[\rho,Q]$ and not as approximations
to $E_\text{xc}[\rho,Q]$. This has important consequences for the construction of such approximate 
exchange--correlation functional. In particular, one has to realize that $E_\text{xc}^{(ss)}[\rho,Q]$
does not fulfill the fractional spin condition discussed in Section~\ref{sec:fractional-spin} and thus this
condition should not be included when constructing approximations to it.

In addition, in spin-restricted KS-DFT the ground-state spin density is not directly available and has to be 
determined after calculating the ground-state electron density by minimizing $E_\text{xc}[\rho,Q]$ with 
respect to $Q$ as discussed in Section~\ref{sec:spindens-restr}. In this step, one has to use $E_\text{xc}[\rho,Q]$
--- which includes the correct spin-density dependence --- instead of $E_\text{xc}^{(ss)}[\rho,Q]$.
Thus, the construction of a different class of approximate exchange--correlation functionals that
include the spin-density dependence by approximating $E_\text{xc}[\rho,Q]$  (or its spin-state specific 
analogue $E^S_\text{xc}[\rho,Q]$) instead of $E_\text{xc}^{(ss)}[\rho,Q]$ would be required.
Of course, the fractional spin condition applies to $E_\text{xc}[\rho,Q]$ and $E^S_\text{xc}[\rho,Q]$, and should 
also be incorporated when constructing such approximations.

In spin-unrestricted KS-DFT, it is not possible to require that the noninteracting reference system is an
eigenfunction of $\hat{\boldsymbol{S}}^2$ \cite{pople_spin-unrestricted_1995,perdew_fundamental_2009}. 
Thus, it is not possible to define a spin-state specific analogue of $T_s^{(u)}[\rho,Q]$. Consequently, the 
spin-state dependence only enters the exchange--correlation functional, which becomes
\begin{equation}
  E_\text{xc}^{(u),S}[\rho,Q] = F_\text{HK}^S[\rho,Q] - T_s^{(u)}[\rho,Q] - J[\rho].
\end{equation}
For constructing approximations to this spin-state specific exchange--correlation functional, usually
a strategy similar to the one in spin-restricted KS-DFT is applied. To this end, the description is restricted to
the case of $M_S=S$, i.e., only the maximal eigenvalue of $\hat{S}_z$ is allowed. Then, the spin-state
specific exchange--correlation functional can be expressed as
\begin{equation}
  E_\text{xc}^{(u),S}[\rho,Q] = E_\text{xc}^{(u,ss)}\big[\rho,Q^{M_S=S}\big] .
\end{equation}
Here, different spin states can be distinguished based on the integral of the spin density. However, the 
functional $E_\text{xc}^{(u),ss}[\rho,Q]$ does not describe the correct spin-density dependence
of $E_\text{xc}^{(u)}[\rho,Q]$ anymore. Neither does $E_\text{xc}^{(u),S}[\rho,Q]$, which is limited to
spin densities corresponding to $M_S=S$. However, the spin densities $Q^{M_S}$ corresponding to
other eigenvalues of $\hat{S}_z$ can be obtained from the scaling relation of Eq.~\eqref{eq:spindens-ms-scaling}.
Again, it is important to realize that the fractional spin condition does not apply to $E_\text{xc}^{(u),ss}[\rho,Q]$.

The idea of using the spin density as a means to distinguish different spin states in a spin-unrestricted
KS-DFT formalism is taken even further in broken-symmetry DFT  \cite{noodleman_valence_1981,
reiher_definition_2007,neese_prediction_2009}, where the requirement that the spin density of 
the noninteracting reference system matches the correct spin density of the fully interacting system 
is sacrificed in favor of obtaining accurate energetics for low-spin states. Consequently, it has been 
suggested that in this case, the spin density in fact serves to describe the (spin-state specific)
on-top pair density \cite{perdew_escaping_1995}. If broken--symmetry DFT calculations are interpreted 
in this way, one would need to determine the spin density in a separate step from the minimization 
of  $E_\text{xc}[\rho,Q]$  (or its spin-state  specific analogue $E^S_\text{xc}[\rho,Q]$), as discussed 
above for spin-restricted KS-DFT.

\section{Spin in Relativistic DFT}
\label{sec:relativistic-dft}

\subsection{Spin and Current in Relativistic Quantum Mechanics}

So far, the discussion has focused on spin in nonrelativistic DFT. For the
sake of completeness, we now consider the generalization to the more fundamental
relativistic regime. The relativistic theory relies on Dirac's semi-classical
theory of the electron (see Ref.~\cite{reiher_relativistic_2009} for a detailed account). 
This quantum theory describes the relativistic motion of the electron in a classical external
electromagnetic field, represented by the scalar and vector potentials $\phi_\text{ext}(\boldsymbol{r})$ and 
$\boldsymbol{A}_\text{ext}(\boldsymbol{r})$. The equation of motion reads in this case (in Gaussian units),
\begin{equation}
\label{eq:dirac-1el}
\hat{h}^D\psi(\boldsymbol{r},t) =\biggl[c\boldsymbol{\alpha}\cdot\hat{\boldsymbol{p}}
+ \beta m_e c^2 +q_e \Bigl(\phi_\text{ext}(\boldsymbol{r}) - \boldsymbol{\alpha}\cdot\boldsymbol{A}_\text{ext}(\boldsymbol{r})\Bigr) \biggr]
\psi(\boldsymbol{r},t) =  {\rm i}\hbar \frac{\partial }{\partial t} \psi(\boldsymbol{r},t),
\end{equation}
which yields an energy eigenvalue equation for determining the stationary states when 
the right hand side is replaced by $E\, \psi(\boldsymbol{r},t)$.
The Dirac Hamiltonian $\hat{h}^D$ consists of the kinetic energy operator
$c\boldsymbol{\alpha}\cdot\hat{\boldsymbol{p}}$, the rest energy term $\beta m_e c^2$, and the
interaction operator $q_e \bigl(\phi_\text{ext}(\boldsymbol{r}) - \boldsymbol{\alpha}\cdot\boldsymbol{A}(\boldsymbol{r})\bigr)$, 
where $c$ is the speed of light and $\hat{\boldsymbol{p}} = -{\rm i}\hbar{\nabla}$ is the momentum operator. 
The Dirac matrices are contained in the parameters $\boldsymbol{\alpha} = (\alpha_x, \alpha_y, \alpha_z)$ 
and $\beta$, which are in the standard representation,
\begin{equation}
  \alpha^i = 
  \begin{pmatrix}
  0 & \sigma_i \\
  \sigma_i & 0
  \end{pmatrix}
  \qquad \text{and} \qquad 
  \beta = 
  \begin{pmatrix}
  1_2 & 0 \\
  0 & -1_2
  \end{pmatrix},
\end{equation}
where $\sigma_i$ are the the Pauli spin matrices $\sigma_x$, $\sigma_y$, and $\sigma_z$
defined in Eq.~\eqref{eq:pauli-matrices}.
A consequence of these four-dimensional operators is that the Dirac Hamiltonian $\hat{h}^D$ is 
a $4 \times 4$ matrix operator with a four-component eigenvector $\psi(\boldsymbol{r}, t)$, 
the so-called 4-spinor. Such a four-component description naturally includes spin, which had 
to be introduced in an \textit{ad hoc} fashion in the nonrelativistic theory.

In analogy to the nonrelativistic case, one can define the relativistic spin operator as
\begin{equation}
\label{eq:s-op-rel}
  \hat{\boldsymbol{s}} = \frac{\hbar}{2} \boldsymbol{\sigma}^{(4)}
  = \frac{\hbar}{2} (\sigma^{(4)}_x,\sigma^{(4)}_y,\sigma^{(4)}_z)^T
  \qquad
  \text{with}
  \quad
  \sigma_i^{(4)} 
  =   
  \begin{pmatrix}
    \sigma_i & 0 \\ 
    0 & \sigma_i \\ 
  \end{pmatrix}.
\end{equation}
This operators still obey the commutation relations of an angular momentum.
However, in the relativistic case $\hat{\boldsymbol{s}}^2$ and $\hat{s}_z$ do not commute
with the Hamiltonian. Therefore, eigenstates of $\hat{h}^D$ cannot be chosen as
eigenfunctions of $\hat{\boldsymbol{s}}^2$ and $\hat{s}_z$ anymore and spin is thus not a
good quantum number in relativistic theory. In spherically symmetric systems such as atoms, 
one can consider the total angular momentum instead. This also has the consequence that,
in contrast to the nonrelativistic case, the expectation value of $\hat{s}_z$ will depend on the
choice of the quantization axis. 

In the case of many electrons, the relativistic wavefunction again assumes a tensor structure, i.e.,
for $N$ electrons, the wavefunction formally has $4^N$ components. It has to fulfill the Pauli
principle by being antisymmetric with respect to the exchange of any two electrons. The relativistic 
many-electron Hamiltonian is then given by (neglecting projectors on positive-energy solutions for 
the sake of brevity),
\begin{equation}
  \hat{H}^D = \sum_{i=1}^N \hat{h}^D(i) + \sum_{i=1}^N \sum_{j=i+1}^N \hat{g}(i,j), 
\end{equation}
where $\hat{h}^D(i)$ is the one-electron Hamiltonian of Eq.~\eqref{eq:dirac-1el} acting on electron $i$
and $\hat{g}(i,j)$ is the operator describing the interaction between electrons $i$ and $j$. 
The form of the exact electron--electron interaction operator can be derived from quantum electrodynamics, 
but usually only approximate forms are employed in practice.
The simplest approximation is to employ the nonrelativistic Coulomb operator. However, the resulting 
Dirac--Coulomb Hamiltonian is not Lorentz invariant. This approximation is improved by the Dirac--Coulomb--Gaunt Hamiltonian, 
which also includes the unretarded magnetic interaction between electrons, whereas the Dirac--Coulomb--Breit 
Hamiltonian approximately describes the retarded electromagnetic interaction (for a detailed discussion, see 
chapter~8 in Ref.~\cite{reiher_relativistic_2009}).

As in nonrelativistic theory, an electron density and a current density can be defined in the relativistic 
many-electron theory such that they fulfill a continuity equation. This results in the definition of the 
electron density as \cite{reiher_relativistic_2009,fux_electron_2012}
\begin{align}
  \rho(\boldsymbol{r}) &= N \int \Psi^\dagger(\boldsymbol{r}, \boldsymbol{r}_2, \dotsc, \boldsymbol{r}_N) \,
                                               \Psi(\boldsymbol{r}, \boldsymbol{r}_2, \dotsc, \boldsymbol{r}_N) \, {\rm d}^3r_2 \dotsm {\rm d}^3r_N
\end{align}
and of the current density as
\begin{align}
  \boldsymbol{j}(\boldsymbol{r}) &= c \, N \int \Psi^\dagger(\boldsymbol{r}, \boldsymbol{r}_2, \dotsc, \boldsymbol{r}_N) \, \boldsymbol{\alpha}_1 \,
                                               \Psi(\boldsymbol{r}, \boldsymbol{r}_2, \dotsc, \boldsymbol{r}_N) \, {\rm d}^3r_2 \dotsm {\rm d}^3r_N
\end{align}
where $\boldsymbol{\alpha}_1$ indicates that the Dirac matrices act on the first electron and the dagger denotes the
transposed and complex conjugate spinor. This definition of the current density still holds in the presence of external 
magnetic fields.

\subsection{Relativistic Current-DFT}

A central aspect of relativistic theories is that all fundamental physical
equations must preserve their form under Lorentz transformations from one
inertial frame of reference to another one. This requires that the equations are
castable in tensorial form. Hence, all quantities are joined to 4-vectors,
which are basic physical quantities in any relativistic theory. For
instance, the electromagnetic potentials are joined to yield the 4-potential
$A^\mu=(\phi,\boldsymbol{A})$. This is the reason why a relativistic theory has
to include both the scalar and the vector potentials simultaneously. The density 
$\rho(\boldsymbol{r})$ is also part of a 4-vector, namely of the 4-current $j^\mu(\boldsymbol{r})$. 
The other three components are given by the current density $\boldsymbol{j}(\boldsymbol{r})$ 
such  that $j^\mu=(c\rho,\boldsymbol{j})$.

Now, a similar decomposition of the energy expectation value as in the nonrelativistic
case can be performed,
\begin{equation}
  E = \langle \Psi | \hat{H}^D | \Psi \rangle 
  = \langle \Psi | \hat{T}^D | \Psi \rangle + \langle \Psi | \hat{V}_\text{ext} | \Psi \rangle + \langle \Psi | \hat{V}_{ee} | \Psi \rangle, 
\end{equation}
where $\hat{T}^D$ is the relativistic ``kinetic energy'' operator, collecting all terms of the one-electron Hamiltonian $\hat{h}^D$ 
containing the Dirac matrices $\boldsymbol{\alpha}$ and $\beta$ (i.e., the kinetic energy and the rest energy terms), 
$\hat{V}_\text{ext}$ consists of the remaining one-electron terms and describes the interaction with the external 
electromagnetic potentials, and $\hat{V}_{ee}$ is the electron--electron interaction operator. The second term 
can be calculated directly from the 4-current, without the need to know the full wavefunction, as 
\begin{align}
  \label{eq:vext-rel}
  \langle \Psi | \hat{V}_\text{ext} | \Psi \rangle 
  &= \frac{q_e}{c} \int j_\mu(\boldsymbol{r}) A^\mu_\text{ext}(\boldsymbol{r}) \, {\rm d}^3r 
  \nonumber \\
   &= q_e \int \rho(\boldsymbol{r}) \phi_\text{ext}(\boldsymbol{r}) \, {\rm d}^3r 
         + \frac{q_e}{c} \int \boldsymbol{j}(\boldsymbol{r}) \cdot \boldsymbol{A}_\text{ext}(\boldsymbol{r}) \, {\rm d}^3r = V_\text{ext}[j^\mu] , 
\end{align}
where Einstein's convention of implicit summation over repeated lower and upper indices has been employed. 
The first term is the same as in the nonrelativistic case and describes the electrostatic interaction of the
electron density with the external potential, whereas the second term accounts for the interaction of the current
density with the external magnetic field.

A 4-current may be understood as the source of a 4-potential which can be calculated from the relativistic 
generalization of the Poisson equation of electrostatics,
\begin{equation}
\square A^\mu(\boldsymbol{r}) = \frac{4\pi}{c} j^\mu(\boldsymbol{r}),
\end{equation}
with the D'Alembertian operator being the Minkowski space generalization of the 
three-dimensional Laplacian, i.e., it is defined as $\square= \frac{1}{c^2} \frac{\partial^2}{\partial t^2 } - \Delta$.
From the solution of this equation, one obtains for the 4-potential generated by the electronic 4-current $j^\mu$,
\begin{equation}
  A^\mu[j^\mu] = \frac{q_e}{c} \int \frac{j^\mu(\boldsymbol{r}')}{|\boldsymbol{r}-\boldsymbol{r}'|}  \, {\rm d}^3r'.
\end{equation}
The classical interaction energy $J[j^\mu]$ of the 4-current $j^\mu$ with itself (i.e., the relativistic analogue of
the classical Coulomb interaction in nonrelativistic theory) is then given by
\begin{align}
\label{eq:rel-coulomb}
  J[j^\mu] &= \frac{q_e^2}{2 c^2} \int \frac{j_\mu(\boldsymbol{r}) j^\mu(\boldsymbol{r}')}{|\boldsymbol{r} - \boldsymbol{r}'|} \, {\rm d}^3r {\rm d}^3r'
  \nonumber \\
  &= \frac{q_e^2}{2} \int \frac{\rho(\boldsymbol{r}) \rho(\boldsymbol{r}')}{|\boldsymbol{r} - \boldsymbol{r}'|} \, {\rm d}^3r {\rm d}^3r'
        + \frac{q_e^2}{2 c^2} \int \frac{\boldsymbol{j}(\boldsymbol{r}) \cdot \boldsymbol{j}(\boldsymbol{r}')}{|\boldsymbol{r} - \boldsymbol{r}'|} \, {\rm d}^3r {\rm d}^3r'.
\end{align}
Here, the first term is the classical electrostatic (Coulomb) interaction $J[\rho]$, whereas the second term accounts for the 
magnetic interaction of the electrons. Note that this second term should only be present if the Gaunt or Breit interaction 
has been included in the Hamiltonian.

The analogy between $V_\text{ext}[j^\mu]$ and $J[j^\mu]$ [Eqs.~\eqref{eq:vext-rel} and~\eqref{eq:rel-coulomb}] and 
their nonrelativistic counterparts $V_\text{ext}[\rho]$ and $J[\rho]$ suggests that the fundamental quantity for a 
relativistic formulation of DFT is actually the 4-current, i.e., the combination of the electronic density $\rho(\boldsymbol{r})$ 
and the current density $\boldsymbol{j}(\boldsymbol{r})$. For this reason, the relativistic generalization of standard DFT 
has been known as current-density functional theory (CDFT) \cite{rel-hk,macdonald_relativistic_1979,engel_relativistic_2002,
engel_relativistic_2003}. 

It can be shown that in relativistic theory, an analogue of the HK theorem exists: The 4-current $j^\mu(\boldsymbol{r})$
uniquely determines the external 4-potential $A^\mu(\boldsymbol{r})$ up to a gauge transformation \cite{rel-hk}. Therefore, 
it is possible to consider the energy $E[j^\mu]$ as a functional of the 4-current $j^\mu(\boldsymbol{r})$. As in the 
nonrelativistic case, one then proceeds by introducing a reference system of noninteracting electrons with 
the relativistic Hamiltonian
\begin{equation}
\hat{H}^D_s = \sum_{i=1}^N c\boldsymbol{\alpha}(i)\cdot\hat{\boldsymbol{p}}_i
+ \beta(i) m_e c^2 +q_e \Bigl(\phi_s(\boldsymbol{r}_i) - \boldsymbol{\alpha}(i)\cdot\boldsymbol{A}_s(\boldsymbol{r}_i)\Bigr).
\end{equation}
The eigenfunctions of this noninteracting Hamiltonian are given by Slater determinants $\Psi_s = |\varphi_1, \varphi_2, \dotsc, \varphi_N|$ 
constructed from 4-spinors $\varphi_i(\boldsymbol{r})$, which solve the relativistic one-electron KS equations
\begin{equation}
  \bigg[ c\boldsymbol{\alpha}\cdot\hat{\boldsymbol{p}}
+ \beta m_e c^2 +q_e \Bigl(\phi_s(\boldsymbol{r}) - \boldsymbol{\alpha}\cdot\boldsymbol{A}_s(\boldsymbol{r})\Bigr)\bigg] 
 \varphi_i(\boldsymbol{r}) = \epsilon_i \varphi_i(\boldsymbol{r}).
\end{equation}
For the resulting single Slater determinant, the electron density is given by
\begin{equation}
\rho(\boldsymbol{r}) = \sum_i^{\rm occ} \varphi_i^\dagger(\boldsymbol{r}) \varphi_i(\boldsymbol{r}), 
\end{equation}
an the relativistic current density reads
\begin{equation}
\label{currentdens}
\boldsymbol{j}(\boldsymbol{r}) = c \sum_i^{\rm occ} \varphi_i^\dagger(\boldsymbol{r}) \, \boldsymbol{\alpha} \, \varphi_i(\boldsymbol{r}).
\end{equation}

The noninteracting kinetic energy $T_s[j^\mu]$ can then be defined as the kinetic energy $T_s=\langle\Psi_s|\hat{T}^D|\Psi_s\rangle$
of such a noninteracting system with the 4-current $j^\mu$, which allows for a similar decomposition of the total energy 
functional as in nonrelativistic DFT, i.e.,
\begin{equation}
  E[j^\mu] = T_s[j^\mu] + V_\text{ext}[j^\mu] + J[j^\mu] + E_\text{xc}[j^\mu].
\end{equation}
Here, the exchange--correlation energy functional is defined as containing all the energy contributions not 
accounted for by the first three terms. In analogy to nonrelativistic KS-DFT, it can then be shown that the 
4-current $j^\mu(\boldsymbol{r})$ of the true interacting system can be determined from the self-consistent 
solution of the relativistic KS equations for a noninteracting system\cite{rel-hk,saue_fourcomponent_2002}, 
with the 4-potential 
\begin{equation}
 A^\mu_s(\boldsymbol{r}) = A^\mu_\text{ext}(\boldsymbol{r}) + A^\mu[j^\mu](\boldsymbol{r}) + \frac{1}{q_e} \frac{\delta E_\text{xc}[j^\mu]}{\delta j^\mu(\boldsymbol{r})}.
\end{equation}
The functional derivatives of the exchange--correlation functional with respect to the components of the 4-current defines 
the exchange--correlation potential $v_\text{xc}^\mu[j^\mu](\boldsymbol{r})$, which now assumes a four-component form.

\subsection{Electron Density and Spin Density in Relativistic DFT}

In view of the previous discussions concerning nonrelativistic KS-DFT, we now face the 
following question: (i) Is it still possible to formulate a relativistic DFT in terms of the electron density 
only, (ii) how is the  fundamental 4-current related to the density and spin density considered as 
fundamental quantities in nonrelativistic DFT, and (iii) how do the nonrelativistic restricted and 
unrestricted formulations of KS-DFT emerge from the relativistic framework?

Regarding the first question, under some additional assumptions it is indeed possible to prove a relativistic
HK theorem for the electron density only \cite{engel_relativistic_1996,saue_fourcomponent_2002,
van_wllen_relativistic_2010}: In the case considered throughout this paper, the external potential
is the {\it electrostatic} potential of all atomic nuclei in a molecule that are at rest (Born--Oppenheimer
approximation). As the nuclei are not moving, they do not create magnetic fields. In addition, we neglect
any magnetic fields that stem from nuclear spins and assume that there are no additional external electromagnetic 
fields. As a consequence, the external electromagnetic 4-potential $A^\mu$ only contains the time-independent 
scalar potential of the atomic nuclei, i.e., $\phi_\text{ext}(\boldsymbol{r}) = v_\text{nuc}(\boldsymbol{r})$ and
$\boldsymbol{A}(\boldsymbol{r}) = 0$. This assumption is also common practice in almost every relativistic 
quantum chemical calculation. Then, it can be shown that the external scalar potential $\phi_\text{ext}(\boldsymbol{r})$ within
this specific reference frame is (up to a constant) uniquely determined by the electron density $\rho(\boldsymbol{r})$ 
only  \cite{engel_relativistic_1996,saue_fourcomponent_2002,van_wllen_relativistic_2010}. Within such a 
framework, the relativistic total energy functional becomes
\begin{equation}
  E[\rho] = T_s[j^\mu[\rho]] + V_\text{ext}[\rho] + J[j^\mu[\rho]] + E_\text{xc}[j^\mu[\rho]].
\end{equation}
Here, the current density $\boldsymbol{j}(\boldsymbol{r})$ is still required, but it is now uniquely determined 
by the electron density (i.e., $\boldsymbol{j} = \boldsymbol{j}[\rho]$), in the same way in which the spin density
is determined by the electron density in nonrelativistic DFT. If a magnetic interaction between the
electron is not included (i.e., the Dirac--Coulomb Hamiltonian is employed), also the analogue of the 
Coulomb interaction reduces to a functional $J[\rho]$ of the density only  \cite{saue_fourcomponent_2002}. Moreover,
it now becomes possible to set up theories in which not the full 4-current is used as fundamental variable, but 
only the parts of it that are related to the total electron density and the spin density. This will be discussed further 
in the following subsection.

In order to understand the relation of the 4-current to the spin density, it
is important to realize that the definition of the current density (naturally) involves a
velocity operator, which is in close analogy to classical mechanics (correspondence principle)
\cite{reiher_relativistic_2009}. The velocity operator in Dirac's theory of the electron follows from the
Heisenberg equation of motion applied to the position operator and turns out
to be $c\boldsymbol{\alpha}$. Hence, as the Dirac $\boldsymbol{\alpha}$ matrices contain the
Pauli spin matrices $\boldsymbol{\sigma}$, we see immediately that the current density
$\boldsymbol{j}$ carries the spin information \cite{fux_electron_2012}.

The relation to the spin density can be made more explicit by invoking a Gordon composition of the 
current density which separates it into a charge- and a spin-related current \cite{rel-hk,engel_density_2011}. 
For the one-electron case, one can carry out this decomposition by rewriting the Dirac eigenvalue equation as,
\begin{equation}
\label{eq:de-gordon}
  \psi(\boldsymbol{r}) = \frac{1}{m_e c^2} \bigg[ - c \, \beta \, \boldsymbol{\alpha} \cdot \Big(\hat{\boldsymbol{p}}
            - \frac{q_e}{c} \boldsymbol{A}_\text{ext}(\boldsymbol{r}) \Big) + \beta \Big( E - q_e \phi_\text{ext}(\boldsymbol{r}) \Big) \bigg] \psi(\boldsymbol{r})
\end{equation}
and by splitting up the definition of the current in a somewhat artificial way as
\begin{equation}
  \boldsymbol{j} = c \, \psi^\dagger(\boldsymbol{r}) \, \boldsymbol{\alpha} \, \psi(\boldsymbol{r}) 
  = \frac{c}{2} \, \psi^\dagger(\boldsymbol{r}) \, \boldsymbol{\alpha} \, \psi(\boldsymbol{r}) 
     + \frac{c}{2} \, \psi^\dagger(\boldsymbol{r}) \, \boldsymbol{\alpha} \, \psi(\boldsymbol{r}).
\end{equation}
Eq.~\eqref{eq:de-gordon} can now be used to replace $\psi$ in the first term and $\psi^\dagger$ in the
last term. As is shown in the Appendix, by exploiting the commutation relations of the Dirac matrices $\{\alpha^k,\beta\}=0$ 
and $\alpha^k\alpha^l = \frac{1}{2}[\alpha^k, \alpha^l] + \frac{1}{2} \{\alpha^k, \alpha^l\} = {\rm i} \varepsilon_{kln} \sigma_n^{(4)} + \delta_{kl}$,
one arrives at \cite{rel-hk,engel_density_2011} (see also pages~552--558 in Ref.~\cite{baym_lectures_1969})
\begin{align}
 \boldsymbol{j}(\boldsymbol{r})
 =& \ \frac{1}{2 m_e} \bigg[ \psi^\dagger(\boldsymbol{r}) \beta  \Big(\hat{\boldsymbol{p}} - \frac{q_e}{c} \boldsymbol{A}_\text{ext}(\boldsymbol{r}) \Big) \psi(\boldsymbol{r}) 
           + \Big(\hat{\boldsymbol{p}}^* - \frac{q_e}{c} \boldsymbol{A}_\text{ext}(\boldsymbol{r}) \Big) \psi^\dagger(\boldsymbol{r})  \beta \psi(\boldsymbol{r}) \bigg]
\nonumber \\
    &+ \ \frac{\hbar}{2 m_e} \boldsymbol{\nabla} \times \psi^\dagger(\boldsymbol{r}) (\beta \boldsymbol{\sigma}^{(4)}) \psi(\boldsymbol{r}),
\end{align}
where the 3-vector $\boldsymbol{\sigma}^{(4)} = (\sigma_x^{(4)}, \sigma_y^{(4)}, \sigma_z^{(4)})^T$ contains the $4\times4$ Pauli
matrices $\sigma_i^{(4)}$ introduced in Eq.~\eqref{eq:s-op-rel}.
In a many-electron system described by the Dirac--Coulomb Hamiltonian, a similar decomposition of the
current density can be performed (even though the derivation becomes slightly more complicated, see Appendix), 
and one obtains,
\begin{align}
  \label{eq:gordon-decomp}
  \boldsymbol{j}(\boldsymbol{r}_1) 
  =&\ N \frac{1}{2 m_e} \int \bigg[ \Psi^\dagger \beta_1  \Big(\hat{\boldsymbol{p}}_1 - \frac{q_e}{c} \boldsymbol{A}_\text{ext}(\boldsymbol{r}_1) \Big) \Psi 
           + \Big(\hat{\boldsymbol{p}}_1^* - \frac{q_e}{c} \boldsymbol{A}_\text{ext}(\boldsymbol{r}_1) \Big) \Psi^\dagger  \beta_1 \Psi \bigg]
              \, {\rm d}^3r_2 \dotsm {\rm d}^3r_N
\nonumber \\
  &+ N \frac{\hbar}{2 m_e} \boldsymbol{\nabla}_1 \times \int \Psi^\dagger \, \beta_1 \boldsymbol{\sigma}_1^{(4)} \, \Psi \, {\rm d}^3r_2 \dotsm {\rm d}^3r_N.
\end{align}
The first line of this expression resembles the definition of the current density in nonrelativistic quantum mechanics,
whereas the second term can be identified as arising from the electron spin. This can be made more apparent
by defining the magnetization (density),
\begin{equation}
  \boldsymbol{m}(\boldsymbol{r}_1) 
  =  N \int  \Psi^\dagger \, \beta_1 \boldsymbol{\sigma}_1^{(4)} \, \Psi \ {\rm d}^3r_2 \dotsm {\rm d}^3r_N.
\end{equation}
Then, the contribution of the second term of Eq.~\eqref{eq:gordon-decomp} to the interaction energy 
with the external electromagnetic potentials [cf. Eq.~\eqref{eq:vext-rel}] becomes
\begin{align}
\label{eq:v-ext-spin}
  V_\text{ext}^\text{spin}[j^\mu] =
  \frac{q_e \hbar}{2 m_e c} \int \big(\boldsymbol{\nabla} \times \boldsymbol{m}(\boldsymbol{r})\big) \cdot \boldsymbol{A}_\text{ext}(\boldsymbol{r}) \,  {\rm d}^3r
  = -\mu_B \int \boldsymbol{m}(\boldsymbol{r}) \cdot \boldsymbol{B}_\text{ext}(\boldsymbol{r}) \, {\rm d}^3r
\end{align}
where $\boldsymbol{B}_\text{ext} = \boldsymbol{\nabla} \times \boldsymbol{A}_\text{ext}$ is the external
magnetic field. The minus sign originates from the negative charge
of the electron. This closely resembles the form of the spin Zeeman interaction in the nonrelativistic
case [cf. Eq.~\eqref{eq:zeeman-op-nonrel}]. If an inhomogeneous magnetic field in $z$-direction is considered, 
Eq.~\eqref{eq:v-ext-spin} reduces to
\begin{align}
  V_\text{ext}^\text{spin}[j^\mu] 
  = -\mu_B \int m_z(\boldsymbol{r}) B_z(\boldsymbol{r}) \, {\rm d}^3r,
\end{align}
and by comparison with Eq.~\eqref{eq:zeeman-nonrel} we notice that $m_z(\boldsymbol{r})$ can be identified
with the spin density $Q(\boldsymbol{r})$ in nonrelativistic theory. However, while in the nonrelativistic
case the spin density is (in the absence of external magnetic fields) independent of the choice of the quantization 
axis, this is not the case in the relativistic theory, where $\hat{\sigma}_z$ does not commute with the Hamiltonian 
because of the presence of spin--orbit interactions.

\subsection{Relativistic Spin-DFT}

In the relativistic CDFT formalism discussed above, the noninteracting reference system is chosen such that it has the
same 4-current $j^\mu(\boldsymbol{r})$ [i.e., the same electron density $\rho(\boldsymbol{r})$ and the same current
density $\boldsymbol{j}(\boldsymbol{r})$] as the true interacting system. This results in a noninteracting kinetic-energy 
functional $T_s[\rho,\boldsymbol{j}]$ that can be defined as
\begin{equation}
  T_s[\rho,\boldsymbol{j}] = \min_{\Psi_s \rightarrow \rho,\boldsymbol{j}} \langle \Psi_s | \hat{T}^D | \Psi_s \rangle,
\end{equation}
where the constrained search has to be restricted to positive-energy wavefunctions $\Psi_s$ to avoid a
variational collaps.
In such a formalism, the KS equations then contain a four-component exchange--correlation potential.

However, we also pointed out that in the case of molecular systems in the absence of external magnetic
fields, a description relying on the electron density $\rho(\boldsymbol{r})$ only as fundamental variable is 
sufficient. Therefore, it is formally also possible to develop a relativistic
``density-only'' KS-DFT that resembles the nonrelativistic restricted KS-DFT formalism. This can be
achieved by only requiring from the noninteracting reference system that it has the same electron density as the
interacting system, and consequently defining the noninteracting kinetic energy as,
\begin{equation}
  T_s^{(d)}[\rho] = \min_{\Psi_s^{(d)} \rightarrow \rho} \big\langle \Psi_s^{(d)} \big| \hat{T} \big| \Psi_s^{(d)} \big\rangle,
\end{equation}
again restricting $\Psi_s$ to positive energy solutions. Then, the total energy functional can be decomposed as,
\begin{equation}
  E[\rho] = T_s^{(d)}[\rho] + V_\text{ext}[\rho] + J[\rho] + E_\text{xc}^{(d)}[\rho],
\end{equation}
where we have neglected the magnetic interactions in $J[\rho]$. Note that such a decomposition
implies a different definition of the exchange--correlation energy. The resulting KS equations then
feature a one-component exchange--correlation potential $v_\text{xc}^{(d)}[\rho] 
= (1/q_e) \, \delta E_\text{xc}^{(d)}[\rho]/\delta\rho(\boldsymbol{r})$. However, in such a formalism the 
current density $\boldsymbol{j}_s(\boldsymbol{r})$ and the magnetization
$\boldsymbol{m}_s(\boldsymbol{r})$ of the noninteracting reference system do not agree with the 
true interacting system. Instead, these are again a functional of the density only, just as the spin density is
in nonrelativistic restricted KS-DFT.

In between full CDFT and relativistic density-only KS-DFT, different intermediate formulations 
of relativistic KS-DFT are now also possible. For a relativistic system of noninteracting electrons, the
magnetization is given by
\begin{equation}
\boldsymbol{m}(\boldsymbol{r}) = \sum_i^{\rm occ} \varphi^\dagger_i(\boldsymbol{r}) \, \beta \boldsymbol{\sigma}^{(4)} \, \varphi_i(\boldsymbol{r}). 
\end{equation}
The noninteracting reference system can then be set up such that, in addition to the electron density, some parts 
of the magnetization density agree with those of the interacting system. For instance, we can require
that the $z$-components of the magnetization $m_z(\boldsymbol{r})$ match, or we can demand that 
the lengths of the magnetization vector $|\boldsymbol{m}(\boldsymbol{r})|$ at each point in space agree.
The choice of what quantity is to be 
reproduced by a relativistic KS-DFT formalism in addition to the density is our freedom of choice. If we choose to also 
reproduce the $z$-component of the magnetization density, this approach is called the collinear [``$(cl)$''] approach because an artificial 
external global quantization axis is introduced for the spin. If we require to reproduce the length of magnetization instead, this is 
a noncollinear [``$(nc)$''] approach because the magnetization is a vector field representing a magnetic dipole moment whose direction 
changes with position \cite{van_wllen_relativistic_1999,van_wllen_spin_2002,saue_fourcomponent_2002,
van_wllen_relativistic_2010,scalmani_new_2012}. 

\begin{figure}
  \caption{Relationship between wave function, total electron density, and the magnetization of the 
  system of fully interacting electrons and of the Kohn--Sham reference systems of noninteracting 
  electrons in different version of relativistic KS-DFT.}
  \label{fig:rel-ks-dft}

  \vspace{5ex}
  \begin{center}
  \includegraphics{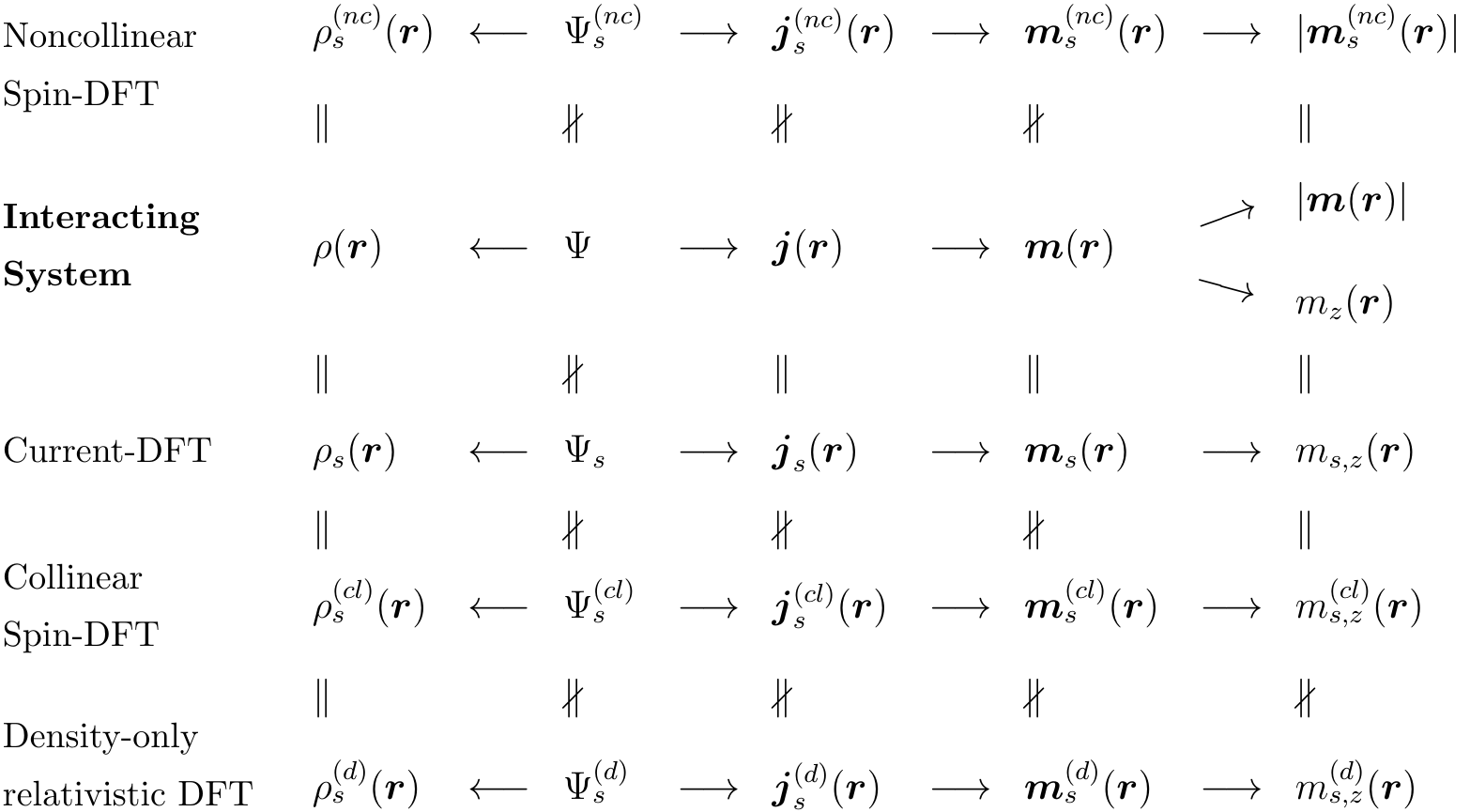}
  \end{center}

\end{figure}

Note that these different choices for the noninteracting reference system each implies a different definition of the noninteracting
kinetic energy functional, $T_s^{(cl)}[\rho,m_z]$ and $T_s^{(nc)}[\rho,|\boldsymbol{m}|]$, and thus also of the exchange--correlation 
energy, $E_\text{xc}^{(cl)}[\rho,m_z]$ and $E_\text{xc}^{(nc)}[\rho,|\boldsymbol{m}|]$, respectively. For the choices mentioned here, 
where in addition to the electron density one additional quantity is reproduced by the KS system, the resulting exchange--correlation 
potential has two components. This is in close analogy to the case of nonrelativistic unrestricted KS-DFT. Therefore, approximate
exchange--correlation functionals developed in the nonrelativistic domain are usually employed in practical applications of such
relativistic spin-DFT schemes.
However, the exchange--correlation potential is defined differently in nonrelativistic unrestricted  KS-DFT and in the relativistic
collinear and noncollinear cases. Consequently, also different exact conditions apply to the exchange--correlation functional. 
The different possible choices for setting up relativistic KS-DFT are summarized in Fig.~\ref{fig:rel-ks-dft}.

Finally, we discuss how the nonrelativistic unrestricted KS-DFT formalism emerges from relativistic spin-DFT.
This is most easily seen for the collinear approach, although the noncollinear one reduces to the same nonrelativistic
limit as well. The  $z$-component of the magnetization density of the KS reference system is given by
\begin{equation}
m_{z}(\boldsymbol{r}) = \sum_i^{\rm occ} \varphi^\dagger_i(\boldsymbol{r}) \, \beta\sigma_z^{(4)} \, \varphi_i(\boldsymbol{r}).
\end{equation}
If spin--orbit coupling is neglected, the Hamiltonian commutes with the spin operators and the KS spinors can
each be expressed as (two-component) spin orbitals $\varphi_{i,\alpha}=\varphi_i^\alpha\alpha$ or $\varphi_{i,\beta}=\varphi_i^\beta\beta$,
where $\varphi_i^\sigma$ are 2-spinors consisting of an upper and a lower component. Then, the spin--orbit 
coupling free (SOfree) $z$-component of the magnetization resembles the nonrelativistic spin density,
\begin{align}
m^{\rm SOfree}_{z}(\boldsymbol{r}) &= \sum_{i,\sigma}^{\rm occ} \varphi^{\dagger}_{i,\sigma}(\boldsymbol{r}) 
\, \beta \sigma_z^{(4)} \, \varphi_{i,\sigma}(\boldsymbol{r})
= \sum_{i}^{\alpha} \varphi_i^{\alpha,\dagger}(\boldsymbol{r}) \, \sigma_z \, \varphi_i^\alpha(\boldsymbol{r})
    - \sum_{i}^{\beta} \varphi_i^{\beta,\dagger}(\boldsymbol{r}) \, \sigma_z \, \varphi_i^\beta(\boldsymbol{r})
\nonumber \\
&= \sum_{i}^\alpha \Big[  | \varphi_i^{\alpha,U}(\boldsymbol{r})|^2 - | \varphi_i^{\alpha,L}(\boldsymbol{r})|^2 \Big]
      - \sum_{i}^\beta \Big[ | \varphi_i^{\beta,U}(\boldsymbol{r})|^2 - | \varphi_i^{\beta,L}(\boldsymbol{r})|^2 \Big]
= Q(\boldsymbol{r}),
\end{align}
where the superscript ``$U$'' and ``$L$'' denote the upper and lower components, respectively.
The neglect of spin--orbit coupling is, of course, an approximation that yields a scalar-relativistic Hamiltonian (which considers
kinematic relativistic effects only) or even a nonrelativistic Hamiltonian if the speed of light is taken to be infinity. 

The two 2-spinors $\varphi_i^\alpha$ and $\varphi_i^\beta$ considered here can each be reduced to one-component spin-orbitals if a 
unitary transformation \cite{peng_exact_2012} is performed to decouple the upper and lower components. This then also
requires a unitary transformation of the operators involved in the calculation of $\boldsymbol{m}_z$ \cite{mastalerz_douglaskrollhess_2008}. 
Still, taking the limit $c \rightarrow \infty$ yields the nonrelativistic theory and relativistic spin-DFT reduces to nonrelativistic unrestricted KS-DFT.

\section{Future Directions for Spin-DFT}
\label{sec:future}

As we have seen, different options exist for setting up KS-DFT for open-shell systems. In the
nonrelativistic case, one has to choose between a spin-restricted and a spin-unrestricted formulation 
of KS-DFT. In the former case, the wavefunction of the noninteracting reference system can always be 
chosen as an eigenfunction of $\hat{\boldsymbol{S}}^2$, but its spin density differs from the correct one.
Alternatively, in spin-unrestricted KS-DFT the noninteracting reference system has the correct spin density,
but cannot be an eigenfunction of $\hat{\boldsymbol{S}}^2$. In relativistic DFT even more options in 
between density-only KS-DFT (with a one-component exchange--correlation potential) and full CDFT
(with a four-component exchange--correlation potential) are possible. In particular, one can choose to
reproduce either a single component or the magnitude of the magnetization vector in collinear and
non-collinear relativistic KS-DFT (which both employ a two-component exchange--correlation 
potential), respectively.

Which choices we make may be determined by our ability to set up a proper approximation to the 
corresponding exchange--correlation energy functional. However, it is important to understand
that these different choices imply different definitions of the noninteracting kinetic energy and the
exchange--correlation functional. Thus, different exact conditions apply to these functionals,
which must be considered when developing such approximations. This is most obvious for the
fractional spin condition in the nonrelativistic case. While in spin-restricted KS-DFT, the constancy
condition directly applies to the exchange--correlation functional, in the spin-unrestricted case it
does not hold for the exchange--correlation functional alone, but to the sum of noninteracting
kinetic and exchange--correlation energy. It can be expected that in the latter case, devising
approximate exchange--correlation functionals that include this condition will be significantly 
more difficult.

While exact DFT should always lead to the correct ground-state --- irrespective of its spin state ---
in practice it appears more useful to rely on a theory that is able to target different spin states
separately. This should also simplify the development of approximate functionals because
it becomes possible to account for
the different exact conditions applying to the exchange--correlation hole for different spin
states \cite{baerends-jpca-holes,perdew_escaping_1995}. Note that 
such a spin-state specific DFT always has to operate within a spin--orbit coupling free framework, 
because in the fully relativistic theory $\hat{\boldsymbol{S}}^2$ does not commute with the Hamiltonian 
and different spin states do not correspond to the lowest state of a specific symmetry anymore. 

Thus, besides finding approximations that accurately account for the spin-density dependence of the 
(nonrelativistic)
exchange--correlation functionals $E_\text{xc}[\rho,Q]$ or $E_\text{xc}^{(u)}[\rho,Q]$, finding ways of 
including the spin-state dependence into these functionals is another important problem for open-shell
systems. In common approximations, it appears that the spin-density dependence is actually
used to model this spin-state dependence (i.e., the integral of the spin density is used to distinguish
spin states). Spin-state and spin-density dependence of the exchange--correlation
functional are intermingled in all available approximate functionals, which manifests itself in 
their violation of the fractional spin condition \cite{cohen_insights_2008,cohen_fractional_2008}.
To make progress in the development of reliable density-functional approximations, we believe it 
will be essential to consider the spin-state and the spin-density
dependence of the exchange--correlation functional separately.

\section*{Acknowledgments}

C.R.J acknowledges funding from the DFG-Center for Functional Nanostructures (CFN).
M.R. is grateful for financial support from the Swiss national science foundation SNF.

\appendix
\section{Gordon Decomposition of the Current Density}
\label{app:gordon}


In the one-electron case, the current density is given by
\begin{equation}
  \boldsymbol{j}(\boldsymbol{r}) = c \, \psi(\boldsymbol{r})^\dagger \boldsymbol{\alpha} \, \psi(\boldsymbol{r}).
\end{equation}
For the $k$-component, we can rewrite this definition in a somewhat artificial way as 
\begin{equation}
  \label{eq:jk-rewritten}
  j^k = c \, \psi^\dagger \alpha^k \psi 
  = \frac{c}{2} \, \psi^\dagger \alpha^k \psi + \frac{c}{2} \, \psi^\dagger \alpha^k \psi,
\end{equation}
where we dropped the dependence of $\psi$ on the spatial coordinate to simplify 
the notation.
The eigenvalue equation of the one-electron Dirac Hamiltonian equation can be
rewritten to obtain an expression for $\psi$,
\begin{equation}
  \psi
      = \frac{1}{m_e c^2} \bigg[ - c \, \beta \, \boldsymbol{\alpha} \cdot \Big(\hat{\boldsymbol{p}}
         - \frac{q_e}{c} \boldsymbol{A} \Big) + \beta \Big( E - q_e \phi \Big) \bigg] \psi,
\end{equation}
and by taking the transpose and complex conjugate also an expression for $\psi^\dagger$ (exploiting $\alpha_k^\dagger = \alpha_k$) 
\begin{align}
 \psi^\dagger
      &= \frac{1}{m_e c^2} \bigg[ -c \Big(\hat{\boldsymbol{p}}^*
         - \frac{q_e}{c} \boldsymbol{A} \Big) \psi^\dagger \cdot \boldsymbol{\alpha} \beta
         + \Big( E - q_e \phi \Big) \psi^\dagger \beta \bigg].
\end{align}

These expressions can now be used to replace $\psi$ in the first term and $\psi^\dagger$ in the second term 
of Eq.~\eqref{eq:jk-rewritten} to obtain,
\begin{align}
  j^k =& \ \frac{1}{2 m_e c} \, \psi^\dagger \alpha^k \bigg[ -c \, \beta \, \boldsymbol{\alpha} \cdot \Big(\hat{\boldsymbol{p}}
              - \frac{q_e}{c} \boldsymbol{A} \Big) + \beta \Big( E - q_e \phi \Big) \bigg] \psi 
\nonumber \\
          &+ \frac{1}{2 m_e c} \bigg[ -c  \Big(\hat{\boldsymbol{p}}^*
             - \frac{q_e}{c} \boldsymbol{A} \Big) \psi^\dagger \cdot \boldsymbol{\alpha} \beta
             + \Big( E - q_e \phi \Big) \psi^\dagger \beta \bigg]  \alpha^k \psi,
\end{align}
and by reordering the different terms we get
\begin{align}
  j^k 
=& \ \frac{1}{2 m_e} \bigg[ -\psi^\dagger \alpha^k \beta \, \boldsymbol{\alpha} \cdot \Big(\hat{\boldsymbol{p}}
              - \frac{q_e}{c} \boldsymbol{A} \Big) \psi 
           - \Big(\hat{\boldsymbol{p}}^* - \frac{q_e}{c} \boldsymbol{A} \Big) \psi^\dagger \cdot \boldsymbol{\alpha} 
              \, \beta \, \alpha^k \psi \bigg]
\nonumber \\
&+ \ \frac{1}{2 m_e c} \Big( E - q_e \phi \Big) \bigg[ \psi^\dagger \alpha^k \beta \psi 
   + \psi^\dagger \beta \alpha^k \psi \bigg].
\end{align}

The last term is zero because of $\{\alpha^k, \beta\} = 0$. For the first term, we use 
\begin{equation}
 \alpha^k\alpha^l = \frac{1}{2}[\alpha^k, \alpha^l] + \frac{1}{2} \{\alpha^k, \alpha^l\} = {\rm i} \varepsilon_{kln} \sigma_n^{(4)} + \delta_{kl}
\end{equation}
to arrive at (employing the convention of implicit summation over repeated indices)
\begin{align}
  j^k 
=& \ \frac{1}{2 m_e} \bigg[ \psi^\dagger \beta \alpha^k \alpha^l \Big(\hat{p}_l
              - \frac{q_e}{c} A_l \Big) \psi 
           + \Big(\hat{p}_l^* - \frac{q_e}{c} A_l \Big)\psi^\dagger   \beta \alpha^l
              \alpha^k \psi \bigg]
\nonumber \\
=& \ \frac{1}{2 m_e} \bigg[ \psi^\dagger \beta ({\rm i} \varepsilon_{kln} \sigma_n^{(4)} + \delta_{kl}) \Big(\hat{p}_l
              - \frac{q_e}{c} A_l \Big) \psi 
           + \Big(\hat{p}_l^* - \frac{q_e}{c} A_l \Big) \psi^\dagger \beta 
           ({\rm i} \varepsilon_{lkn} \sigma_n^{(4)} + \delta_{kl}) \psi \bigg].
\end{align}           
After regrouping the different terms, we obtain
\begin{align}
j^k
=& \ \frac{1}{2 m_e} \bigg[ \psi^\dagger \beta \delta_{kl} \Big(\hat{p}_l - \frac{q_e}{c} A_l \Big) \psi 
           + \Big(\hat{p}^*_l - \frac{q_e}{c} A_l \Big) \psi^\dagger  \beta \delta_{kl} \psi \bigg]
\nonumber \\
&+ \ \frac{1}{2 m_e} \bigg[ \psi^\dagger \beta {\rm i} \varepsilon_{kln} \sigma_n^{(4)} \, \hat{p}_l \, \psi 
           +  \hat{p}^*_l \psi^\dagger  \beta {\rm i} \varepsilon_{lkn} \sigma_n^{(4)}  \psi \bigg]
\nonumber \\
&+ \ \frac{1}{2 m_e} \bigg[ \psi^\dagger \beta {\rm i} \varepsilon_{kln} \sigma_n^{(4)} \Big(- \frac{q_e}{c} A_l \Big) \psi 
           + \Big(- \frac{q_e}{c} A_l \Big) \psi^\dagger \beta {\rm i} \varepsilon_{lkn} \sigma_n^{(4)}  \psi \bigg],
\end{align}
where the last term is zero because $\varepsilon_{kln} = - \varepsilon_{lkn}$. Hence, we find
\begin{align}
j^k =& \ \frac{1}{2 m_e} \bigg[ \psi^\dagger \beta \Big(\hat{p}_k - \frac{q_e}{c} A_k \Big) \psi 
           + \Big(\hat{p}^*_k - \frac{q_e}{c} A_k \Big) \psi^\dagger  \beta \psi \bigg]
\nonumber \\
&+ \ \frac{1}{2 m_e} \bigg[ \psi^\dagger \beta {\rm i} \varepsilon_{kln} \sigma_n^{(4)} \Big( -{\rm i}\hbar \frac{\partial}{\partial r_l} \Big) \psi 
           +  \Big({\rm i}\hbar \frac{\partial}{\partial r_l} \Big) \psi^\dagger  \beta {\rm i} \varepsilon_{lkn} \sigma_n^{(4)}  \psi \bigg]
\nonumber \\
=& \ \frac{1}{2 m_e} \bigg[ \psi^\dagger \beta  \Big(\hat{p}_k - \frac{q_e}{c} A_k \Big) \psi 
           + \Big(\hat{p}^*_k - \frac{q_e}{c} A_k \Big) \psi^\dagger  \beta \psi \bigg]
\nonumber \\    
&+ \ \frac{\hbar}{2 m_e} \varepsilon_{kln} \, \frac{\partial}{\partial r_l} \big[  \psi^\dagger (\beta \sigma_n^{(4)}) \psi \big]
\end{align}
and combining the different components of $\boldsymbol{j}$ again gives
\begin{align}
 \label{eq:gordon-decomp-1e}
 \boldsymbol{j}
 =& \frac{1}{2 m_e} \bigg[ \psi^\dagger \beta  \Big(\hat{\boldsymbol{p}} - \frac{q_e}{c} \boldsymbol{A} \Big) \psi 
           + \Big(\hat{\boldsymbol{p}}^* - \frac{q_e}{c} \boldsymbol{A} \Big) \psi^\dagger  \beta \psi \bigg]
    + \ \frac{\hbar}{2 m_e} \boldsymbol{\nabla} \times \psi^\dagger (\beta \boldsymbol{\sigma}^{(4)}) \psi.
\end{align}


For an $N$-electron system, we start from the definition of the many-electron current density \cite{fux_electron_2012},
\begin{equation}
  \label{eq:app-cd-many-el}
  \boldsymbol{j}(\boldsymbol{r}_1) = c \, N \int \Psi(\boldsymbol{r}_1, \boldsymbol{r}_2, \dotsc, \boldsymbol{r}_N)^\dagger \, \boldsymbol{\alpha}_1 \,
                                               \Psi(\boldsymbol{r}_1, \boldsymbol{r}_2, \dotsc, \boldsymbol{r}_N) \, {\rm d}^3r_2 \dotsm {\rm d}^3r_N,
\end{equation}
and the wavefunction is an eigenfunction of the many-electron Dirac--Coulomb Hamiltonian
\begin{equation}
  \hat{H}^D = \hat{h}^D(1) + \sum_{i=2}^N h^D(i) + \hat{V}_\text{ee},
\end{equation}
projected onto the electronic (positive-energy) states.

We can now rewrite the corresponding eigenvalue equation  in a similar fashion as in the one-electron case to
obtain for the wavefunction
\begin{equation}
\label{eq:app-psi-many}
  \Psi
      = \frac{1}{m_e c^2} \bigg[ - c \, \beta_1 \, \boldsymbol{\alpha}_1 \cdot \Big(\hat{\boldsymbol{p}}_1
         - \frac{q_e}{c} \boldsymbol{A}(\boldsymbol{r}_1) \Big) + \beta_1 \Big( E - q_e \phi(\boldsymbol{r}_1) \Big) 
         - \beta_1 \sum_{i=2}^N h^D(i) - \beta_1 \hat{V}_\text{ee}\bigg] \Psi
\end{equation}
and for its transpose and complex conjugate
\begin{equation}
\label{eq:app-psi-dag-many}
  \Psi^\dagger
      = \frac{1}{m_e c^2} \bigg[ - c \Big(\hat{\boldsymbol{p}}_1^* - \frac{q_e}{c} \boldsymbol{A}(\boldsymbol{r}_1) 
      \Big)  \Psi^\dagger  \boldsymbol{\alpha}_1 \beta_1 + \Big( E - q_e \phi(\boldsymbol{r}_1) \Big) \Psi^\dagger  \beta_1 
         - \sum_{i=2}^N \big(h^D(i) \Psi\big)^\dagger  \beta_1  - (\hat{V}_\text{ee} \Psi)^\dagger  \beta_1 \bigg].
\end{equation}

Now, we can rewrite the $k$-component of the integrand in Eq.~\eqref{eq:app-cd-many-el} as
\begin{equation}
  c \, \Psi^\dagger \alpha^k_1 \Psi 
  = \frac{c}{2} \, \Psi^\dagger \alpha^k_1 \Psi + \frac{c}{2} \, \Psi^\dagger \alpha^k_1 \Psi.
\end{equation}
After substituting Eq.~\eqref{eq:app-psi-many} for $\Psi$ in the first term and Eq.~\eqref{eq:app-psi-dag-many} for 
$\Psi^\dagger$ in the second term, we obtain, in addition to the term already present in the one-electron 
case [cf. Eq.~\eqref{eq:gordon-decomp-1e}],
\begin{align*}
  c \, \Psi^\dagger \alpha^k_1 \Psi 
 =& \ \eqref{eq:gordon-decomp-1e} - \frac{1}{2 m_e c} \, \Psi^\dagger \alpha^k_1 \Big[\beta_1 \sum_{i=2}^N \big(h^D(i) \Psi \big)^\dagger + \beta_1 \hat{V}_\text{ee}\Psi \Big] 
\nonumber \\
    &\hspace{1.2cm}- \frac{1}{2 m_e c} \, \Big[\sum_{i=2}^N \big(h^D(i) \Psi \big)^\dagger \beta_1 + (\hat{V}_\text{ee} \Psi)^\dagger  \beta_1 \Big] \alpha^k_1 \Psi
\nonumber \\
 =& \ \eqref{eq:gordon-decomp-1e} - \frac{1}{2 m_e c} \Big[ \Psi^\dagger \alpha^k_1 \beta_1 \sum_{i=2}^N \big( h^D(i) \Psi \big)
         + \sum_{i=2}^N \big( h^D(i) \Psi \big)^\dagger \beta_1 \alpha^k_1 \Psi  \Big]
\nonumber \\
 & \hspace{1.2cm}- \frac{1}{2 m_e c}  \Big[ \Psi^\dagger \alpha^k_1 \beta_1 \hat{V}_\text{ee}\Psi
         + (\hat{V}_\text{ee} \Psi)^\dagger  \beta_1\alpha^k_1 \Psi  \Big].
\end{align*}
With a multiplicative operator acting equally on all components of the wavefunction for $\hat{V}_\text{ee}$, the last 
term is zero because of $\{\alpha^k, \beta\} = 0$. 
However, if the Gaunt or Breit interaction is included, the corresponding term is not zero and gives an 
additional contribution to the current that is not present in the one-electron case. This term contains the
spin--spin interactions.

For the remaining terms, we now consider one of the terms separately, for instance $h^D(2)$, and note that
\begin{align}
 &- \frac{1}{2 m_e c} \Big[ \Psi^\dagger \alpha^k_1 \beta_1 \big( h^D(2) \Psi \big) + \big( h^D(2) \Psi \big)^\dagger \beta_1 \alpha^k_1 \Psi  \Big]
 \nonumber \\
 =&- \frac{1}{2 m_e c} \Big[ \Psi^\dagger \alpha^k_1 \beta_1 \big( h^D(2) \Psi \big) -  \big( h^D(2) \Psi \big)^\dagger \alpha^k_1 \beta_1 \Psi  \Big],
\end{align}
because of $\{\alpha_1^k, \beta_1\}=0$. Integrating the expression in square brackets over $\boldsymbol{r}_2$ we now find
\begin{align}
 & \ \int \Psi^\dagger \alpha^k_1 \beta_1 \big( h^D(2) \Psi \big) \ {\rm d}^3r_2
  - \int \big( h^D(2) \Psi \big)^\dagger \alpha^k_1 \beta_1 \Psi \ {\rm d}^3r_2
\nonumber \\
  =& \ \int \Psi^\dagger \alpha^k_1 \beta_1 \big( h^D(2) \Psi \big) \ {\rm d}^3r_2
  - \int \Psi^\dagger \alpha^k_1 \beta_1 \big( h^D(2) \Psi \big) \ {\rm d}^3r_2 = 0
\end{align}
where for the second term we exploited that $h^D(2)$ is hermitian and commutes 
with $\alpha^k_1$ and $\beta_1$ because these act on different electrons.
Combining all of these results, we find that (with the Dirac--Coulomb Hamiltonian),
\begin{align}
  \boldsymbol{j}(\boldsymbol{r}_1) 
  =&\ c \, N \int \Psi(\boldsymbol{r}_1, \boldsymbol{r}_2, \dotsc, \boldsymbol{r}_N)^\dagger \, \boldsymbol{\alpha}_1 \,
                                               \Psi(\boldsymbol{r}_1, \boldsymbol{r}_2, \dotsc, \boldsymbol{r}_N) \, {\rm d}^3r_2 \dotsm {\rm d}^3r_N
\nonumber \\
  =&\ \frac{1}{2 m_e} \, N \int \bigg[ \Psi^\dagger \beta_1  \Big(\hat{\boldsymbol{p}}_1 - \frac{q_e}{c} \boldsymbol{A}(\boldsymbol{r}_1) \Big) \Psi 
           + \Big(\hat{\boldsymbol{p}}_1^* - \frac{q_e}{c} \boldsymbol{A}(\boldsymbol{r}_1) \Big) \Psi^\dagger  \beta_1 \Psi \bigg]
              \, {\rm d}^3r_2 \dotsm {\rm d}^3r_N
\nonumber \\
  &+ \frac{\hbar}{2 m_e} \, N \int  \boldsymbol{\nabla} \times (\Psi^\dagger \, \beta_1 \boldsymbol{\sigma}_1^{(4)} \, \Psi) \, {\rm d}^3r_2 \dotsm {\rm d}^3r_N
\end{align}

\clearpage

\section*{Author Biographies}

\linespread{1.2}\selectfont

\subsection*{Christoph R. Jacob}

\vspace{-3ex}
\begin{wrapfigure}{l}{5.5cm}
\vspace{-3ex}
\includegraphics[width=5cm]{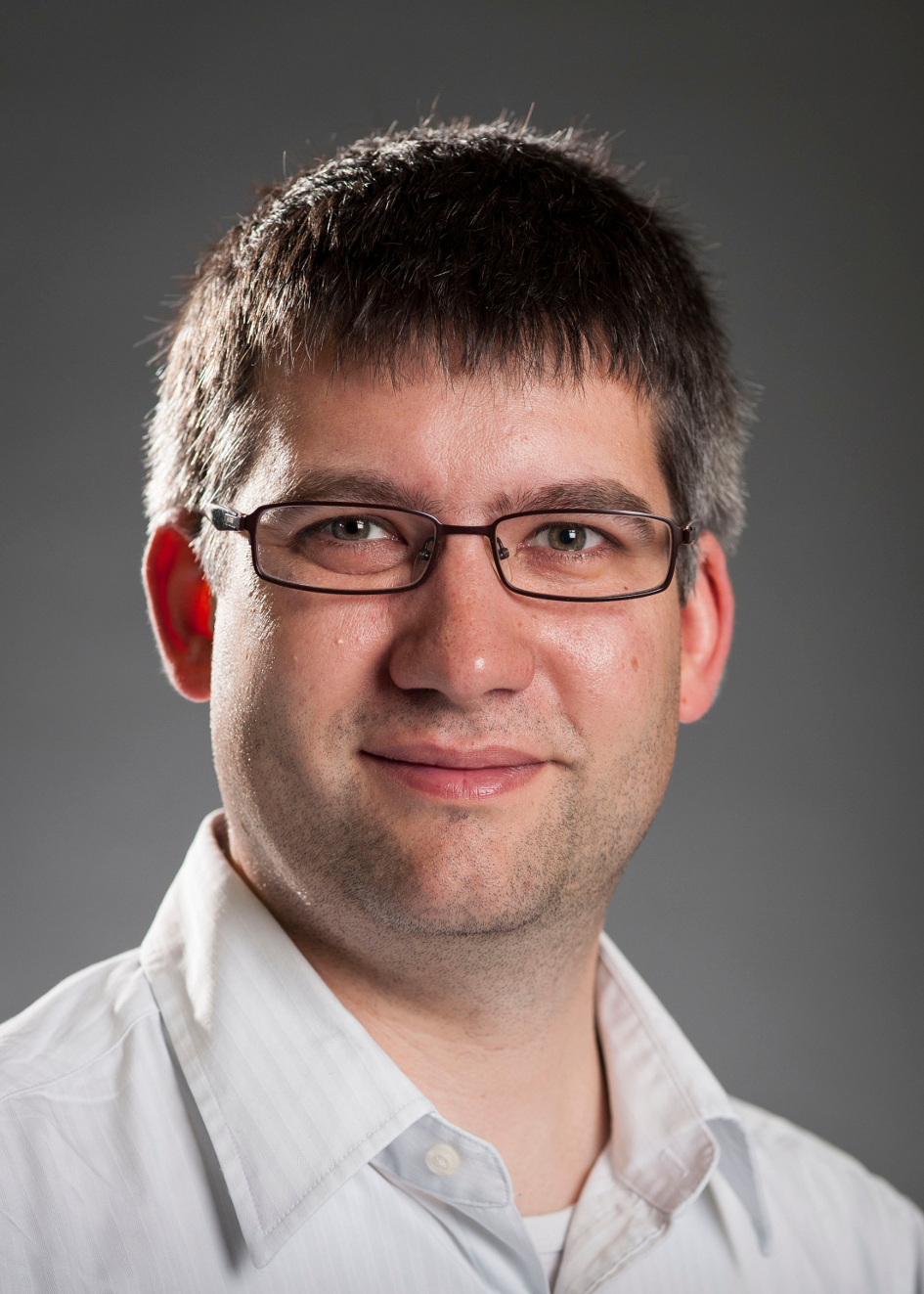}
\end{wrapfigure}
Christoph R.\ Jacob heads an independent ``Young Scientist Group'' in theoretical chemistry
at the  Karlsruher Institute of Technology (KIT). He studied chemistry and mathematics
at Philipps University Marburg and at the University of Karlsruhe. Following a one-year stay in the group of
Peter Schwerdtfeger at Auckland University in 2004, he joined the group of  Lucas Visscher und 
Evert-Jan Baerends at VU University Amsterdam, from where he obtained his PhD degree in 2007. 
He then moved on to the group of Markus Reiher at ETH Zurich, before taking up his current position 
at KIT in 2010. His main research interests are the development of quantum-chemical method for
complex systems, which includes the development of subsystem and embedding methods as
well as fundamental work in density-functional theory, and theoretical spectroscopy of large
chemical systems, ranging from vibrational spectroscopy of biomolecules to optical and X-ray 
spectroscopy of nanostructures.

\subsection*{Markus Reiher}

\vspace{-3ex}
\begin{wrapfigure}{l}{5.5cm}
\vspace{-3ex}
\includegraphics[width=5cm]{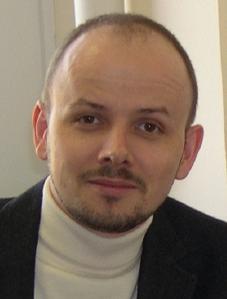}
\end{wrapfigure}
Markus Reiher is professor for theoretical chemistry at ETH Zurich since 2006. After studying chemistry, he received his PhD
in theoretical chemistry from the University of Bielefeld working with J\"urgen Hinze in 1998. In 2002, he finished his habilitation 
thesis in the group of Bernd Artur Hess at the University of Erlangen and continued as a private docent first in Erlangen and 
then at the University of Bonn.
In 2005, he accepted an offer for a professorship in physical chemistry from the University of Jena, where he worked until 
he moved to ETH Zurich. His research covers many different areas in theo\-retical chemistry and ranges from relativistic 
quantum chemistry, (vibrational) spectroscopy, density functional theory, transition metal catalysis and bioinorganic 
chemistry to the development of new electron-correlation theories and smart algorithms for inverse quantum chemistry.

\clearpage

\linespread{1.66}\selectfont

\listoffigures

\clearpage

\clearpage

\clearpage

\clearpage

\listoftables

\clearpage

\begin{table}
  \caption{Comparison of the spin-restricted and spin-unrestricted formulations of KS-DFT. 
  "Correct" indicates that the quantity calculated for the noninteracting reference system 
  agrees with the corresponding one of the fully interacting system.}
  \label{tab:comparison-properties}
  \begin{center}
  {\linespread{1.2}\selectfont
  \begin{tabular}{p{5cm}cc}
    \hline\hline
      & spin-restricted & spin-unrestricted \\
      & KS-DFT & KS-DFT \\
    \hline \\[-1.5ex]
     correct $\rho_s(\boldsymbol{r})$?					&  Yes 	&  Yes	\\[1ex]
     correct $Q_s(\boldsymbol{r})$?					&  No 	&  Yes	\\[1ex]
     $\Psi_s$ is eigenfunction of $\hat{\boldsymbol{S}}^2$?	&  Yes 	&  No	\\[1ex]
     correct $\langle \hat{\boldsymbol{S}}^2 \rangle$?		&  Maybe 	&  No	\\[1ex]
     $\Psi_s$ is eigenfunction of $\hat{S_z}$?			&  Yes 	&  Yes	\\[1ex]
     correct $\langle \hat{S_z} \rangle$?				&  Maybe 	&  Yes	\\[1ex]
    \hline\hline 
  \end{tabular}}
  \end{center}
\end{table}

\clearpage

\begin{sidewaystable}
  \caption{Definition of the noninteracting kinetic energy, exchange--correlation energy, and 
  exchange--correlation potential in the spin-restricted and spin-unrestricted formulations of KS-DFT.}
  \label{tab:comparison-functionals}
  \begin{center}
  \begin{tabular}{p{6cm}cp{1cm}c}
    \hline\hline
      & spin-restricted KS-DFT  & & spin-unrestricted KS-DFT  \\
    \hline \\[-1.5ex]
     noninteracting kinetic energy	& $T_s[\rho] =  \min\limits_{\Psi_s \rightarrow \rho} \langle \Psi_s | \hat{T} | \Psi_s \rangle$ 	
     						&	& $T_s^{(u)}[\rho,Q]     =  \min\limits_{\Psi_s^{(u)} \rightarrow \rho,Q} \big\langle \Psi_s^{(u)} \big| \hat{T} \big| \Psi_s^{(u)} \big\rangle$ 
							\\[2ex]
     decomposition of HK functional	& $F_\text{HK}[\rho,Q] = T_s[\rho] + J[\rho] + E_\text{xc}[\rho,Q]$
     						&	& $F_\text{HK}[\rho,Q] = T_s^{(u)}[\rho,Q] + J[\rho] + E_\text{xc}^{(u)}[\rho,Q]$
							\\[2ex]
     exchange--correlation energy	& $E_\text{xc}[\rho,Q] = F_\text{HK}[\rho,Q] - T_s[\rho] - J[\rho]$
     						&	& $E_\text{xc}^{(u)}[\rho,Q] = F_\text{HK}[\rho,Q] - T_s^{(u)}[\rho,Q] - J[\rho]$
							\\[2ex]
     exchange--correlation potential	& $v_\text{xc}[\rho] = \dfrac{1}{q_e} \dfrac{\delta E_\text{xc}[\rho,Q]}{\delta\rho(\boldsymbol{r})}$ 
     						&	& $v_\text{xc}^\text{tot}[\rho,Q] = \dfrac{1}{q_e} \dfrac{\delta E_\text{xc}^\text{(u)}[\rho,Q]}{\delta\rho(\boldsymbol{r})}$ \\[3ex]		
					&	& 	&  $v_\text{xc}^\text{spin}[\rho,Q] = \dfrac{1}{q_e} \dfrac{\delta E_\text{xc}^\text{(u)}[\rho,Q]}{\delta Q(\boldsymbol{r})}$ \\[1ex]					
    \hline\hline 
  \end{tabular}
  \end{center}
\end{sidewaystable}


\begin{thebibliography}{100}
\expandafter\ifx\csname urlstyle\endcsname\relax
  \providecommand{\doi}[1]{DOI \discretionary{}{}{}#1}\else
  \providecommand{\doi}{DOI \discretionary{}{}{}\begingroup
  \urlstyle{rm}\Url}\fi
\providecommand{\bibAnnoteFile}[1]{%
  \IfFileExists{#1}{\begin{quotation}\noindent\textsc{Key:} #1\\
  \textsc{Annotation:}\ \input{#1}\end{quotation}}{}}
\providecommand{\bibAnnote}[2]{%
  \begin{quotation}\noindent\textsc{Key:} #1\\
  \textsc{Annotation:}\ #2\end{quotation}}

\bibitem{jeschke_distance_2007}
G.~Jeschke, Y.~Polyhach, \emph{Phys. Chem. Chem. Phys.} \textbf{2007},
  \emph{9}, 1895--1910.
\input{jeschke_distance_2007}

\bibitem{herrmann_organic_2010}
C.~Herrmann, G.~C. Solomon, M.~A. Ratner, \emph{J. Am. Chem. Soc.}
  \textbf{2010}, \emph{132}, 3682--3684.
\input{herrmann_organic_2010}

\bibitem{herrmann_designing_2011}
C.~Herrmann, G.~C. Solomon, M.~A. Ratner, \emph{J. Chem. Phys.} \textbf{2011},
  \emph{134}, 224306.
\input{herrmann_designing_2011}

\bibitem{timco_engineering_2009}
G.~A. Timco, S.~Carretta, F.~Troiani, F.~Tuna, R.~J. Pritchard, C.~A. Muryn,
  E.~J.~L. McInnes, A.~Ghirri, A.~Candini, P.~Santini, G.~Amoretti,
  M.~Affronte, R.~E.~P. Winpenny, \emph{Nature Nanotech.} \textbf{2009},
  \emph{4}, 173--178.
\input{timco_engineering_2009}

\bibitem{mcevoy_water-splitting_2006}
J.~P. McEvoy, G.~W. Brudvig, \emph{Chem. Rev.} \textbf{2006}, \emph{106},
  4455--4483.
\input{mcevoy_water-splitting_2006}

\bibitem{vignais_occurrence_2007}
P.~M. Vignais, B.~Billoud, \emph{Chem. Rev.} \textbf{2007}, \emph{107},
  4206--4272.
\input{vignais_occurrence_2007}

\bibitem{stiebritz_hydrogenases_2012}
M.~T. Stiebritz, M.~Reiher, \emph{Chem. Sci.} \textbf{2012}, \emph{3}, 1739.
\input{stiebritz_hydrogenases_2012}

\bibitem{hu_decoding_2010}
Y.~Hu, M.~W. Ribbe, \emph{Acc. Chem. Res.} \textbf{2010}, \emph{43}, 475--484.
\input{hu_decoding_2010}

\bibitem{schrder_two-state_2000}
D.~Schr\"{o}der, S.~Shaik, H.~Schwarz, \emph{Acc. Chem. Res.} \textbf{2000},
  \emph{33}, 139--145.
\input{schrder_two-state_2000}

\bibitem{helgaker-book}
T.~Helgaker, P.~J{\o}rgensen, J.~Olsen, \emph{{M}olecular {E}lectronic
  {S}tructure {T}heory}, John Wiley \& Sons, Chichester, \textbf{2000}.
\input{helgaker-book}

\bibitem{koch-holthausen}
W.~Koch, M.~C. Holthausen, \emph{{A} {C}hemist's {G}uide to {D}ensity
  {F}unctional {T}heory}, 2nd Ed., Wiley-VCH, Weinheim, \textbf{2001}.
\input{koch-holthausen}

\bibitem{reiher_theoretical_2009}
M.~Reiher, \emph{Chimia} \textbf{2009}, \emph{63}, 140--145.
\input{reiher_theoretical_2009}

\bibitem{roos_not_2008}
B.~O. Roos, V.~Veryazov, J.~Conradie, P.~R. Taylor, A.~Ghosh, \emph{J. Phys.
  Chem. B} \textbf{2008}, \emph{112}, 14099--14102.
\input{roos_not_2008}

\bibitem{rado_binding_2008}
M.~Rado\'{n}, K.~Pierloot, \emph{J. Phys. Chem. A} \textbf{2008}, \emph{112},
  11824--11832.
\input{rado_binding_2008}

\bibitem{rado_electronic_2010}
M.~Rado\'{n}, E.~Broclawik, K.~Pierloot, \emph{J. Phys. Chem. B} \textbf{2010},
  \emph{114}, 1518--1528.
\input{rado_electronic_2010}

\bibitem{sala_water-oxidation_2010}
X.~Sala, M.~Z. Ertem, L.~Vigara, T.~K. Todorova, W.~Chen, R.~C. Rocha,
  F.~Aquilante, C.~J. Cramer, L.~Gagliardi, A.~Llobet, \emph{Angew. Chem. Int.
  Ed.} \textbf{2010}, \emph{49}, 7745--7747.
\input{sala_water-oxidation_2010}

\bibitem{planas_electronic_2011}
N.~Planas, L.~Vigara, C.~Cady, P.~Mir\'{o}, P.~Huang, L.~Hammarstr\"{o}m,
  S.~Styring, N.~Leidel, H.~Dau, M.~Haumann, L.~Gagliardi, C.~J. Cramer,
  A.~Llobet, \emph{Inorg. Chem.} \textbf{2011}, \emph{50}, 11134--11142.
\input{planas_electronic_2011}

\bibitem{vigara_experimental_2012}
L.~Vigara, M.~Z. Ertem, N.~Planas, F.~Bozoglian, N.~Leidel, H.~Dau, M.~Haumann,
  L.~Gagliardi, C.~J. Cramer, A.~Llobet, \emph{Chem. Sci.} \textbf{2012},
  \emph{3}, 2576--2586.
\input{vigara_experimental_2012}

\bibitem{chan_introduction_2008}
G.~K.-L. Chan, J.~J. Dorando, D.~Ghosh, J.~Hachmann, E.~Neuscamman, H.~Wang,
  T.~Yanai, in \emph{Frontiers in Quantum Systems in Chemistry and Physics},
  vol.~18, (Edited by S.~Wilson, P.~J. Grout, J.~Maruani, G.~Delgado-Barrio,
  P.~Piecuch), 1st Ed., Springer, Dordrecht, \textbf{2008}, 49--65,
  arXiv:0711.1398 [cond-mat.str-el].
\input{chan_introduction_2008}

\bibitem{marti_density_2010}
K.~H. Marti, M.~Reiher, \emph{Z. Phys. Chem.} \textbf{2010}, \emph{224},
  583--599.
\input{marti_density_2010}

\bibitem{marti_new_2011}
K.~H. Marti, M.~Reiher, \emph{Phys. Chem. Chem. Phys.} \textbf{2011},
  \emph{13}, 6750--6759.
\input{marti_new_2011}

\bibitem{frenking_nature_2000}
G.~Frenking, N.~Fr\"{o}hlich, \emph{Chem. Rev.} \textbf{2000}, \emph{100},
  717--774.
\input{frenking_nature_2000}

\bibitem{ziegler_theoretical_2005}
T.~Ziegler, J.~Autschbach, \emph{Chem. Rev.} \textbf{2005}, \emph{105},
  2695--2722.
\input{ziegler_theoretical_2005}

\bibitem{neese_prediction_2009}
F.~Neese, \emph{Coord. Chem. Rev.} \textbf{2009}, \emph{253}, 526--563.
\input{neese_prediction_2009}

\bibitem{podewitz_spin_2010}
M.~Podewitz, M.~Reiher, \emph{Adv. Inorg. Chem.} \textbf{2010}, \emph{62},
  177--230.
\input{podewitz_spin_2010}

\bibitem{podewitz_density_2011}
M.~Podewitz, T.~Weymuth, M.~Reiher, in \emph{Modeling of Molecular Properties},
  (Edited by P.~Comba), Wiley-VCH, Weinheim, \textbf{2011}, 137--163.
\input{podewitz_density_2011}

\bibitem{ghosh_just_2006}
A.~Ghosh, \emph{J. Biol. Inorg. Chem.} \textbf{2006}, \emph{11}, 671--673.
\input{ghosh_just_2006}

\bibitem{cramer_density_2009}
C.~J. Cramer, D.~G. Truhlar, \emph{Phys. Chem. Chem. Phys.} \textbf{2009},
  \emph{11}, 10757.
\input{cramer_density_2009}

\bibitem{reiher_reparameterization_2001}
M.~Reiher, O.~Salomon, B.~A. Hess, \emph{Theor. Chem. Acc.} \textbf{2001},
  \emph{107}, 48--55.
\input{reiher_reparameterization_2001}

\bibitem{reiher_theoretical_2002}
M.~Reiher, \emph{Inorg. Chem.} \textbf{2002}, \emph{41}, 6928--6935.
\input{reiher_theoretical_2002}

\bibitem{ghosh_high-level_2003}
A.~Ghosh, P.~R. Taylor, \emph{Curr. Opin. Chem. Biol.} \textbf{2003}, \emph{7},
  113--124.
\input{ghosh_high-level_2003}

\bibitem{harvey_dft_2004}
J.~N. Harvey, \emph{Struct. Bond.} \textbf{2004}, \emph{112}, 151--184.
\input{harvey_dft_2004}

\bibitem{herrmann_spin_2006}
C.~Herrmann, L.~Yu, M.~Reiher, \emph{J. Comput. Chem.} \textbf{2006},
  \emph{27}, 1223--1239.
\input{herrmann_spin_2006}

\bibitem{swart_accurate_2008}
M.~Swart, \emph{J. Chem. Theory Comput.} \textbf{2008}, \emph{4}, 2057--2066.
\input{swart_accurate_2008}

\bibitem{ye_accurate_2010}
S.~Ye, F.~Neese, \emph{Inorg. Chem.} \textbf{2010}, \emph{49}, 772--774.
\input{ye_accurate_2010}

\bibitem{swart_spin_2012}
M.~Swart, \emph{Int. J. Quantum Chem.} \textbf{2012}, in press, {DOI}:
  10.1002/qua.24255.
\input{swart_spin_2012}

\bibitem{conradie_dft_2007}
J.~Conradie, A.~Ghosh, \emph{J. Phys. Chem. B} \textbf{2007}, \emph{111},
  12621--12624.
\input{conradie_dft_2007}

\bibitem{boguslawski_can_2011}
K.~Boguslawski, {\relax Ch}.~R. Jacob, M.~Reiher, \emph{J. Chem. Theory
  Comput.} \textbf{2011}, \emph{7}, 2740--2752.
\input{boguslawski_can_2011}

\bibitem{boguslawski_accurate_2012}
K.~Boguslawski, K.~H. Marti, O.~Legeza, M.~Reiher, \emph{J. Chem. Theory
  Comput.} \textbf{2012}, in press, {DOI}: 10.1021/ct300211j.
\input{boguslawski_accurate_2012}

\bibitem{noodleman_valence_1981}
L.~Noodleman, \emph{J. Chem. Phys.} \textbf{1981}, \emph{74}, 5737--5743.
\input{noodleman_valence_1981}

\bibitem{jonkers_broken_1982}
G.~Jonkers, C.~A. de~Lange, L.~Noodleman, E.~J. Baerends, \emph{Mol. Phys.}
  \textbf{1982}, \emph{46}, 609--620.
\input{jonkers_broken_1982}

\bibitem{noodleman_models_1985}
L.~Noodleman, J.~G. Norman, J.~H. Osborne, A.~Aizman, D.~A. Case, \emph{J. Am.
  Chem. Soc.} \textbf{1985}, \emph{107}, 3418--3426.
\input{noodleman_models_1985}

\bibitem{noodleman_ligand_1986}
L.~Noodleman, E.~R. Davidson, \emph{Chem. Phys.} \textbf{1986}, \emph{109},
  131--143.
\input{noodleman_ligand_1986}

\bibitem{reiher_definition_2007}
M.~Reiher, \emph{Faraday Discuss.} \textbf{2007}, \emph{135}, 97--124.
\input{reiher_definition_2007}

\bibitem{noodleman_orbital_1995}
L.~Noodleman, C.~Y. Peng, D.~A. Case, J.~M. Mouesca, \emph{Coord. Chem. Rev.}
  \textbf{1995}, \emph{144}, 199--244.
\input{noodleman_orbital_1995}

\bibitem{van_wllen_broken_2009}
C.~van W\"{u}llen, \emph{J. Phys. Chem. A} \textbf{2009}, \emph{113},
  11535--11540.
\input{van_wllen_broken_2009}

\bibitem{pantazis_new_2009}
D.~A. Pantazis, M.~Orio, T.~Petrenko, S.~Zein, E.~Bill, W.~Lubitz,
  J.~Messinger, F.~Neese, \emph{Chem.--Eur. J.} \textbf{2009}, \emph{15},
  5108--5123.
\input{pantazis_new_2009}

\bibitem{schinzel_density_2010}
S.~Schinzel, J.~Schraut, A.~V. Arbuznikov, P.~E.~M. Siegbahn, M.~Kaupp,
  \emph{Chem.--Eur. J.} \textbf{2010}, \emph{16}, 10424--10438.
\input{schinzel_density_2010}

\bibitem{cohen_challenges_2012}
A.~J. Cohen, P.~Mori-S\'{a}nchez, W.~Yang, \emph{Chem. Rev.} \textbf{2012},
  \emph{112}, 289--320.
\input{cohen_challenges_2012}

\bibitem{yang-parr}
R.~G. Parr, W.~Yang, \emph{{D}ensity-{F}unctional {T}heory of {A}toms and
  {M}olecules}, Oxford University Press, Oxford, \textbf{1989}.
\input{yang-parr}

\bibitem{gross_density_1990}
E.~K.~U. Gross, R.~M. Dreizler, \emph{{D}ensity {F}unctional {T}heory: {A}n
  {A}pproach to the {Q}uantum {M}any-{B}ody {P}roblem}, Springer, Berlin,
  \textbf{1990}.
\input{gross_density_1990}

\bibitem{fiolhais_primer_2003}
C.~Fiolhais, F.~Nogueira, M.~A.~L. Marques, \emph{{A} {P}rimer in {D}ensity
  {F}unctional {T}heory}, Lecture Notes in Physics, Springer, Berlin,
  \textbf{2003}.
\input{fiolhais_primer_2003}

\bibitem{engel_density_2011}
E.~Engel, R.~M. Dreizler, \emph{{D}ensity {F}unctional {T}heory: {A}n
  {A}dvanced {C}ourse}, Springer, Heidelberg, \textbf{2011}.
\input{engel_density_2011}

\bibitem{pauli_zur_1927}
W.~Pauli, \emph{Z. Phys.} \textbf{1927}, \emph{43}, 601--623.
\input{pauli_zur_1927}

\bibitem{cohen-tannoudji-1}
C.~Cohen-Tannoudji, B.~Diu, F.~Laloe, \emph{{Q}uantum {M}echanics, {V}ol. 1},
  Wiley, New York, \textbf{1978}.
\input{cohen-tannoudji-1}

\bibitem{mcweeny_spins_2004}
R.~McWeeny, \emph{{S}pins in {C}hemistry}, Dover Publications, Mineola, N.Y.,
  \textbf{2004}.
\input{mcweeny_spins_2004}

\bibitem{reiher_relativistic_2009}
M.~Reiher, A.~Wolf, \emph{{R}elativistic {Q}uantum {C}hemistry: {T}he
  {F}undamental {T}heory of {M}olecular {S}cience}, Wiley-VCH, Weinheim,
  \textbf{2009}.
\input{reiher_relativistic_2009}

\bibitem{szabo-ostlund}
A.~Szabo, N.~S. Ostlund, \emph{{M}odern {Q}uantum {C}hemistry}, Dover
  Publications, Mineola, N.Y., \textbf{1996}.
\input{szabo-ostlund}

\bibitem{dirac_quantum_1929}
P.~A.~M. Dirac, \emph{Proc. Roy. Soc. Ser. A} \textbf{1929}, \emph{123},
  714--733.
\input{dirac_quantum_1929}

\bibitem{mcweeny_methods_1969}
R.~McWeeny, B.~T. Sutcliffe, \emph{{M}ethods of {M}olecular {Q}uantum
  {M}echanics}, Academic Press, New York, \textbf{1969}.
\input{mcweeny_methods_1969}

\bibitem{lwdin_quantum_1955}
P.-O. L\"{o}wdin, \emph{Phys. Rev.} \textbf{1955}, \emph{97}, 1490.
\input{lwdin_quantum_1955}

\bibitem{heisenberg-pauli-1}
W.~Heisenberg, \emph{Z. Phys.} \textbf{1926}, \emph{38}, 411--426.
\input{heisenberg-pauli-1}

\bibitem{heisenberg-pauli-2}
W.~Heisenberg, \emph{Z. Phys.} \textbf{1926}, \emph{39}, 499--518.
\input{heisenberg-pauli-2}

\bibitem{matsen_spin-free_1964}
F.~A. Matsen, \emph{Adv. Quantum Chem.} \textbf{1964}, \emph{1}, 59--114.
\input{matsen_spin-free_1964}

\bibitem{pauncz_spin_1979}
R.~Pauncz, \emph{{S}pin {E}igenfunctions}, Plenum Press, New York,
  \textbf{1979}.
\input{pauncz_spin_1979}

\bibitem{pauncz_symmetric_1995}
R.~Pauncz, \emph{{T}he {S}ymmetric {G}roup in {Q}uantum {C}hemistry},
  CRC-Press, Boca Raton, FL, \textbf{1995}.
\input{pauncz_symmetric_1995}

\bibitem{jeschke-book}
A.~Schweiger, G.~Jeschke, \emph{{P}rinciples of {P}ulse {E}lectron
  {P}aramagnetic {R}esonance}, Oxford University Press, \textbf{2001}.
\input{jeschke-book}

\bibitem{kaupp-book}
M.~Kaupp, M.~B\"{u}hl, V.~G. Malkin, \emph{{C}alculation of {N}{M}{R} and
  {E}{P}{R} {P}arameters. {T}heory and {A}pplications}, Wiley-VCH, Weinheim,
  \textbf{2004}.
\input{kaupp-book}

\bibitem{rastrelli_predicting_2009}
F.~Rastrelli, A.~Bagno, \emph{Chem.--Eur. J.} \textbf{2009}, \emph{15},
  7990{\textendash}8004.
\input{rastrelli_predicting_2009}

\bibitem{autschbach_calculation_2011}
J.~Autschbach, S.~Patchkovskii, B.~Pritchard, \emph{J. Chem. Theory Comput.}
  \textbf{2011}, \emph{7}, 2175--2188.
\input{autschbach_calculation_2011}

\bibitem{aquino_scalar_2012}
F.~Aquino, B.~Pritchard, J.~Autschbach, \emph{J. Chem. Theory Comput.}
  \textbf{2012}, \emph{8}, 598--609.
\input{aquino_scalar_2012}

\bibitem{gillet_determination_2007}
J.-M. Gillet, \emph{Acta Cryst. A} \textbf{2007}, \emph{63}, 234--238.
\input{gillet_determination_2007}

\bibitem{zheludev_spin_1994}
A.~Zheludev, V.~Barone, M.~Bonnet, B.~Delley, A.~Grand, E.~Ressouche, P.~Rey,
  R.~Subra, J.~Schweizer, \emph{J. Am. Chem. Soc.} \textbf{1994}, \emph{116},
  2019--2027.
\input{zheludev_spin_1994}

\bibitem{baron_spin-density_1996}
V.~Baron, B.~Gillon, O.~Plantevin, A.~Cousson, C.~Mathoni\`{e}re, O.~Kahn,
  A.~Grand, L.~\"{O}hrstr\"{o}m, B.~Delley, \emph{J. Am. Chem. Soc.}
  \textbf{1996}, \emph{118}, 11822--11830.
\input{baron_spin-density_1996}

\bibitem{pontillon_magnetization_1999}
Y.~Pontillon, A.~Caneschi, D.~Gatteschi, R.~Sessoli, E.~Ressouche,
  J.~Schweizer, E.~Lelievre-Berna, \emph{J. Am. Chem. Soc.} \textbf{1999},
  \emph{121}, 5342--5343.
\input{pontillon_magnetization_1999}

\bibitem{claiser_combined_2005}
N.~Claiser, M.~Souhassou, C.~Lecomte, B.~Gillon, C.~Carbonera, A.~Caneschi,
  A.~Dei, D.~Gatteschi, A.~Bencini, Y.~Pontillon, E.~Leli\`{e}vre-Berna,
  \emph{J. Phys. Chem. B} \textbf{2005}, \emph{109}, 2723--2732.
\input{claiser_combined_2005}

\bibitem{zaharko_spin-density_2010}
O.~Zaharko, P.~J. Brown, M.~Mys{\textquoteright}kiv, \emph{Phys. Rev. B}
  \textbf{2010}, \emph{81}, 172405.
\input{zaharko_spin-density_2010}

\bibitem{mcweeny_density_1961}
R.~McWeeny, Y.~Mizuno, \emph{Proc. Roy. Soc. Ser. A} \textbf{1961}, \emph{259},
  554--577.
\input{mcweeny_density_1961}

\bibitem{davidson_reduced_1976}
E.~R. Davidson, \emph{{R}educed {D}ensity {M}atrices in {Q}uantum {C}hemistry},
  Academic Press, New York, \textbf{1976}.
\input{davidson_reduced_1976}

\bibitem{mazziotti_two-electron_2011}
D.~A. Mazziotti, \emph{Chem. Rev.} \textbf{2011}, \emph{112}, 244--262.
\input{mazziotti_two-electron_2011}

\bibitem{hohenberg-kohn-1964}
P.~Hohenberg, W.~Kohn, \emph{Phys. Rev.} \textbf{1964}, \emph{136}, B864--B871.
\input{hohenberg-kohn-1964}

\bibitem{levy-1979}
M.~Levy, \emph{Proc. Natl. Acad. Sci. U. S. A.} \textbf{1979}, \emph{76},
  6062--6065.
\input{levy-1979}

\bibitem{levy-1982}
J.~P. Perdew, R.~G. Parr, M.~Levy, J.~L. Balduz~Jr., \emph{Phys. Rev. Lett.}
  \textbf{1982}, \emph{49}, 1691.
\input{levy-1982}

\bibitem{lieb_density_1983}
E.~H. Lieb, \emph{Int. J. Quantum Chem.} \textbf{1983}, \emph{24}, 243--277.
\input{lieb_density_1983}

\bibitem{van_leeuwen_density_2003}
R.~van Leeuwen, \emph{Adv. Quantum Chem.} \textbf{2003}, \emph{43}, 25--94.
\input{van_leeuwen_density_2003}

\bibitem{eschrig_fundamentals_2003}
H.~Eschrig, \emph{{T}he {F}undamentals of {D}ensity {F}unctional {T}heory}, 2nd
  Ed., Eagle, Ed. am Gutenbergplatz, Leipzig, \textbf{2003}.
\input{eschrig_fundamentals_2003}

\bibitem{kohn_density_1985}
W.~Kohn, in \emph{Highlights of Condensed-Matter Theory}, (Edited by
  F.~Bassani, F.~Fumi, M.~P. Tosi), Elsevier, Amsterdam, \textbf{1985}, 1--15.
\input{kohn_density_1985}

\bibitem{perdew_fundamental_2009}
J.~P. Perdew, A.~Ruzsinszky, L.~A. Constantin, J.~Sun, G.~I. Csonka, \emph{J.
  Chem. Theory Comput.} \textbf{2009}, \emph{5}, 902--908.
\input{perdew_fundamental_2009}

\bibitem{perdew_self-interaction_1981}
J.~P. Perdew, A.~Zunger, \emph{Phys. Rev. B} \textbf{1981}, \emph{23},
  5048--5079.
\input{perdew_self-interaction_1981}

\bibitem{von_barth_local_1972}
U.~von Barth, L.~Hedin, \emph{J. Phys. C: Solid State Phys.} \textbf{1972},
  \emph{5}, 1629--1642.
\input{von_barth_local_1972}

\bibitem{ayers_legendre-transform_2006}
P.~W. Ayers, W.~Yang, \emph{J. Chem. Phys.} \textbf{2006}, \emph{124}, 224108.
\input{ayers_legendre-transform_2006}

\bibitem{holas_comment_2006}
A.~Holas, R.~Balawender, \emph{J. Chem. Phys.} \textbf{2006}, \emph{125},
  247101.
\input{holas_comment_2006}

\bibitem{eschrig_density_2001}
H.~Eschrig, W.~E. Pickett, \emph{Solid State Commun.} \textbf{2001},
  \emph{118}, 123--127.
\input{eschrig_density_2001}

\bibitem{capelle_nonuniqueness_2001}
K.~Capelle, G.~Vignale, \emph{Phys. Rev. Lett.} \textbf{2001}, \emph{86}, 5546.
\input{capelle_nonuniqueness_2001}

\bibitem{gl_derivative_2010}
T.~G\'{a}l, P.~Geerlings, \emph{Phys. Rev. A} \textbf{2010}, \emph{81}, 032512.
\input{gl_derivative_2010}

\bibitem{gl_energy_2010}
T.~G\'{a}l, P.~Geerlings, \emph{J. Chem. Phys.} \textbf{2010}, \emph{133},
  144105.
\input{gl_energy_2010}

\bibitem{gidopoulos_potential_2007}
N.~I. Gidopoulos, \emph{Phys. Rev. B} \textbf{2007}, \emph{75}, 134408--8.
\input{gidopoulos_potential_2007}

\bibitem{gl_differentiability_2007}
T.~G\'{a}l, \emph{Phys. Rev. B} \textbf{2007}, \emph{75}, 235119--5.
\input{gl_differentiability_2007}

\bibitem{gl_nonuniqueness_2009}
T.~G\'{a}l, P.~W. Ayers, F.~De~Proft, P.~Geerlings, \emph{J. Chem. Phys.}
  \textbf{2009}, \emph{131}, 154114.
\input{gl_nonuniqueness_2009}

\bibitem{YZA-2000}
W.~Yang, Y.~Zhang, P.~W. Ayers, \emph{Phys. Rev. Lett.} \textbf{2000},
  \emph{84}, 5172.
\input{YZA-2000}

\bibitem{cohen_insights_2008}
A.~J. Cohen, P.~Mori-Sanchez, W.~Yang, \emph{Science} \textbf{2008},
  \emph{321}, 792--794.
\input{cohen_insights_2008}

\bibitem{cohen_fractional_2008}
A.~J. Cohen, P.~Mori-Sanchez, W.~Yang, \emph{J. Chem. Phys.} \textbf{2008},
  \emph{129}, 121104.
\input{cohen_fractional_2008}

\bibitem{gunnarsson_exchange_1976}
O.~Gunnarsson, B.~I. Lundqvist, \emph{Phys. Rev. B} \textbf{1976}, \emph{13},
  4274--4298.
\input{gunnarsson_exchange_1976}

\bibitem{carter-kin-review}
Y.~A. Wang, E.~A. Carter, in \emph{Theoretical Methods in Condensed Phase
  Chemistry}, (Edited by S.~D. Schwartz), Kluwer, Dordrecht, \textbf{2000},
  117--184.
\input{carter-kin-review}

\bibitem{xia_can_2012}
J.~Xia, C.~Huang, I.~Shin, E.~A. Carter, \emph{J. Chem. Phys.} \textbf{2012},
  \emph{136}, 084102.
\input{xia_can_2012}

\bibitem{kohn-sham-1965}
W.~Kohn, L.~J. Sham, \emph{Phys. Rev.} \textbf{1965}, \emph{140}, A1133--A1138.
\input{kohn-sham-1965}

\bibitem{levy_electron_1982}
M.~Levy, \emph{Phys. Rev. A} \textbf{1982}, \emph{26}, 1200, copyright (C) 2009
  The American Physical Society; Please report any problems to prola@aps.org,
  {DOI}: 10.1103/PhysRevA.26.1200.
\input{levy_electron_1982}

\bibitem{englisch_exact_1984}
H.~Englisch, R.~Englisch, \emph{Phys. Status Solidi B} \textbf{1984},
  \emph{124}, 373--379.
\input{englisch_exact_1984}

\bibitem{schipper_one_1998}
P.~R.~T. Schipper, O.~V. Gritsenko, E.~J. Baerends, \emph{Theor. Chem. Acc.}
  \textbf{1998}, \emph{99}, 329--343.
\input{schipper_one_1998}

\bibitem{morrison_electron_2002}
R.~C. Morrison, \emph{J. Chem. Phys.} \textbf{2002}, \emph{117}, 10506--10511.
\input{morrison_electron_2002}

\bibitem{katriel_study_2004}
J.~Katriel, S.~Roy, M.~Springborg, \emph{J. Chem. Phys.} \textbf{2004},
  \emph{121}, 12179--12190.
\input{katriel_study_2004}

\bibitem{pople_spin-unrestricted_1995}
J.~A. Pople, P.~M.~W. Gill, N.~C. Handy, \emph{Int. J. Quantum Chem.}
  \textbf{1995}, \emph{56}, 303--305.
\input{pople_spin-unrestricted_1995}

\bibitem{chipman_theoretical_1983}
D.~M. Chipman, \emph{J. Chem. Phys.} \textbf{1983}, \emph{78}, 3112--3132.
\input{chipman_theoretical_1983}

\bibitem{chipman_spin_1992}
D.~M. Chipman, \emph{Theor. Chem. Acc.} \textbf{1992}, \emph{82}, 93--115.
\input{chipman_spin_1992}

\bibitem{wang_evaluation_1995}
J.~Wang, A.~D. Becke, V.~H. Smith~Jr., \emph{J. Chem. Phys.} \textbf{1995},
  \emph{102}, 3477--3480.
\input{wang_evaluation_1995}

\bibitem{cohen_evaluation_2007}
A.~J. Cohen, D.~J. Tozer, N.~C. Handy, \emph{J. Chem. Phys.} \textbf{2007},
  \emph{126}, 214104.
\input{cohen_evaluation_2007}

\bibitem{daul_density_1994}
C.~Daul, \emph{Int. J. Quantum Chem.} \textbf{1994}, \emph{52}, 867--877.
\input{daul_density_1994}

\bibitem{daul_calculation_1995}
C.~A. Daul, K.~G. Doclo, A.~C. St\"{u}ckl, in \emph{Recent Advances in Density
  Functional Methods, Part 2}, (Edited by D.~P. Chong), World Scientific,
  Singapore, \textbf{1995}, 61--113.
\input{daul_calculation_1995}

\bibitem{filatov_spin-restricted_1998}
M.~Filatov, S.~Shaik, \emph{Chem. Phys. Lett.} \textbf{1998}, \emph{288},
  689--697.
\input{filatov_spin-restricted_1998}

\bibitem{filatov_spin-restricted_1999}
M.~Filatov, S.~Shaik, \emph{Chem. Phys. Lett.} \textbf{1999}, \emph{304},
  429--437.
\input{filatov_spin-restricted_1999}

\bibitem{illas_spin_2006}
F.~Illas, I.~Moreira, J.~Bofill, M.~Filatov, \emph{Theor. Chem. Acc.}
  \textbf{2006}, \emph{116}, 587--597.
\input{illas_spin_2006}

\bibitem{frank_molecular_1998}
I.~Frank, J.~Hutter, D.~Marx, M.~Parrinello, \emph{J. Chem. Phys.}
  \textbf{1998}, \emph{108}, 4060--4069.
\input{frank_molecular_1998}

\bibitem{grimm_restricted_2003}
S.~Grimm, C.~Nonnenberg, I.~Frank, \emph{J. Chem. Phys.} \textbf{2003},
  \emph{119}, 11574--11584.
\input{grimm_restricted_2003}

\bibitem{nonnenberg_restricted_2003}
C.~Nonnenberg, S.~Grimm, I.~Frank, \emph{J. Chem. Phys.} \textbf{2003},
  \emph{119}, 11585--11590.
\input{nonnenberg_restricted_2003}

\bibitem{della_sala_open-shell_2003}
F.~Della~Sala, A.~G\"{o}rling, \emph{J. Chem. Phys.} \textbf{2003}, \emph{118},
  10439--10454.
\input{della_sala_open-shell_2003}

\bibitem{vitale_open-shell_2005}
V.~Vitale, F.~Della~Sala, A.~G\"{o}rling, \emph{J. Chem. Phys.} \textbf{2005},
  \emph{122}, 244102.
\input{vitale_open-shell_2005}

\bibitem{bagus_singlettriplet_1975}
P.~S. Bagus, B.~I. Bennett, \emph{Int. J. Quantum Chem.} \textbf{1975},
  \emph{9}, 143--148.
\input{bagus_singlettriplet_1975}

\bibitem{ziegler_calculation_1977}
T.~Ziegler, A.~Rauk, E.~J. Baerends, \emph{Theor. Chim. Acta} \textbf{1977},
  \emph{43}, 261--271.
\input{ziegler_calculation_1977}

\bibitem{perdew_escaping_1995}
J.~P. Perdew, A.~Savin, K.~Burke, \emph{Phys. Rev. A} \textbf{1995}, \emph{51},
  4531.
\input{perdew_escaping_1995}

\bibitem{fux_electron_2012}
S.~Fux, M.~Reiher, \emph{Struct. Bond.} \textbf{2012}, \emph{147}, 99--142.
\input{fux_electron_2012}

\bibitem{rel-hk}
A.~K. Rajagopal, J.~Callaway, \emph{Phys. Rev. B} \textbf{1973}, \emph{7},
  1912.
\input{rel-hk}

\bibitem{macdonald_relativistic_1979}
A.~H. MacDonald, S.~H. Vosko, \emph{J. Phys. C: Solid State Phys.}
  \textbf{1979}, \emph{12}, 2977--2990.
\input{macdonald_relativistic_1979}

\bibitem{engel_relativistic_2002}
E.~Engel, in \emph{Relativistic Electronic Structure Theory Part 1:
  Fundamentals}, (Edited by P.~Schwerdtfeger), Elsevier, Amsterdam,
  \textbf{2002}, 523--621.
\input{engel_relativistic_2002}

\bibitem{engel_relativistic_2003}
E.~Engel, R.~M. Dreizler, S.~Varga, B.~Fricke, in \emph{Relativistic Effects in
  Heavy-Element Chemistry and Physics}, (Edited by B.~A. Hess), Wiley,
  Chichester, \textbf{2003}, 123--161.
\input{engel_relativistic_2003}

\bibitem{saue_fourcomponent_2002}
T.~Saue, T.~Helgaker, \emph{J. Comput. Chem.} \textbf{2002}, \emph{23},
  814--823.
\input{saue_fourcomponent_2002}

\bibitem{engel_relativistic_1996}
E.~Engel, R.~Dreizler, \emph{Top. Curr. Chem.} \textbf{1996}, \emph{181},
  1--80.
\input{engel_relativistic_1996}

\bibitem{van_wllen_relativistic_2010}
C.~Van~W\"{u}llen, in \emph{Relativistic Methods for Chemists}, (Edited by
  M.~Barysz, Y.~Ishikawa), Springer, Dordrecht, Challenges and Advances in
  Computational Chemistry and Physics, \textbf{2010}, 191--214.
\input{van_wllen_relativistic_2010}

\bibitem{baym_lectures_1969}
G.~Baym, \emph{{L}ectures {O}n {Q}uantum {M}echanics}, Benjamin--Cummings, New
  York, \textbf{1969}.
\input{baym_lectures_1969}

\bibitem{van_wllen_relativistic_1999}
C.~Van~W\"{u}llen, \emph{J. Comput. Chem.} \textbf{1999}, \emph{20}, 51--62.
\input{van_wllen_relativistic_1999}

\bibitem{van_wllen_spin_2002}
C.~Van~W\"{u}llen, \emph{J. Comput. Chem.} \textbf{2002}, \emph{23}, 779--785.
\input{van_wllen_spin_2002}

\bibitem{scalmani_new_2012}
G.~Scalmani, M.~J. Frisch, \emph{J. Chem. Theory Comput.} \textbf{2012},
  \emph{8}, 2193--2196.
\input{scalmani_new_2012}

\bibitem{peng_exact_2012}
D.~Peng, M.~Reiher, \emph{Theor. Chem. Acc.} \textbf{2012}, \emph{131}, 1081.
\input{peng_exact_2012}

\bibitem{mastalerz_douglaskrollhess_2008}
R.~Mastalerz, R.~Lindh, M.~Reiher, \emph{Chem. Phys. Lett.} \textbf{2008},
  \emph{465}, 157--164.
\input{mastalerz_douglaskrollhess_2008}

\bibitem{baerends-jpca-holes}
E.~J. Baerends, O.~V. Gritsenko, \emph{J. Phys. Chem. A} \textbf{1997},
  \emph{101}, 5383--5403.
\input{baerends-jpca-holes}

\end{thebibliography}
\end{document}